\documentclass[11pt,a4paper]{article}
\usepackage{jheppub_kim}
\usepackage{rotating}
\usepackage{graphicx,epsfig}
\usepackage{amsmath}
\usepackage {amssymb}
\usepackage{subfigure}

\usepackage{relsize}

\usepackage{array,multirow}
\usepackage{soul}
\usepackage{subfigure}
\usepackage{dsfont}
\usepackage{hyperref}
\usepackage{txfonts}
\usepackage{newlfont}
\usepackage{times}
\newcommand\overcirc[1]{{\mathop{#1}\limits^{\circ}}}

\title{\boldmath Phase Portraits of general $f(T)$ Cosmology}

\author[a,b]{A. Awad}
\author[c,d]{W. El Hanafy}
\author[c,d,e]{G.G.L. Nashed}
\author[f,g,h]{Emmanuel N. Saridakis}

\affiliation[a]{Department of Physics, School of Sciences and Engineering, American
University in
Cairo, P.O. Box 74, AUC Avenue New Cairo, Cairo, Egypt}
\affiliation[b]{Physics Department, Faculty of Science, Ain Shams University, Cairo 11566,
Egypt}

\affiliation[c]{Centre for Theoretical Physics, the British University in Egypt, P.O. Box
43, El
Sherouk City, Cairo 11837, Egypt}
\affiliation[d]{Egyptian Relativity Group (ERG), Cairo University, Giza 12613, Egypt}

\affiliation[e]{Mathematics Department, Faculty of Science, Ain Shams University, Cairo
11566,Egypt}
\affiliation[f]{Chongqing University of Posts \& Telecommunications, Chongqing,
400065,China}
\affiliation[g]{Department of Physics, National Technical University of Athens, Zografou
Campus GR
157 73, Athens, Greece}
\affiliation[h]{CASPER, Physics Department, Baylor University, Waco, TX 76798-7310, USA}

\emailAdd{adel.awad@bue.edu.eg}
\emailAdd{waleed.elhanafy@bue.edu.eg}
\emailAdd{nashed@bue.edu.eg}
\emailAdd{Emmanuel\_Saridakis@baylor.edu}

\abstract{
We use dynamical system methods to explore the general behaviour of $f(T)$ cosmology.
In contrast to the standard  applications of dynamical analysis, we present a way to
transform the equations into a one-dimensional autonomous system, taking advantage of the
crucial property that the torsion scalar in flat FRW geometry is just a function of the
Hubble function, thus the field equations include only up to first derivatives of it,
and therefore
in a general $f(T)$ cosmological scenario  every quantity is expressed only in terms of
the Hubble function. The great advantage is that for one-dimensional systems it is
easy to construct the phase space portraits,  and thus extract information and explore in
detail the  features and possible behaviours of  $f(T)$ cosmology. We utilize the phase
space portraits and we show that $f(T)$ cosmology can describe the universe evolution in
agreement with observations, namely starting from a Big Bang singularity, evolving into
the subsequent thermal history and the matter domination, entering into a late-time
accelerated  expansion, and resulting to the de Sitter phase in the far future.
Nevertheless, $f(T)$ cosmology can present a rich class of more exotic behaviours, such
as the cosmological bounce and turnaround, the phantom-divide crossing, the Big Brake
and the Big Crunch, and it may exhibit various singularities, including the non-harmful
ones of type II
and type IV. We study the phase space of three specific viable $f(T)$ models offering a
complete picture. Moreover, we present a new model of $f(T)$ gravity that can lead to a
universe in agreement with observations, free of perturbative instabilities, and applying
the  Om(z) diagnostic test we confirm that it is in agreement with the combination of
SNIa, BAO and CMB data at 1$\sigma$ confidence
level.}

\begin{document}
\maketitle
\flushbottom
\section{Introduction}\label{Sec1}
Current astrophysical and cosmological observations are essential to consummate and
sharpen our
knowledge of the fundamental constituents of the universe. Cosmic microwave background
(CMB)
anisotropies observations provides us with a substantial information on the physics of
the
primordial universe, which is essential in understanding, as well as constraining,
physical models
explaining early-time cosmology. The CMB anisotropies have been measured by COBE, WMAP and
Planck
satellites with high precisions
\cite{Smoot:1992td,Spergel:2003ApJS,Bennett:2013ApJS,Ade:2015lrj,
Adam:2015rua}. The observations have shown that baryonic matter constitutes only a small
portion ($\sim 5\%$) of the universe's energy content. However, observations of the
rotation galaxy curves \cite{Rubin:1978ApJL}, galaxy clustering \cite{Faber:1979ARAA} and
galaxy X-ray emission \cite{Fabricant:1980ApJ} have shown that $\sim 26 \%$ of the
universe's total energy is dark matter. Dark
matter is needed for forming clusters and large scale structures early enough, since
baryonic
matter alone cannot explain the existence of these structures at high redshifts. On the
other hand, the luminosity distance observations \cite{Riess:1998AJ, Perlmutter:1999ApJ}
of type Ia supernovae (SNIa), affirmed by CMB observations
\cite{Spergel:2003ApJS,Bennett:2013ApJS,Ade:2015lrj,Adam:2015rua},
 have come
up with another unexpected result, namely that the cosmic expansion has passed from
deceleration to
acceleration a few billion years ago, i.e at redshift $z_{tr} \sim 0.6-0.8$
\cite{Farooq:2016zwm}.
Since none of the known matter fields can explain this accelerated expansion, cosmologists
have
assumed new obscure cosmic species, that are collectively called dark energy, to explain
it. This
component represents $\sim 69 \%$ of the universe's total energy.

The simplest and basic scenario which incorporates the dark energy component is a
cosmological
constant ($\Lambda$) universe. With the addition of cold dark matter (CDM) and the
consideration of
a flat Friedmann-Robertson-Walker (FRW) geometry (since the spatial flat geometry gives a
good
agreement with temperature power spectrum of Planck observations), this scenario is the
so-called $\Lambda$CDM one. Although the $\Lambda$CDM paradigm is the one that fits the
data in the most
efficient way, the possibility of the dynamical nature of the dark energy component, as
well as
possible tensions relating to the direct observations of the Hubble function
\cite{DiValentino:2017gzb,Freedman:2017yms,DiValentino:2016hlg}, might indicate towards
dynamical dark energy models,
unless unknown uncertainties will be discovered \cite{Zhao:2017cud,DiValentino:2017zyq}.
Dynamical
dark energy models could explain the late accelerated expansion by introducing various
extra fields
(for reviews see \cite{Copeland:2006wr,Cai:2009zp}), or by assuming fluids with exotic
equation of
state, such as the Chaplygin fluid.

On the other hand, modified gravity may provide an alternative approach to interpret the
accelerated
 expansion. In particular, one wishes to construct a gravitational modification, that
includes
extra degrees of freedom that can drive the universe acceleration, at early or late times,
which
however still possesses general relativity as a particular limit. Although most of the
works in
modified gravity start from the standard curvature-based formulation (for reviews see
\cite{Nojiri:2006ri,Capozziello:2011et}),
one can alternatively construct modified gravities based on torsion \cite{Cai:2015emx}.
In
particular, starting from the Teleparallel Equivalent of General Relativity (TEGR)
\cite{Pereira.book}, in which the Lagrangian is the torsion scalar $T$,
one can extend $T$ to $f(T)$ resulting to $f(T)$ gravity
\cite{Bengochea:2008gz,Linder:2010py}.
Note that although TEGR is completely equivalent with general relativity at the level of
equations,
$f(T)$ gravity is different from $f(R)$ gravity. Thus, $f(T)$ cosmology proves to be very
interesting both for early-time
\cite{Ferraro:2006jd,Ferraro:2008ey,Bamba:2014zra,Bamba:2016wjm}
as well as late-time
\cite{Chen:2010va,Tsyba:2010ji,Yang:2010hw,Bamba:2010iw,
Myrzakulov:2010tc,Bamba:2010wb,Dent:2011zz,Cai:2011tc,Sharif:2011bi,Wei:2011mq,Wu:2011xa,
Karami:2011np,
Jamil:2011mc,Karami:2012if,Setare:2012vs,Dong:2012en,Tamanini:2012hg,Daouda:2012wt,
Banijamali:2012nx,Darabi:2012zh,Liu:2012fk,Setare:2012ry,Cardone:2012xq,
Chattopadhyay:2012eu,
Bamba:2012ka,
Setare:2013xh,Li:2013xea,Camera:2013bwa,Bamba:2013fta,Basilakos:2013rua,Qi:2014yxa,
Nashed:2014baa,
Darabi:2014dla,
Nashed:2015pda,Fayaz:2015yka,Abedi:2015cya,Krssak:2015oua,Pan:2016jli,
Paliathanasis:2016vsw,Geng:2011aj,Xu:2012jf,Kofinas:2014owa,Haro:2014wha,Hanafy:2014ica,
Sharif:2014CoTPh,Hanafy:2015lda,Capozziello:2015rda,Nunes:2016qyp,Nunes:2016plz,
Qi:2017xzl,
Awad:2017ign,
Oikonomou:2017isf,Jawad:2017ApSS,Capozziello:2017bxm,Karpathopoulos:2017arc}
universe evolution, while the black hole solutions in this framework also lead to
interesting features
\cite{Miao:2011ki,Capozziello:2012zj,
Bhadra:2013IJTP,Rodrigues:2013IJMPD,Aftergood:2014wla,Paliathanasis:2014iva,
Nashed:2015JPSJ,
Junior:2015dga,Nashed:2015EPJP,Junior:2015fya,Nashed:2016tbj,Ahmed:2016cuy,
Rodrigues:2016uor,Nashed:2017GrCo,Mai:2017riq,Awad:2017tyz,Capozziello:2017uam}.

One of the important features of both general relativity, as well as modified gravity, is
that the
highly nonlinear nature of the corresponding cosmological equations in general does not
allow for
the extraction of analytical solutions. Nevertheless, one can apply the dynamical systems
method
\cite{Copeland:1997et,Ferreira:1997au,Coley:2003mj,Leon2011} which allows to extract the
global
behaviour of the scenario, independently of the initial conditions and bypassing the
complexity of
the equations. The dynamical systems method
serves as an important tool to describe, analyze and classify various features of
cosmological
models, and additionally it makes more transparent the  relation between different
branches of
solutions. In particular, in cosmological applications we are interested in the following
set of
ordinary differential equations
\begin{equation}\label{ODE}
\frac{d\mathbf{x}}{dt}=\mathbf{f(x)},
\end{equation}
where $\mathbf{x}= \left(x_{1}, \cdots, x_{n}\right)$ are the variables needed to
characterize the
system and the functions $\mathbf{f(x)}= \left(f_{1}(x_{1},\cdots x_{n})\right.$,
$\cdots$, $\left.
f_{n}(x_{1}, \cdots, x_{n})\right)$ are determined by the system. In general, in most
nonlinear systems analytical solutions cannot be extracted. When there is no explicit
dependence of $\mathbf{f(x)}$ on $t$, the system is \textit{autonomous}, and then the
geometric approach which has been developed is
applied in order to study the qualitative behaviour and the stability. In this
framework one
uses topological and geometrical   procedures in order to determine the
properties of the set of all solutions, by visualizing it as \textit{trajectories} in a
\textit{phase space}. In this sense, the
phase of the system at instant time can be described by a \textit{phase vector}
$\mathbf{x} \in \mathbf{X} \subseteq \mathds{R}^{n}$, where $\mathbf{X}$ is an
$n$-dimensional phase space  and the
maps $\mathbf{f}: \mathbf{X}\to\mathbf{X}$ are \textit{vector fields} on
$\mathds{R}^{n}$. Hence, one visualizes the phase space with frozen trajectories
on it, and by using geometrical reasoning he can extract essential information
about the system even without solving it explicitly.

In general applications of dynamical system methods in given cosmological scenarios, one
results to a multi-dimensional system that needs to be investigated
\cite{Coley:2003ASSL,Ellis:2005CUP,Chen:2008ft,Leon2011,awad2013,Boehmer:2014vea,
Kofinas:2014aka,
Odintsov:2017icc}. Analyzing multi-dimensional autonomous systems can be a complicated
task, since they incorporate a large amount of information, with many solution
branches and possible behaviours. However, $f(T)$ cosmology is a very interesting
exception, since in a flat FRW geometry its corresponding dynamical system (\ref{ODE})
can be reduced to a \textit{one-dimensional (or first order) autonomous system}, i.e
having $n=1$. The reason behind this crucial and very helpful property is the special
feature of $f(T)$ gravity, namely that the torsion scalar in flat FRW geometry is just
$T=-6H^2$, i.e it can be used interchangeably with the Hubble function $H$ (in turn this
is a result from the most general feature of $f(T)$ gravity, namely that it has
second-order field equations, in contrast with most models of modified gravity which
include higher-order field equations). Hence, a general $f(T)$ cosmological scenario
can be described by a one-dimensional autonomous system, where everything is expressed
in terms of the Hubble function, and thus its possible behaviours
can be extracted and investigated in huge detail.

In the present work we are interested in using the above property of $f(T)$ cosmology to
transform the cosmological equations into a one-dimensional autonomous system, and then
explore its features in general. We mention here that the phase space analysis of $f(T)$
cosmology has been performed in the literature
\cite{Wu:2010xk,Skugoreva:2014ena,Carloni:2015lsa,Bamba:2016gbu,Awad:2017sau,
ElHanafy:2017sih, ElHanafy:2017xsm}, however it was based on the usual approach and thus
it was performed only for particular specific $f(T)$ models (see also
\cite{Hohmann:2017jao,Mirza:2017vrk}, where a general analysis is performed but in the
framework of multi-dimensional systems).
On the other hand, in the present work, due to the above formalism, we are able to
explore
the phase space portraits in full detail and for general $f(T)$ cosmology.
After this general analysis, we proceed to the investigation of specific viable $f(T)$
models, reproducing the results of the literature, providing a unified and full picture
of $f(T)$ cosmology.

We organize the manuscript as follows: In Sec. \ref{Sec2}, we review briefly the
essential background of $f(T)$ gravity, and we apply it in a cosmological framework.
In Sec. \ref{Sec3}, we present the theory of cosmological phase portraits and we show that
$f(T)$ cosmology has the crucial property to result in a one-dimensional autonomous
system. Then we study the basic features of the phase portraits in general $f(T)$
cosmology. In Sec. \ref{Sec4}, we explore the phase space portraits of three viable
specific $f(T)$ models, investigating all possible cosmological evolutions. In Sec.
\ref{Sec5}, taking into account the information gained through the phase space portraits,
we propose a new $f(T)$ model and we show that it can successfully describe the universe
evolution in agreement with observations. Finally, in Sec. \ref{Sec6} we summarize the
obtained results.

\section{$f(T)$ gravity and cosmology}\label{Sec2}
In this section we briefly review $f(T)$ gravity and we apply it in a cosmological
framework. We start from a 4-dimensional smooth manifold $\mathcal{M}$, and a given a
set of tetrad fields $e_{a\mu}$ defined on $\mathcal{M}$, where the Latin letters denote
the local Lorentz (tangent space) indices, and the Greek letters denote the tensor
(spacetime) indices. The local Lorentz and the
tensor
indices can be interchanged using
\begin{equation}\label{orthonormal}
e^{a}{_\nu}e_{a}{^\mu}=\delta^{\mu}_{\nu},\quad e^{a}{_\mu}e_{b}{^\mu}=\delta^{a}_{b}.
\end{equation}
One can construct the metric tensor
\begin{equation}\label{metric}
g_{\mu\nu}=\eta_{ab}e^{a}{_\mu}e^{b}{_\nu},
\end{equation}
where $\eta_{ab}=diag(1,-1,-1,-1)$ is the metric tensor of the tangent space.
In the above manifold one can introduce the Levi-Civita connection
\begin{equation}\label{Levi-Civita}
\overcirc{\Gamma}{^{\alpha}}{_{\mu\nu}}= \frac{1}{2} g^{\alpha
\sigma}\left(\partial_{\nu}g_{\mu \sigma}+\partial_{\mu}g_{\nu
\sigma}-\partial_{\sigma}g_{\mu \nu}\right),
\end{equation}
which satisfies the metricity condition
$
\overcirc{\nabla}_{\sigma}g_{\mu\nu}=0$,
where the operator $\overcirc{\nabla}$ denotes the covariant derivative of the
Levi-Civita
connection. The corresponding torsion tensor $\overcirc{T}{^\alpha}{_{\mu\nu}}$ vanishes
identically, since the connection $\overcirc{\Gamma}{^{\alpha}}{_{\mu\nu}}$ is
symmetric, while its
curvature
tensor is the usual one, namely
$
\overcirc{R}{^\alpha}{_{\sigma\mu\nu}}=\partial_{\mu}\overcirc{\Gamma}{^{\alpha}}{_{
\sigma\nu}}
-\partial_{\nu}\overcirc{\Gamma}{^{\alpha}}{_{\sigma\mu}}
+\overcirc{\Gamma}{^{\alpha}}{_{\lambda\mu}} \overcirc{\Gamma}{^{\lambda}}{_{\sigma\nu}}
-\overcirc{\Gamma}{^{\alpha}}{_{\lambda\nu}}
\overcirc{\Gamma}{^{\lambda}}{_{\sigma\mu}}$, and non-vanishing. As usual, in general
relativity one constructs the Ricci scalar $\overcirc{R}$, through contractions of this
curvature tensor, and uses it as the Lagrangian that describes the gravitational field.

On the other hand, one can alternatively introduce the Weitzenb\"{o}ck
connection, constructed directly from the
tetrad fields as
\begin{equation}
\label{W-connection}
\Gamma^{\alpha}{_{\mu\nu}}=e_{a}{^\alpha}\partial_{\nu}e^{a}{_\mu}=-e^{a}{_\mu}\partial_{
\nu}e_{a}{
^\alpha}.
\end{equation}
This connection defines auto-parallelism, where the covariant derivative of
the tetrad fields vanishes
\begin{equation}\label{AP-condition}
\nabla_{\sigma}e^{a}{_\mu}=0,
\end{equation}
($\nabla$ denotes the covariant derivative corresponding to the Weitzenb\"{o}ck
connection), and thus we directly deduce that the Weitzenb\"{o}ck
connection is a metric one:
\begin{equation}\label{W-metricity}
\nabla_{\sigma}g_{\mu\nu}=0.
\end{equation}
The important feature of this connection is that it has vanishing curvature tensor
$R^{\alpha}{_{\sigma\mu\nu}}=0$, while its
torsion tensor is given by
\begin{equation}\label{Torsion}
T^{\alpha}{_{\mu\nu}}=\Gamma^{\alpha}{_{\nu\mu}}-\Gamma^{\alpha}{_{\mu\nu}}=e{_a}{^\alpha}
\left(\partial_{\mu} e{^a}{_\nu}-\partial_{\nu} e{^a}{_\mu}\right).
\end{equation}
Additionally, we can define the contortion
tensor
$
K^{\alpha}{_{\mu\nu}}=\Gamma^{\alpha}{_{\mu\nu}}-\overcirc{\Gamma}{^\alpha}{_{\mu\nu}}=e_{
a}{^\alpha}
\overcirc{\nabla}_{\nu}e^{a}{_\mu}$,
which can be expressed as
\begin{equation}
\label{Torsion-Contortion}
K^{\alpha}{_{\mu\nu}}=\frac{1}{2}\left(T_{\mu}{^\alpha}{_\nu}+T_{\nu}{^\alpha}{_\mu}-T^{
\alpha}{_{\mu\nu}}\right).
\end{equation}
Using contractions of the torsion tensor one can construct the teleparallel torsion scalar
as
\begin{equation}\label{torsionscalar1}
T=\frac{1}{4}T^{\alpha}{_{\mu\nu}}T_{\alpha}{^{\mu\nu}}+\frac{1}{2}
T^{\alpha}{_{\mu\nu}}T^{\nu\mu}{_\alpha}-T_{\mu}T^{\mu}\equiv S_{\alpha}{^{\mu
\nu}}T^{\alpha}{_{\mu \nu}},
\end{equation}
where for convenience we have introduced the superpotential
\begin{equation}\label{superpotential}
S_{\alpha}{^{\mu \nu}}=\frac{1}{4}\left(T_{\alpha}{^{\mu
\nu}}+T{^\mu}{_\alpha}{^\nu}-T{^\nu}{_
\alpha}{^\mu}\right)
+\frac{1}{2}\left(\delta^{\nu}_{\alpha}T^{\mu}-\delta^{\mu}_{\alpha}T^{\nu}\right),
\end{equation}
which is skew symmetric in the last pair of indices. One can straightforwardly see that
\begin{equation}\label{identity}
e\overcirc{R}\equiv -eT+2\partial_{\mu}(eT^{\mu}),
\end{equation}
with $e=\sqrt{-g}=\det\left(e{^a}{_\mu}\right)$, and thus when $T$ is used as a
gravitational Lagrangian it will lead to the same equations with the use of the Ricci
scalar $\overcirc{R}$. That is why the torsional theory characterized by the action
\begin{equation}\label{TEGR-action}
\mathcal{S}_{TEGR}=\frac{1}{2\kappa^2}\int d^{4}x~ e \left(T+\Lambda\right),
\end{equation}
with
$\kappa^2=8\pi G$ and $\Lambda$ a (cosmological) constant,
is called Teleparallel Equivalent of General
Relativity (TEGR).

One can be inspired by the $f(\overcirc{R})$ extensions of the Einstein-Hilbert action,
and extend $T$ to $f(T)$, i.e use the action \cite{Cai:2015emx}
\begin{equation}\label{action}
{\mathcal S}=\int d^{4}x~ e\left[\frac{1}{2 \kappa^2}f(T)\right],
\end{equation}
resulting to the $f(T)$ gravity.
Adding also the action for the matter sector, and varying the action with
respect to the vierbein, gives rise to the field equations, namely \cite{BF09,L10}
\begin{equation}\label{field_eqns}
\frac{1}{e} \partial_\mu \left( e S_a^{\verb| |\mu\nu} \right) f_{T}-e_a^\lambda
T^\rho_{\verb| |\mu \lambda} S_\rho^{\verb| |\nu\mu}f_{T}+S_a^{\verb| |\mu\nu}
\partial_\mu T f_{
TT}
+\frac{1}{4} e_a^\nu f(T)=\frac{{\kappa}^2}{2} e_a^\mu \mathfrak{T}_\mu^{\verb| |\nu},
\end{equation}
with $f_{T}:=\frac{df}{dT}$ and $f_{TT}:=\frac{d^{2}f}{dT^2}$, and where
${\mathfrak{T}_{\mu}}^{\nu}=e{
^a}{_\mu}\left(-\frac{1}{e}\frac{\delta \mathcal{L}_{m}}{\delta e{^a}{_\nu}}\right)$ is
the
energy-momentum tensor of the matter fields.

$f(T)$ gravity exhibits interesting properties. In particular, through a  Hamiltonian
analysis one can show that in  $D$ spacetime dimensions it has $D - 1$ extra
degrees of freedom, corresponding to one massive vector field or one
massless vector field with one scalar field \cite{Li:2011rn}. Furthermore, although
the theory has second-order field equations at the background level, at the perurbation
level instabilities could arise, and thus one must impose specific conditions for the
absence of ghost and Laplacian instabilities \cite{Chen:2010va,Dent:2011zz}.

In order to apply $f(T)$ gravity in a cosmological framework we impose the
homogeneous and isotropic geometry
\begin{equation}\label{tetrad}
{e_{\mu}}^{a}=\textmd{diag}(1,a(t),a(t),a(t)),
\end{equation}
which corresponds to the flat Friedmann-Robertson-Walker (FRW) metric
\begin{equation}
ds^2= dt^2-a^2(t)\, \delta_{ij} dx^i dx^j,
\end{equation}
where $a(t)$ is the scale factor.
Calculating the torsion scalar $T$ from (\ref{torsionscalar1}) for the vierbein choice
(\ref{tetrad}), gives
\begin{equation}\label{TorHubble}
T=-6H^2,
\end{equation}
where $H\equiv \dot{a}/a$ is the Hubble function and dots denote derivative with respect
to $t$. Hence, in $f(T)$ gravity there is a relation that directly relates
$T$ (whose arbitrary function is the Lagrangian of the theory) to $H$. As we
mentioned in the Introduction, this relation is crucial for the present work.

Inserting the vierbein choice (\ref{tetrad}) into the general field equations
(\ref{field_eqns}), we acquire the two modified Friedmann equations, namely
\begin{eqnarray}
{H}^2& =& \frac{\kappa^2}{3} \left( \rho+ \rho_{ T} \right), \label{MFR1}\\
2 \dot{{H}} + 3{H}^2&=& - \kappa^2 \left(p+p_{ T }\right),\label{MFR2}
\end{eqnarray}
where $\rho$ and $p$ are respectively the energy density and pressure of the matter
sector, considered to correspond to a perfect fluid with equation-of-state parameter
$w\equiv p/\rho$. In the above expressions
we have defined
\begin{eqnarray}
\rho_{T} &=& \frac{1}{2\kappa^2}\left[2Tf_T-T-f(T)\right],
\label{Tor-density}\\
p_{T} &=&
\frac{1}{2\kappa^2}\left[\frac{f(T)-Tf_T+2T^2f_{TT}}{f_{T}+2Tf_{TT}}\right],
\label{Tor-press}
\end{eqnarray}
i.e $\rho_T$ and $p_T$ incorporate the effects of torsional modifications.
The equations close by considering the standard matter conservation equation
\begin{equation}\label{continuity1}
\dot{\rho}+3H(\rho+p)=0,
\end{equation}
in which case one obtains additionally the conservation equation of the torsional fluid,
namely
\begin{equation}\label{continuity2}
\dot{\rho}_T+3H(\rho_T+p_T)=0,
\end{equation}
while its equation-of-state parameter is $w_T\equiv p_T/\rho_T$. If the above formulation
is applied at late times, then this torsional fluid will constitute the dark
energy sector with equation-of-state parameter
\begin{equation}\label{torsion_EoS}
w_{DE}\equiv w_{T}=\frac{p_{T}}{\rho_{T}}=-1+\frac{[f(T)-2T
f_{T}](f_{T}+2Tf_{TT}-1)}{[f(T)+T-2Tf_{T}]
(f_{T}+2Tf_{TT})}.
\end{equation}
Finally, it proves convenient to define the effective, i.e the total, fluid of the
universe through
\begin{eqnarray}\label{rhoeff}
&&\rho_{eff}\equiv \rho+\rho_T\\
&&p_{eff}=p+p_T,
\label{peff}
\end{eqnarray}
as well as the effective
equation-of-state
parameter
\begin{equation}\label{eff_EoS0}
w_{eff}\equiv \frac{p_{eff}}{\rho_{eff}}=-1-\frac{2}{3}\frac{\dot{H}}{H^2},
\end{equation}
where the last equality arises straightforwardly from the Friedmann equations
(\ref{MFR1}), (\ref{MFR2}). This effective
equation-of-state
parameter is useful since it is straightforwardly related to the deceleration parameter
\begin{equation}
q\equiv -1-\frac{\dot{H}}{H^2}=\frac{1+3 w_{eff}}{2}.
\label{deccelmodI}
\end{equation}

\section{General features of cosmological phase portraits of $f(T)$ cosmology}\label{Sec3}

In this section we investigate the general phase space portraits and the behaviour of
general $f(T)$ cosmology. In order to perform this analysis we take advantage of the
special property of $f(T)$ cosmology that the torsion scalar $T$ can be expressed as a
quadratic function of the Hubble function, namely relation (\ref{TorHubble}). Hence, both
torsional energy density and pressure, given in (\ref{Tor-density}), (\ref{Tor-press}) are
functions of $H$, and then from the first Friedmann equation (\ref{MFR1}) we deduce that
the matter energy density $\rho$ can also be expressed as a function of $H$. Thus, for a
barotropic fluid with $p=p(\rho)$ the pressure is also a function of $H$, and then from
the second Friedmann equation (\ref{MFR2}) it is implied that $\dot{H}$ is a function of
$H$ too, as well as $w_{eff}$ through (\ref{eff_EoS0}). This feature, namely that
$\dot{H}$ in $f(T)$ cosmology can be expressed as a function of $H$, namely
\begin{equation}
\label{basicportr0}
\dot{H}=\mathcal{F}(H),
\end{equation}
is central in the present work, and it allows for the reduction
of the cosmological equations into a one-dimensional autonomous system, which can be
explored in detail.
Finally, from the conservation equation (\ref{continuity1}), taking into account the
above features, we deduce that for a
barotropic fluid the scale factor itself can be expressed as a function of $H$.

In summary, in $f(T)$ cosmology in a flat FRW universe every quantity can be expressed as
a
function of the Hubble function $H$, which allows for the use of one-dimensional
autonomous system methods (note that in case of non-flat geometry the various quantities
would have an additional explicit dependence on the scale factor, e.g
$\dot{H}=\mathcal{F}(H,a(t))$, which is a non-autonomous system and thus the present
methods could not be used).

\subsection{Flow, fixed points and stability}\label{Sec3.1}

\begin{figure}[ht]
\begin{center}
\includegraphics[scale=0.4]{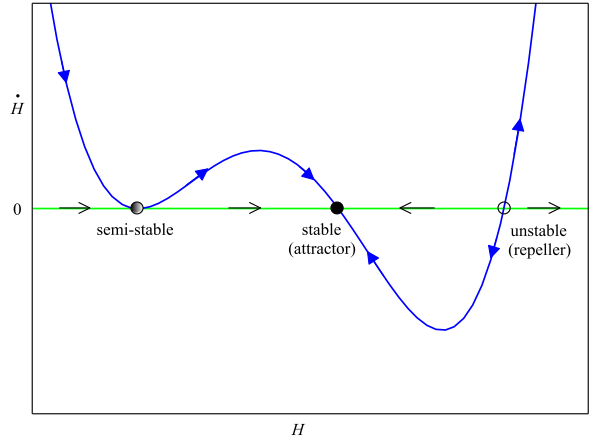}
\caption{{\it{ A toy-example of the phase space portrait that arises from
(\ref{basicportr0}) in $f(T)$ cosmology for notation clarification. The flow is towards
the right
when $\dot{H}>0$ and towards the left when $\dot{H}<0$. Moreover, the flow terminates at
fixed
points, which are the zero's of $ \mathcal{F}(H)$ and are de Sitter solutions
(\ref{basicportr0}). The fixed points are classified
to be unstable, stable or semi-stable according to the sign of
their tangents. Unstable fixed points are represented by open circles with
arrows emanating out of them, while stable
points (attractors) are represented by closed circles with arrows pointing towards them.
Finally, semi-stable points are represented by half-filled
circles, and are stable from one side and unstable from the other side.}}}
\label{Fig:fixed-points}
\end{center}
\end{figure}

We can interpret the differential equation (\ref{basicportr0}) as a vector field on
a line, introducing one of the basic techniques of dynamical analysis. In this view, we
can draw the phase space diagram of $H$, namely draw $\dot{H}$
versus $H$, which is the basic tool to analyze the cosmic evolution in a clear and
transparent way, without the need to know exact analytic solutions.
For clarity, in Fig. \ref{Fig:fixed-points} we present a toy-example of such a portrait,
since portraits like this will be used extensively in the following.
In this framework, one focuses on two features of this analysis,
which constitute a complete description of the evolution of $H$ as a
function of time.

The first feature is the
evolution/flow of the vector field $H$, which is represented by an
arrow showing the direction of the change of $H$ along the x-axis.
This is determined from the sign of $\dot{H}$, since a small change in
$H$ is given by $\delta H= \mathcal{F}(H) \delta t$. In other words, if
$\mathcal{F}(H)>0$ then $H$ flows towards the right (increasing) direction as time $t$
increases. On the contrary, if $\mathcal{F}(H)<0$ then $H$ flows towards the left
(decreasing) direction as time increases.

The second feature is the existence of fixed points or singularities which terminate the
flow.
In particular, solutions of the first-order differential equation
$\dot{H}=\mathcal{F}(H)$
are subject to the initial conditions
$H(t_i)=H_i$ at $t=t_i$ and can be divided into different branches, if the equation has
fixed points. The fixed points are the zero's of $ \mathcal{F}(H)$, which are just
de Sitter solutions, since at these points the Hubble parameter $H=H^*$
is constant and $\dot{H}=0$. If the flow starts from a fixed point, it
will remain eternally at this point, that is the universe will be always in this
de Sitter solution. A small fluctuation away from $H^*$ might derive the solution
towards or away from these points, depending on the type of the fixed point.
Thus, the fixed points are classified
to be \textit{unstable}, \textit{stable} or \textit{semi-stable} according to the sign of
their tangents. Unstable fixed points (repellers) are represented by open circles with
arrows emanating out of them, as can be seen in the example Fig.
\ref{Fig:fixed-points}. Stable
points (attractors) are represented by closed circles with arrows pointing toward them.
Finally, semi-stable are represented by half-filled
circles, and are stable from one side and unstable from the other side.\footnote{This
classification is closely related to the traditional way of linearizing the system
$\dot{H}=\mathcal{F}(H)$ around a fixed point $H^{*}$, allowing for a small
perturbation $\delta H(t)=H(t)-H^{*}$ around $H^*$, resulting to a
differential equation for $\delta H$ in order to check whether the perturbation
decays or grows
\cite{Coley:2003ASSL,Ellis:2005CUP,Chen:2008ft,awad2013,Boehmer:2014vea,Kofinas:2014aka,
Odintsov:2017icc}.} In particular, differentiation with respect to time yields
$
\nonumber \frac{d}{dt}({\delta H})=\frac{d}{dt}\left(H(t)-H^{*}\right)=\dot{H},
$
since $H^*$ is constant, and thus the perturbation in the Hubble space around $H^{*}$
propagates with a rate
\begin{equation}\label{pertubation-rate}
\frac{d}{dt}({\delta H}) =\dot{H} =\mathcal{F}(H) =\mathcal{F}(H^*+ \delta H).
\end{equation}
Applying a Taylor expansion around the fixed point $H^*$ as
$ \mathcal{F}(H^{*}+ \delta H)= \mathcal{F}(H^{*}) + \delta H~
\mathcal{F}'(H^{*})
+ O(\delta H^2)$,
where $\mathcal{F}'(H^{*})=\left.\frac{d}{dH}\mathcal{F}(H)\right|_{H^*}$ (we have used
that $\mathcal{F}(H^*) = 0$ since $H^*$ is
a fixed
point), we finally acquire
$\frac{d}{dt}({\delta H}) \thickapprox \delta H~ \mathcal{F}'(H^*)$.
Therefore, the linearization of $\delta H$ around $H^*$ leads to the solution
\begin{equation}
\nonumber \delta H(t) \propto e^{\mathcal{F}'(H)t}.
\end{equation}
The above equation shows that the slope $\mathcal{F}'(H^*)$ at the fixed point determines
its stability. If $\mathcal{F}'(H^*) > 0$, all small disturbances $\delta H(t)$ grow
exponentially and the fixed point in this case is unstable (repeller or source). If
$\mathcal{F}'(H^*) < 0$, then all sufficiently small disturbances decay exponentially and
the fixed point in this case is stable (attractor or sink). Finally, if the slope
$\mathcal{F}'(H^*)$ alters its sign at the fixed point
(thus the fixed point is an extremum of $ \mathcal{F}(H)$), it is a semi-stable phase
point at which the solution is stable from one side and unstable from the other side.
Lastly, in order to fix our notations we follow \cite{book:Steven} and we call equation
$\dot{H}=\mathcal{F}(H)$ the \textit{phase portrait}, while its solution is the
\textit{phase trajectory}.

An important issue that needs to be mentioned is how fixed points (or singularities) in
one-dimensional systems split solutions into different branches. In particular, one of
the main features of a fixed point is that it can be reached only after infinite time, as
long as $ \mathcal{F}$ is differentiable at that point (this can be easily seen by
expanding $\mathcal{F}$ around the fixed point and integrate $\dot{H}=\mathcal{F}(H)$ in
order to acquire the time
needed to reach that point), a condition that is satisfied in most models. An
important consequence of this statement is that if a solution lies between two fixed
points then it has to start from $t=-\infty$ at one point and reach the other fixed point
at $t=+\infty$, which constitutes a solution branch by itself.
In general, if a model has $N$ fixed points, then we have $N+1$ of the above regions
and thus we obtain $N+1$ different branches of
solutions. Amongst others, and following \cite{awad2013}, we can see that
if $ \mathcal{F}(H)$ is continuous and differentiable and there exists a
future and a past fixed point, the solution is free from types I, II and III
singularities classified in \cite{Nojiri:2005sx}. Additionally, in this case it is easy
to show that there exists a no-go theorem stating that $H(t)$ cannot cross a fixed
point, or $w_{eff}=-1$, at a finite time, since the time to reach this point is infinite.

Here we make a comment on the continuity and differentiability of $\mathcal{F}(H)$. Given
an
initial condition $H(t_0)=H_0$, the continuity of $\mathcal{F}(H)$ guarantees the
existence of a
solution, while its differentiability guarantees the uniqueness of the solution, see for
example \cite{book:Steven}. Nevertheless, from this important theorem one can deduce the
limitations of the
phase-space analysis, namely at the points where $\mathcal{F}(H)\rightarrow\infty$. In
these cases
there is no unique solution locally, and therefore it is not clear how the system will
evolve in later
times \cite{book:Steven}.

In summary, the above formalism provides a qualitative description of the
behaviour of a cosmological scenario, without knowing any exact solution.
In the following, we obtain the phase portraits of $f(T)$ cosmology in
the ($\dot{H}-{H}$) \textit{phase space}, where each point is a \textit{phase point} and
could serve as an initial condition.

Let us now proceed in the application of the above dynamical-system method in the case
of general $f(T)$ cosmology. As we mentioned, we can express the matter energy density
and pressure as functions of $H$, that is using (\ref{MFR1}),
(\ref{Tor-density}), (\ref{Tor-press}) and (\ref{continuity1}), we can write
\begin{eqnarray}
\rho &=& \frac{1}{2\kappa^2}\left[\tilde{f}(H)-H \tilde{f}_{H}\right], \label{FR1H}\\
p &=&\frac{1}{6\kappa^2}\dot{H} \tilde{f}_{HH}-\rho,\label{FR2H}
\end{eqnarray}
where $\tilde{f}(H)=f(-6H^2)$, and
$\tilde{f}_{H}:=\frac{d\tilde{f}}{dH}$ and
$\tilde{f}_{HH}:=\frac{d^{2}\tilde{f}}{dH^2}$.
Then, for a general barotropic matter fluid with $p=w \rho$, equations
(\ref{FR1H}) and (\ref{FR2H}) give
\begin{equation}
\dot{H}= 3(1+w)\left[
\frac{\tilde{f}(H)-H \tilde{f}_{H}}{\tilde{f}_{HH}}\right]=\mathcal{F}(H)
\label{ps}.
\end{equation}
Note that this equation is valid only if $\tilde{f}_{HH}\neq0$. The case where
$\tilde{f}_{HH}=0$ at all times corresponds to $f(T)=\alpha \sqrt{-T}+\beta$, which is
the known ``trivial'' $f(T)$ form for which the $f(T)$ effect is completely eliminated
from
the equations and the theory becomes trivial \cite{Cai:2015emx}. In this work we will not
consider this trivial case. However, one should pay attention to the fact that for a
general $f(T)$ model, $\tilde{f}_{HH}$ may become zero at a specific time, i.e
$\mathcal{F}(H)\rightarrow\infty$. These points correspond to
sudden singularities \cite{Nojiri:2005sx}, and in order to fully analyze their
properties one should propose a specific spacetime extension, which extend non-spacelike
curves
beyond these soft singularities as done in
\cite{Barrow:1986,Barrow:2004xh,Keresztes:2012zn,
awad2013,Awad:2015syb,Awad:2017sau}.
For completeness, note that the above method is fully applicable in the case where $f(T)$
gravity
becomes TEGR, that is general relativity, i.e for $f(T)= T+\Lambda$, since in this case we
always
have $\tilde{f}_{HH}=-12\neq0$.

Equation (\ref{ps}) is the main equation of the one-dimensional autonomous system in
general $f(T)$ cosmology. First of all, it determines the existence of fixed points and
sudden
singularities. In particular, the fixed points are obtained for
$\tilde{f}=H\,\tilde{f}_H$ and they are reached after infinite time if
$\mathcal{F}(H)$ is differentiable at this point. On the other hand, sudden singularities
are points where $\tilde{f}_{HH}$ vanishes. Points of sudden singularities are reached at
finite time and it is generally possible to extend the spacetime
beyond these points, as well as the curves of physical test particles, as has been shown
in \cite{Awad:2017sau}. Note that these singularities are not related to the divergence
of the energy density and/or pressure.

Finally, let us translate the basic observational constraints to a simple
list of requirements about phase space diagrams which describe realistic cosmological
models.
In particular:
\begin{itemize}
\item [(i)] The late time acceleration might be modeled as flowing towards a future fixed
point.
\item [(ii)] The universe crossed from deceleration to acceleration at a redshift
$z_{tr}\geq 0.6$ (the redshift is given by
$z=\frac{a_{0}}{a}-1$, where the current value of the scale factor $a_{0}$ is set to
1). According to (\ref{deccelmodI}) this ``zero acceleration curve'' corresponds
to $q=0$, i.e to $w_{eff}=-1/3$.
\item [(iii)] The universe has passed from  radiation and matter eras in
the past.
\end{itemize}
\subsection{Phase portraits of standard cosmological evolution}\label{nonsigulevol}
Let us now investigate the phase space portraits of $f(T)$ cosmology in which no
singularities are involved, namely the universe behaves in the standard way.
In order to incorporate more efficiently the three observational requirements described
above, and in particular point (ii), we re-write (\ref{eff_EoS0}) as
\begin{equation}\label{effective-phase-portrait}
\dot{H}=-\frac{3}{2}\left(1+w_{eff}\right)H^{2}.
\end{equation}
Hence, one can study this equation in terms of the values of $w_{eff}$. In the following
paragraphs we study the case of constant and varying $w_{eff}$ separately.
\subsubsection{Fixed effective equation of state }\label{Sec3.3.1}
\begin{figure}[ht]
\begin{center}
\includegraphics[scale=0.55]{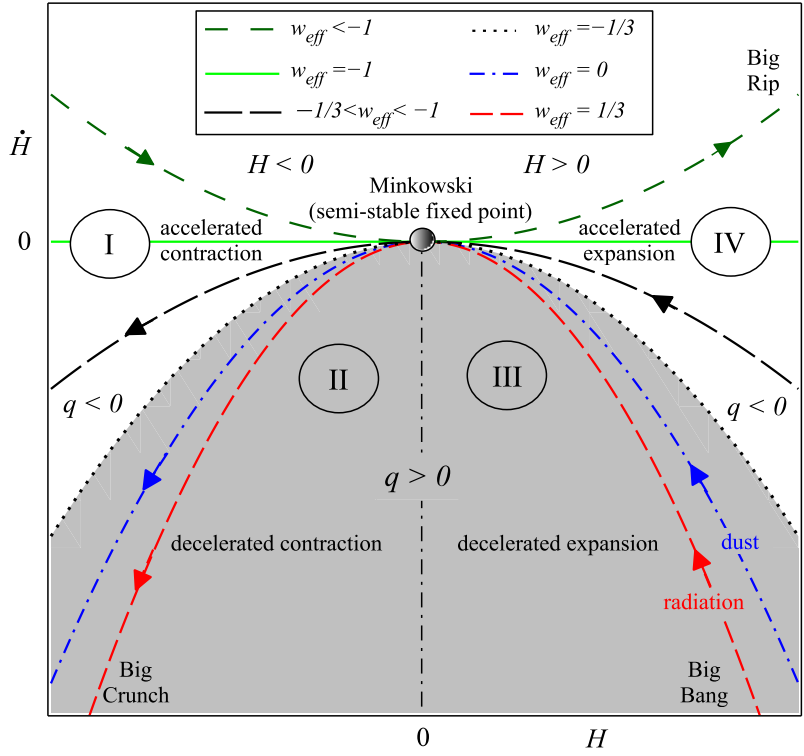}
\caption{{\it{Phase space portraits according to
(\ref{effective-phase-portrait}),
for various values of constant $w_{eff}$. The point at the
origin corresponds to a semi-stable Minkowski universe. The dotted
curve represents the zero acceleration boundary and it divides the phase space into two
regions, namely the decelerated (shaded) and the accelerated (unshaded) region.
The labels (I)-(IV) mark the regions with the four possible behaviours described in the
text. The graph has scale invariance, and every shortening or extension which includes
($0,0$) will look the
same.
}}}
\label{Fig:phase space}
\end{center}
\end{figure}

If $w_{eff}$ is constant then (\ref{effective-phase-portrait}) admits
\textit{only} the trivial solution ($\dot{H},H$) $=$ ($0,0$). Geometrically this
null solution represents the origin of the phase space, which is also a Minkowskian fixed
point. Thus, this null solution splits the phase space into two separate
patches: In the first patch, corresponding to $H>0$, the universe is expanding, while in
the second patch, where $H<0$, the universe is contracting. Notably, as we analyzed above,
transitions between these two patches through the Minkowskian origin cannot
be achieved in a finite time. In order to present the above features in a more
transparent way, in Fig. \ref{Fig:phase space} we construct the corresponding phase
portrait, following the notation of the toy-example Fig. \ref{Fig:fixed-points}.

The phase space, in general, contains four different dynamical regions according to the
values of the Hubble function $H$ and of the deceleration parameter $q\equiv
-1-\dot{H}/H^2$.
Accordingly, we proceed to the following classification:
\begin{itemize}
\item [(i)] The unshaded region in the negative Hubble patch represents an accelerated
contraction phase, since $H<0$ and $q<0$. We label this region as (I). The
$dH/dt<0$ region corresponds to $-1<w_{eff}<-1/3$, while the $dH/dt>0$ corresponds
to $w_{eff}<-1$.
\item [(ii)] The shaded region in the negative Hubble patch represents a decelerated
contraction phase, since $H<0$ and $q>0$. We label this region as (II). In this region
the universe evolves towards a future finite time singularity (Big Crunch).
\item [(iii)] The shaded region in the positive Hubble patch represents a decelerated
expansion phase, since $H>0$ and $q>0$. We label this region as (III). In this region
the universe begins with a finite time singularity (Big Bang).

\item [(iv)] The unshaded region in the positive Hubble patch represents an accelerated
expansion phase, since $H>0$ and
$q<0$. We label this region as (IV). The
$dH/dt<0$ region corresponds to $-1<w_{eff}<-1/3$ while the
$dH/dt>0$ corresponds to $w_{eff}<-1$.
\end{itemize}

Interestingly, we can unify two or more of these
behaviours, obtaining a transition from expansion to contraction if we add a
negative cosmological constant.
In particular, the addition of a negative cosmological constant moves the phase
portrait and the Minkowski fixed point towards the lower part of the figure, making the
transition from expansion to contraction realizable.
On the other hand, under the addition of a positive cosmological constant the phase
portrait will be shifted vertically towards the upper part of the figure, and
thus the Minkowskian fixed point will be moved upwards providing two new fixed
points, which will allow for the transition from contraction to expansion, that is for
the bounce realization \cite{ElHanafy:2017sih}.

In the rest of the manuscript we will follow the color notation of
Fig. \ref{Fig:phase space}, corresponding to the above four regions.
Finally, we mention here that in the case of a fixed $w_{eff}$, the one-dimensional
autonomous systems do not allow for oscillating solutions, since crossing the phantom
divide line $w_{eff}=-1$ requires an infinite time.

\subsubsection{Dynamical effective equation of state}\label{Sec3.3.2}
\begin{figure}[ht]
\begin{center}
\includegraphics[scale=0.7]{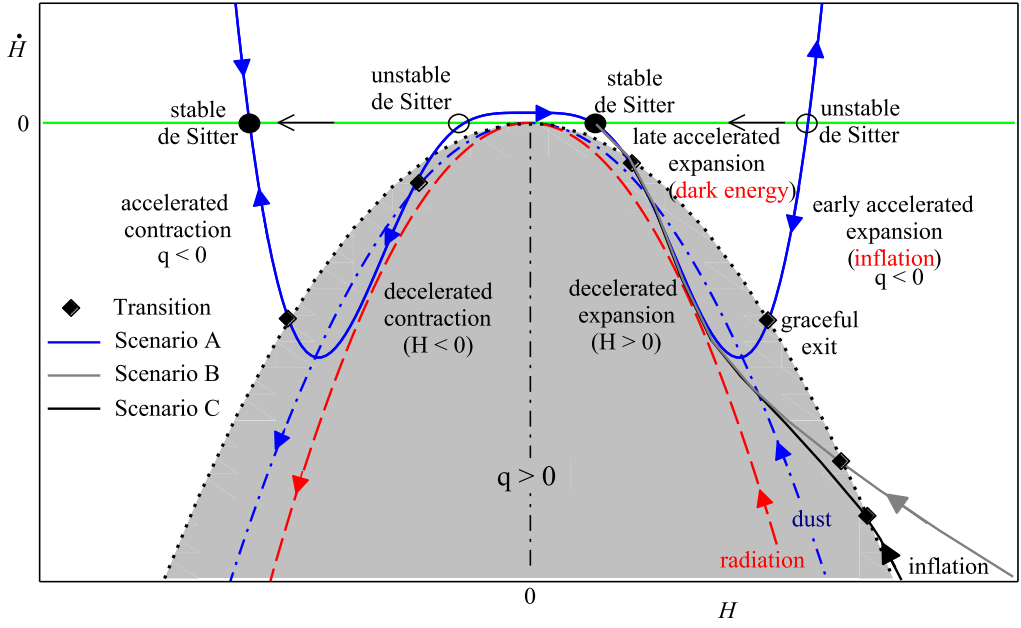}
\caption{{\it{
Phase space portraits according to (\ref{effective-phase-portrait}), in the general case
where $w_{eff}\equiv w_{eff}(H)$. The three scenarios, the transition points, and the
behaviour at different Hubble regimes are described in the text. The dust and radiation
curves are the same with Fig. \ref{Fig:phase space} and are drawn for convenience. The
scale of the graph is determined by the unstable de Sitter point at
$H^\star=H_{Planck}=1/l_{Planck}$. }} }
\label{Fig:unified-history}
\end{center}
\end{figure}

In a general and more realistic case of $f(T)$ cosmology $w_{eff}$ is not constant, but
it is a function of $H$, as was described in detail in the beginning of this section.
In this case, one must fulfill the basic observational requirements listed above.
In Fig. \ref{Fig:unified-history} we present schematic phase space portraits of some
generic and realistic cosmological models, which can in principle arise within the
framework of the $f(T)$ cosmology. Let us analyze some of their features, examining for
simplicity separately the regimes that have different Hubble function values.

\begin{itemize}
\item {Large Hubble function regime.}

Imposing the known observational constraints, we list all possible $f(T)$ cosmologies
according to their behaviour at large $H$. In order to obtain an early
accelerated expansion epoch (inflation), the phase portrait at large $H$ should lie
within region IV. This constraint leads us to distinguish between three different
scenarios depending on the behaviour of $\dot{H}$ at early times:
\begin{equation}\label{Large-H}
t=\int_{H>0}^{H^\star}\dot{H}^{-1}dH=\left\{
\begin{array}{ll}
\infty, & \hbox{$H^\star$= finite, $\dot{H}=0$ (scenario A);} \\
\infty, & \hbox{$H^\star \to \infty$, $\dot{H}\to \infty$ (scenario B);} \\
finite, & \hbox{$H^\star \to \infty$, $\dot{H}\to \infty$ (scenario C).}
\end{array}
\right.
\end{equation}
In scenario A, as shown in Fig. \ref{Fig:unified-history}, the phase portrait
begins with a fixed point in the past, i.e $\dot{H}$= 0, with a very large value of
$H$ (close to the Planck value, $H^\star=H_{Planck}=1/l_{Planck}$). Eq. (\ref{Large-H})
implies that the time required to reach this point is infinite, which enables a
non-singular description of the universe evolution and can in principle describe
inflation. In scenario B, the phase portrait
asymptotically exhibits a linear (or slower) behaviour. The phase portrait in this case
is characterized by $\dot{H}\propto H^{s}$, where $s\leq 1$ as $H\to \infty$, and then
the integral in (\ref{Large-H}) diverges. In
this scenario the universe has a Big Bang singularity, however it is pushed back to
infinite time. In scenario C, the universe
begins with a Big Bang singularity with an asymptotic effective equation-of-state
parameter $-1 < w_{eff} < -1/3$ for large $H$. The phase
portrait in this case is characterized by $\dot{H}\propto H^{s}$, where $s > 1$ as $H\to
\infty$, and then the integral in (\ref{Large-H}) is finite. In these three different
scenarios, the phase portraits should intersect the zero acceleration curve into the
radiation curve at region (III) where Hubble function acquires smaller values. At the
intersection point
$\dot{H}=-H^2$, the universe terminates its accelerated expansion phase.
Therefore, the universe can   exit into the FRW decelerated expansion, and thus the
Hubble parameter should be chosen $H \sim 10^{7}$ GeV in order to obtain a graceful exit.

\item {Intermediate Hubble function regime.}

This regime characterizes the standard cosmological era, and the $f(T)$ models can give
rise to a successful thermal history just as predicted by standard cosmology. Hence, the
phase portraits of scenarios A - C of Fig. \ref{Fig:unified-history} in this regime
correspond to the radiation era.

\item {Small Hubble function regime.}

The phase portraits intersect the zero acceleration curve, namely $\dot{H}=-H^2$, for the
second time, allowing the universe to transform into a late-time accelerated
expansion phase. This point imposes a further constraint, since as we mentioned above,
the late-time transition to acceleration is expected to be at redshift $z_{tr}\gtrsim
0.6$,
i.e
at $H_{tr}\gtrsim 100$ km/s/Mpc. Additionally, in this regime the three scenarios A - C
exhibit a similar behaviour.

\item {Fate of the universe.}

In the future, when $H^{2}$ becomes comparable to the
cosmological constant, the phase portrait evolves towards a future de Sitter fixed point
with
$w_{eff}\sim -1$, as can be seen in Fig. \ref{Fig:unified-history}. However, in some
cases
the universe may evolve towards Minkowski
instead of de Sitter (we did not include this case in Fig.
\ref{Fig:unified-history} since it is in tension with $\Lambda$CDM paradigm).
\end{itemize}

We close this paragraph by mentioning that in order to draw
Fig. \ref{Fig:unified-history} we did not use explicit $f(T)$ forms, since our purpose
is to show how the Hubble-rate flow in phase space is
constrained by basic observational requirements, and moreover how these constraints allow
for three qualitatively different scenarios at large $\dot{H}$. Nevertheless, by
considering an explicit form of $\dot{H}\equiv \mathcal{F} (H)$, which reproduces any of
the scenarios given in Fig. \ref{Fig:unified-history}, the
corresponding $f(T)$ gravity can be obtained from Eq. (\ref{ps}). On the
other hand, the observational requirements obtained from crossing the zero acceleration
curve, either at large or small Hubble function regimes, can be used as usual to
determine the free parameters of the $f(T)$ theory.

\subsection{Phase portraits of finite-time singularities of Type II and IV}\label{Sec3.4}
In the previous subsection we investigated the phase space portraits of $f(T)$ cosmology,
focusing on the standard cosmological evolution, namely in the absence of singularities.
Hence, the only singularities that might possibly appear were those characterized
by $\dot{H}\to \pm \infty$ as $H\to \pm \infty$, namely the asymptotic behaviour of the
parabolic phase portrait, which correspond to the finite-time singularities of Type
III of \cite{Nojiri:2005sx}, such as the Big Bang and Big Crunch.
In the present subsection we wish to extend our investigation of the phase space
portraits
in order to explore more exotic cosmological evolutions such as the
phantom-divide crossing, the bounce realization, the Big Brake and the cosmological
turnaround. Such evolutions may incorporate the appearance of soft finite-time
singularities of type II and IV \cite{Nojiri:2005sx,awad2013,ElHanafy:2017sih}.
As it is known, such exotic cosmological behaviours are impossible in the framework of
general relativity, since in this case the various energy conditions violations that are
necessary for their realization cannot be obtained \cite{Cai:2011bs,Nojiri:2013ru} .
However, it is also known that they are possible in the framework of modified gravity. We
mention that at the observational level, neither Type II nor IV represent harmful
singularities \cite{Nojiri:2005sx}.
In the following, examining the phase space portraits, we show that they can be realized
in the framework of $f(T)$ cosmology.

\subsubsection{Non-singular bounce}

In Fig. \ref{fig:nonsingbounce} we present a schematic phase space portrait of a
non-singular
bounce {\cite{Novello:2008ra,Cai:2009in}. In such a phase portrait the flow is
clockwise,
and the system passes by the four regions of the phase space. This pattern has been
studied in detail in \cite{ElHanafy:2017sih}. The universe in this scenario has an
(eternal in the past) Minkowskian fixed-point origin, and then it enters into region
(II),
that is into a decelerated contraction phase. The phase portrait then intersects the zero
curve acceleration and the universe enters into an accelerated contraction, namely into
region (I), where $\dot{H}_{-}<0$. However,
the phase portrait crosses the phantom divide line $\dot{H}=0$, entering into an
effectively phantom phase where $\dot{H}_{+}>0$.
\begin{figure}[ht]
\begin{center}
\includegraphics[scale=0.5]{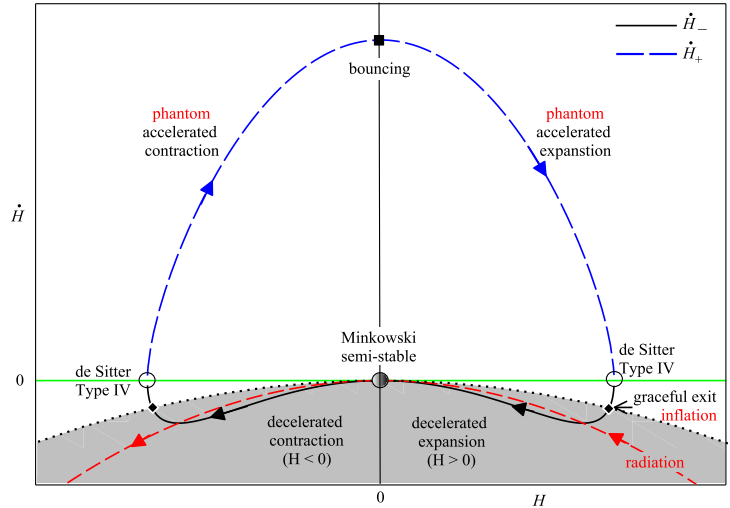}
\caption{ {\it{ Schematic phase space portrait for non-singular bounce realization. The
blue-dashed curve corresponds to $\dot{H}_{+}>0$, while the black-solid to
$\dot{H}_{-}<0$. Similarly to the previous figures, the black-dotted and the red-dashed
curves are respectively the zero acceleration boundary and the radiation
curve, and are drawn for convenience. The
scale of the graph is determined by the Type IV de Sitter singular point at $H=H_{f}$.
}}}
\label{fig:nonsingbounce}
\end{center}
\end{figure}

Although crossing the phantom divide
line is realized through a de Sitter fixed point, the time needed to reach it is not
infinite. In order for this exceptional case to be realized, the following
conditions at the fixed point must be fulfilled \cite{awad2013}:
\begin{itemize}
\item [(i)] $\left.\dot{H}\right|_{H= H_{f}}=0$,
\item [(ii)] $\left.d\dot{H}/dH\right|_{H=H_{f}}$ diverges,
\item [(iii)] $t=\displaystyle{\int}_{H}^{H_{f}} \dot{H}^{-1} dH$= finite,
\item [(iv)] the phase portrait is double-valued around the fixed point $H_{f}$.
\end{itemize}
The first three conditions are required in order for the fixed point to be reached in a
finite time, while the fourth one is necessary for the crossing. It is not difficult to
show that the fixed point in this case is in fact a Type IV singular point according to
the finite
time singularity classification \cite{Nojiri:2005sx}, where all the quantities $a$, $H$
and $\dot{H}
$ are finite but the
second derivative $\ddot{H}=\dot{H}\left(\frac{d\dot{H}}{dH}\right)$ diverges at
$H=H_{f}$. Therefore, we call this type of fixed points a Type IV de Sitter point. On the
phase portrait such points can be recognized, since on them the portrait has
an infinite slope.

In the phantom phase, in the $\dot{H}_{+}>0$ branch, the universe transits from region
(III), which is a phase of accelerated contraction ($H<0$), into region (IV), which is
a phase of accelerated expansion ($H>0$), and throughout this procedure we have
$\dot{H}>0$. Thus, the bouncing point\footnote{At the bounce point the comoving
Hubble horizon $R_{H}=\frac{1}{a H}$ is infinite as $H$ is null, and thus all
comoving modes $k$ are subhorizon. After the bounce, $R_{H}$ shrinks to a minimal value
and therefore some modes are allowed to exit the horizon and transform to
classical modes. Additionally, the scale invariant power spectrum can be obtained just as
in inflationary scenario. Nevertheless, in non-singular bounce the trans-Planckian
problems of inflation can be avoided \cite{Novello:2008ra}.} is
the point ($H=0$, $\dot{H}>0$). In the end of the phantom phase, the universe crosses the
phantom divide line through a Type IV de Sitter point, as indicated by the phase portrait
of Fig. \ref{fig:nonsingbounce}.

Finally, note that  the phase portrait intersects the zero acceleration curve smoothly,
entering into the FRW decelerated expansion phase, i.e region (III). During this era  the
phase portrait matches standard cosmology, and thus it leads to the usual thermal history.
However, the universe evolves towards the Minkowskian origin without exhibiting the
late-time
acceleration phase.

\subsubsection{Singular bounce}

In Fig. \ref{fig:crossing} we present an example of a phase portrait that can
smoothly cross the phantom divide line $\dot{H}=0$ just as in non-singular bounce.
However, $\dot{H}>0$ diverges at $H=0$ producing a Type II ``sudden" finite-time
singularity associated with the bouncing point. The interesting feature is that in this
case the scale factor and its first derivative at this singularity type are finite, and
consequently the Christoffel symbols are finite too. In case of sudden singularities in
the framework of general relativity it has been shown that
physical geodesics are in principle extendible
\cite{Fern:2006AIPC,Fern:2006PRD,Barrow:2013PRD}. In subsequent works
some specific extensions have been proposed, in
addition of applying suitable junction conditions at
$H=0$ to ensure   consistency between the geodesic extensions and the field equations
\cite{Keresztes:2012zn,Awad:2015syb}. In the case of singular bounces the comoving Hubble
radius $R_{H}$ is finite at the bounce point and not all the comoving modes are
subhorizon. Hence, such models are phenomenologically different from the
non-singular bounce \cite{Novello:2008ra}.
\begin{figure}[ht]
\begin{center}
\includegraphics[scale=0.5]{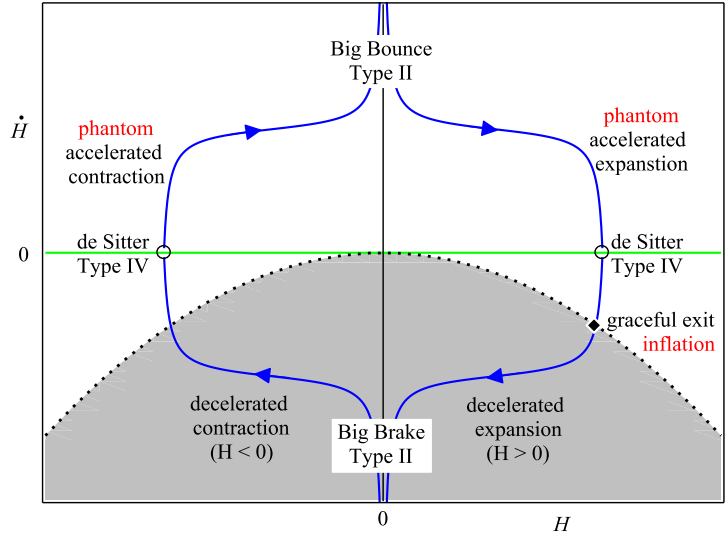}
\caption{ {\it{Schematic phase space portrait for singular bounce realization, in which
the bounce happens through a Type II (sudden) singularity. Similarly to the previous
figures, the black-dotted curve marks
the zero acceleration boundary. The
scale of the graph is determined by the Type IV de Sitter singular point at $H=H_{f}$.
}}}
\label{fig:crossing}
\end{center}
\end{figure}

On the other hand, if $\dot{H}<0$ diverges at $H=0$ then the phase portrait
exhibits the Big Brake cosmology. The general relativistic version of this model suffers
from the so-called ``soft singularity crossing paradox''
\cite{Keresztes:2012zn,Kamenshchik:2013ink,Gergely:2013via},
where the matter density does not vanish at the braking point. However, as we can now
see, instead of adding exotic matter such as anti-Chaplygin gas or tachyon fields to
overcome this problem, one can use $f(T)$ gravity. The reconstruction of the $f(T)$
gravity
which generates phase portraits corresponding to the Big Brake realization was given
in \cite{ElHanafy:2017xsm}. Furthermore, the junction condition in $f(T)$ gravity has
been applied in Big Brake scenario in \cite{Awad:2017sau}.
Finally, as we observe from Fig. \ref{fig:crossing}, if the phase
portrait has a cusp at $H=0$ with finite value of $\dot{H}$, then the bounce/Big Brake
point is associated with a finite-time singularity of Type IV.

In summary, in order for a phase portrait to cross the phantom divide line smoothly
it must be through a de Sitter fixed point of Type IV. However, in order for a phase
portrait to cross between contraction and expansion in singular cosmology, it must be
through a finite time singularity of Type II or IV.

\section{Cosmological phase portraits of specific $f(T)$ models}\label{Sec4} 

In the previous section we investigated the phase space portraits of general $f(T)$
cosmology, and we extracted the general
features and behaviours without specifying to individual models. In the present section
we apply the general method to the specific viable models that have been studied in the
literature. As we can see, we can extract the results already obtained in the literature
using the standard dynamical system methods
\cite{Wu:2010xk,Skugoreva:2014ena,Carloni:2015lsa}, and moreover we can provide
additional information concerning the global behaviour of the universe, focusing on their
differences and similarities.

In particular, in the following three subsections we will separately study
the power-law, the square-root exponential, and the exponential $f(T)$ models. These three
models are the viable ones, since they are in the best agreement with
cosmological observations and Solar System constraints
\cite{Nesseris:2013jea,Cardone:2012xq,Nunes:2016qyp,
Nunes:2016plz,Qi:2017xzl,Iorio:2012cm},
 and
they are
characterized by two parameters, one of which is independent.

\subsection{$f_{1}$CDM model: $f(T)=T+\alpha(-T)^{b}$}\label{Sec4.1}
The power-law $f(T)$ model (hereafter $f_{1}$CDM model) reads as
\cite{BF09}
\begin{equation}\label{Benghochea_fT}
f(T)=T+\alpha(-T)^{b},
\end{equation}
with $\alpha$ and $b$ the two model parameters (the former is dimensionful with units
of [length]$^{2(b-1)}$, while the later is dimensionless). Inserting
(\ref{Benghochea_fT}) into (\ref{Tor-density}) and then into the first Friedmann equation
(\ref{MFR1})
at current time we obtain the relation between $\alpha$ and $b$, namely
\begin{equation}\label{alphaBengochea}
\alpha=(6H_{0}^{2})^{1-b}\frac{1-\Omega_{m,0}}{2b-1},
\end{equation}
where $\Omega_{m,0}=\frac{\kappa^2 \rho_{0}}{3H_0^2}$ is the current value of the matter
density parameter, namely at scale factor $a_0=1$ (the subscript ``0'' marks the
current value of a quantity). In the case $b=0$ the model at hand coincides with TEGR,
i.e general relativity, with a cosmological constant, that is to $\Lambda$CDM cosmology.

Inserting (\ref{Benghochea_fT}) into (\ref{ps}) we acquire
\begin{equation}\label{Bengochea_phase_space}
\dot{H}=-\frac{3}{2}(1+w )H^2\left[\frac{1-\alpha (2b-1) (6H^{2})^{b-1} }{~1 - \alpha
b (2b-1) (6H^{2})^{b-1}}\right].
\end{equation}
Since we are interested in the dark energy era, we focus on the case of dust matter and
we use $w =0$. Additionally, we impose $\Omega_{m,0}=0.318$ and $H_{0}=76.11$ km/s/Mpc in
agreement with observations \cite{Ade:2015xua}.
\begin{figure}[ht]
\begin{center}
\includegraphics[scale=0.7]{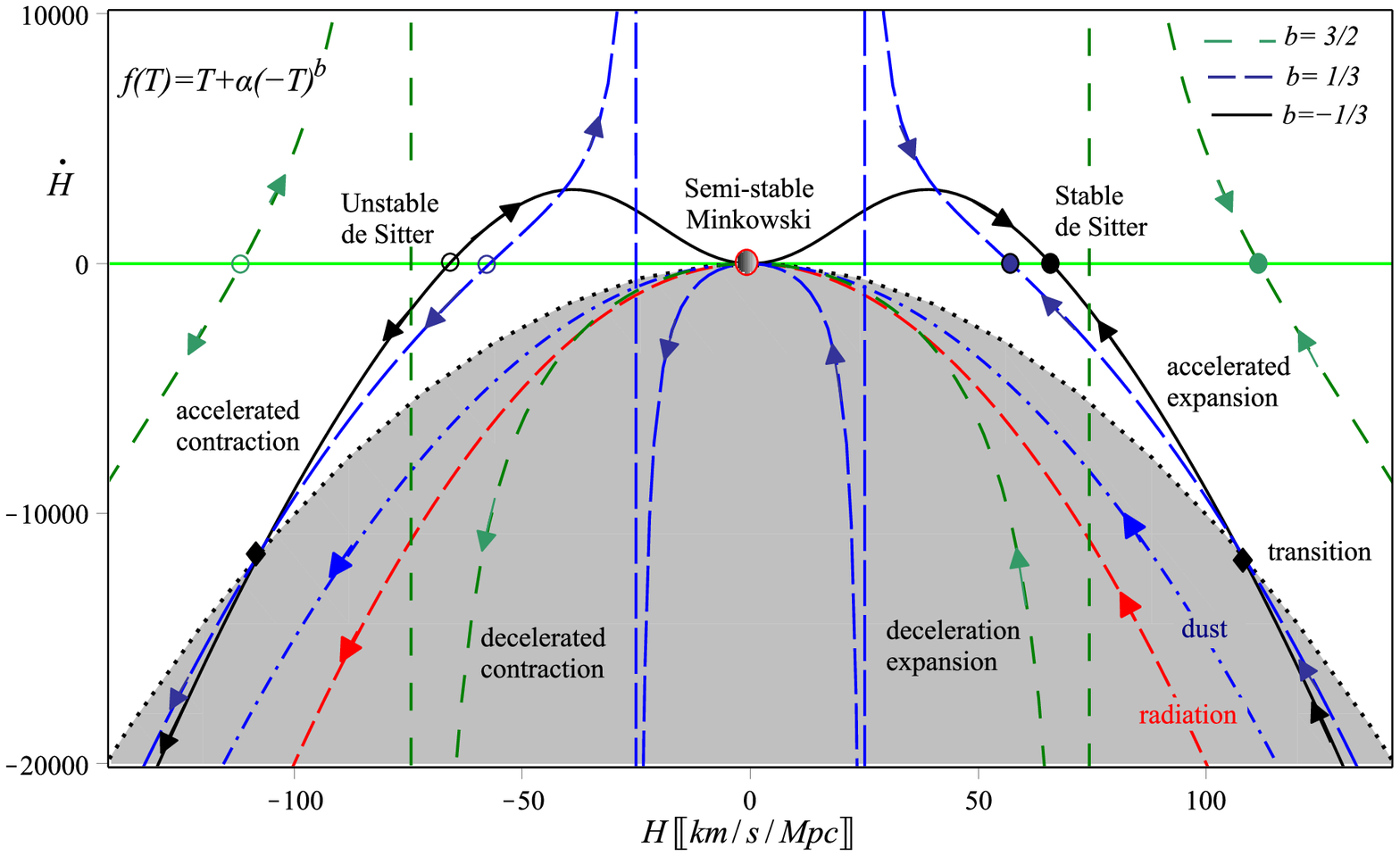}
\caption{{\it{ Phase space portraits for the power-law $f_{1}$CDM model of
(\ref{Benghochea_fT}), according to (\ref{Bengochea_phase_space}), for three values of
the model parameter $b$. Similarly to the previous figures, the zero acceleration
boundary and the dust and radiation
curves, are drawn for convenience. The
scale of the graph is determined by the value of $T_0=-6H_0^2$, and thus by
the imposition of $H_{0}=76.11$ km/s/Mpc. }}}
\label{Fig:phasespace1}
\end{center}
\end{figure}

In Fig. \ref{Fig:phasespace1} we present the phase space portraits of the power-law
$f(T)$ model (\ref{Benghochea_fT}), for three choices of the parameter $b$.
In the $H>0$ half-plane we identify the fixed points by setting $\dot{H}=0$ in
(\ref{Bengochea_phase_space}), and in the examples of the figure they correspond to
$H^{*}=65.95,~57.15$
and $111.51$ km/s/Mpc for $b=-1/3,~1/3$ and $3/2$, respectively. These points represent
stable de Sitter future attractors, and the time required to approach them is
infinite since $\dot{H}$ is finite.
We are interested in the ($H>H^*$)-patch, where the transition from decelerated to
accelerated expansion is realized. The phase trajectory in this region goes towards the
left, i.e towards the decreasing $H$ direction.
The phase portrait shows that the universe begins with an initial singularity
(Big Bang), it exhibits a matter-dominated era, and then it enters into the
late-time acceleration phase. The realization of this epoch sequence favors $b=-1/3,~
1/3$, in which the acceleration transition occurs at $H_{tr}=107.36$ km/s/Mpc  (for
$b=-1/3$) and at  $H_{tr}=107.89$  km/s/Mpc (for $b=1/3$),
i.e at redshift $z_{tr}\sim 0.6$ (for a
discussion on the transition redshift in various $f(T)$ cosmological models see
\cite{Capozziello:2015rda}). Moreover, the model parameter $b$ can
be constrained by knowing the transition time more precisely.
This can
be done by determining the intersection of the phase portrait with the zero acceleration
curve by setting $\dot{H}=-H^{2}$ at $H=H_{tr}$. Using (\ref{Bengochea_phase_space}) we
find
\begin{equation}
\label{alpha2}
\alpha=-\frac{(6H_{tr}^{2})^{1-b}}{(2b-1)(2b-3)},
\end{equation}
and comparing with (\ref{alphaBengochea}) we extract the useful relation
\begin{equation}
\label{b-parameter}
H_{tr}=\frac{H_{0}}{(2b-3)^{\frac{1}{2(b-1)}}(\Omega_{m,0}-1)^{\frac{1}{2(b-1)}}},
\end{equation}
which predicts the Hubble parameter value at the transition as a function of $b$. Some
specific values are presented in Table \ref{Table1}. It is worth mentioning
that relation (\ref{b-parameter})
restricts the parameter $b$ to be less than $3/2$ if we impose the physical requirement
$\Omega_{m,0}<1$.
\begin{table}[ht]
\begin{tabular*}{\columnwidth}{@{\extracolsep{\fill}}cccc@{}}
\hline
$b$ & $H_{tr}$ [km/s/Mpc] & $z_{r}$\\
\hline
$0$ & $\sim 108.91$ & $\sim 0.61$\\
$0.05$ or $0.17$ & $\sim 109.01$ & $\sim 0.63$\\
$0.07$ or $0.15$ & $\sim 109.04$ & $\sim 0.63$\\
$-0.36$ or $0.38$& $\sim 107.18$ & $\sim 0.60$\\
$-0.92$ or $0.5$ & $\sim 103.92$ & $\sim 0.55$\\
\hline
\end{tabular*}
\caption{Values of the Hubble parameter and of the redshift at the transition, for the
power-law $f_{1}$CDM model of (\ref{Benghochea_fT}), according to (\ref{b-parameter}), for
various values of the model parameter $b$.
}
\label{Table1}
\end{table}

In order to cover all possible scenarios we discuss the case where $b=-1/3$ and the
initial Hubble value $0<H_{i}<H^{*}$. From Fig. \ref{Fig:phasespace1} we observe that the
universe interpolates smoothly between semi-stable Minkowski and stable de Sitter at
$-\infty\leq t\leq \infty$. Thus, the universe is non-singular and evolves effectively in
a phantom-like regime as $\dot{H}>0$. In addition, we can also see two other possible
behaviours in contraction phases ($H<0$), where
$-H^{*}<H_{i}<0$ and $H_{i}<H^{*}$.

We close the analysis of this model by examining the fulfillment of basic observational 
requirements. As we mentioned above, the power-law $f(T)$ model (\ref{Benghochea_fT}) 
reduces to $\Lambda$CDM cosmology for $b=0$, and thus we expect that the favored 
$b$-values of the model will be around this value. Indeed, confrontation with 
observations yields that the best fit on the parameter $b$, as measured from the 
combined cosmic chronometers (CC) + H$_0$ +  Supernovae Type I (SNIa) + Baryon Acoustic 
Oscillations (BAO) observational data, is $b=0.05536$ \cite{Nunes:2016qyp}. Then, the 
parameter $\alpha$ from (\ref{alphaBengochea}) is found to be $\alpha\sim 
-1.0543\times 
10^{-43}\textmd{ km}^{-1.8893}$ (the 
units of  $\alpha$  are   km$^{-2(1-b)}$). These values are confirmed by 
different data sets too, and in general at 3$\sigma$ one obtains $-1/3\leq 
b\leq 1/3$ \cite{Nesseris:2013jea,Nunes:2016plz,Qi:2017xzl}.

Finally, apart from the correct 
behaviour at late times we need to ensure that 
the model can realize the matter era too, and it proves that the phase portrait
analysis can be a powerful tool for this goal. In particular, examining the asymptotic
behaviour of phase portrait (\ref{Bengochea_phase_space}) at early times, namely at $H \gg 
H_{tr}$,
for $b\leq 1$ we find that $\dot{H}=-\frac{3}{2}(1+w )H^2$, while for $b>1$ we obtain
$\dot{H}=-\frac{3}{2}\frac{(1+w )}{b} H^2$, where the equation of state has to be set to 
$w=0$ 
since we focus on the dust case. Therefore, it is obvious that the matter dominated era 
can be obtain
only in the regime $b\leq 1$, which is consistent with the constraint of obtaining
late-time acceleration.

In summary, as we see from the application of the phase space portraits, the power-law
$f(T)$ model can lead to interesting phenomenology.

\subsection{$f_{2}$CDM model: $f(T)=T+\alpha
T_{0}\left(1-e^{-p\sqrt{T/T_{0}}}\right)$}\label{Sec4.
2}
In this model the $f(T)$ has a square-root exponential form
\cite{L10}
\begin{equation}\label{Linder_fT}
f(T)=T+\alpha T_{0}\left(1-e^{-p\sqrt{T/T_{0}}}\right),
\end{equation}
where $\alpha$ and $p$ are the two model parameters.
Inserting (\ref{Linder_fT})
into (\ref{Tor-density}) and then into the first Friedmann equation (\ref{MFR1})
at current time we acquire
\begin{equation}\label{alphaLinder}
\alpha=\frac{1-\Omega_{m,0}}{1-(1+p)e^{-p}}.
\end{equation}
This model reduces to $\Lambda$CDM cosmology for $p \rightarrow +\infty$.
Inserting (\ref{Linder_fT}) into the $f(T)$ phase portrait
(\ref{ps}) we obtain
\begin{equation}\label{Linder_phase_space}
\dot{H}=-\frac{3}{2}(1+w)H^{2}
\left\{\frac{1-\alpha
\left(\frac{H_{0}}{H}\right)^{2}\left[1-\left(1+p H/H_{0}\right)e^{-p
H/H_{0}}\right]}{1-\frac{1}{2}\alpha p^{2} e^{-p H/H_{0}}}\right\}.
\end{equation}

Focusing on the dust matter case $w =0$, in Fig. \ref{Fig:phasespace2} we present the
phase space portrait of the square-root exponential model (\ref{Linder_fT}) for some
choices of the parameter $p$. The fixed points correspond to $H^{*}=137.10,~56.89$ and
$-12.69$ km/s/Mpc, for $p=-2/3,~3$ and $1/2$, respectively.
As we observe, the phase portrait for $p=1/2$ reproduces the standard cosmology in $H>0$
region, however it does not possess a Minkowskian fate. Since its fixed point is
$H^{*}<0$, the universe will result into a contracting phase at asymptotically late
times, i.e the cosmological turnaround will be realized. On the other hand, the phase
portrait for $p=3$ leads to a universe evolution in agreement with standard cosmology
one. Moreover, the deceleration-to-acceleration transition occurs at $H=112.78$
km/s/Mpc, namely at $z\sim 0.68$, in agreement with   observations
\cite{Ade:2015xua}.

\begin{figure}[ht]
\begin{center}
\includegraphics[scale=0.7]{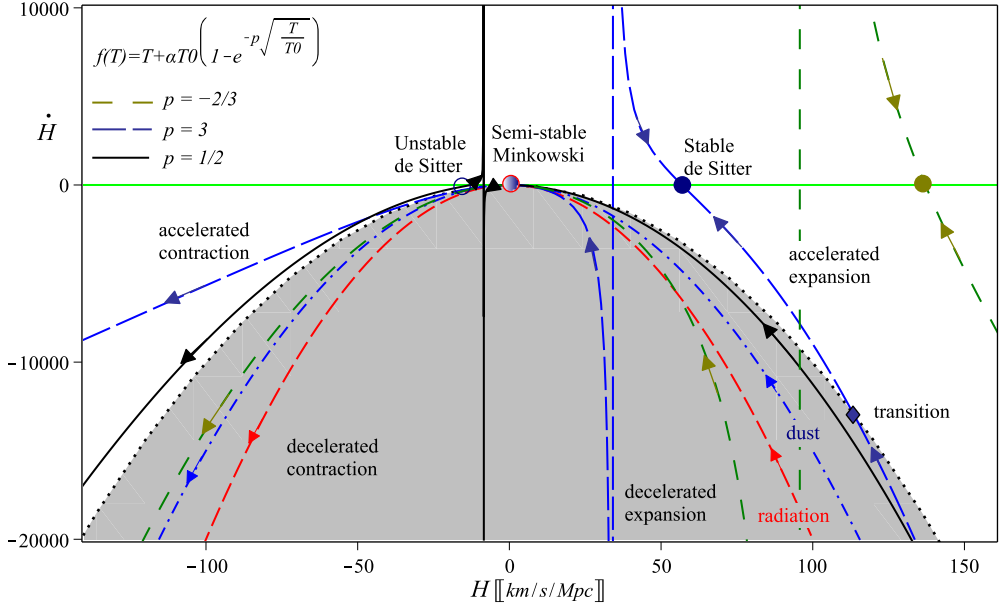}
\caption{{\it{
Phase space portraits for the square-root exponential $f_{2}$CDM model of
(\ref{Linder_fT}), according to (\ref{Linder_phase_space}), for three values of
the model parameter $p$. Similarly to the previous figures, the zero acceleration
boundary and the dust and radiation
curves, are drawn for convenience. The
scale of the graph is determined by the value of $T_0=-6H_0^2$, and thus by
the imposition of $H_{0}=76.11$ km/s/Mpc.
}}}
\label{Fig:phasespace2}
\end{center}
\end{figure}

Let us now examine for which values of the parameter $p$ the model is closer to the
observed evolution. Concerning late times we saw that the larger the $p$ is, the better
is the behaviour. Examining the asymptotic behaviour of
phase portrait (\ref{Linder_phase_space}) at early times, namely at $H \gg H_{tr}$,
for $p< 0$ we find that $
\dot{H}=\frac{3(1+w )}{p}H_{0} H$, while for $p>0$ we obtain
$\dot{H}=-\frac{3}{2}(1+w )H^2$, where $w$ has to be set to 0 since we focus
on the dust case. Therefore, for $p<0$ the phase portrait does not accept the correct
matter era (nevertheless, as can be seen from Fig. \ref{Fig:phasespace2}, the phase
portrait grows linearly as $H\to \infty$ and thus it has no initial finite-time
singularity, since it can be shown that if $\dot{H}\propto H^{r}$ ($r\leq 1$)
asymptotically then the initial singularity is absent \cite{awad2013}).
On the other hand, for $p>0$ we obtain the correct matter era. These intervals
are in agreement with the fact that $\Lambda$CDM cosmology is obtained for $p \rightarrow 
+\infty$. 
In particular, the combined CC + H$_0$ + SNIa + BAO 
observational data yield a best-fit value $b=0.04095$, for $b=1/p$  \cite{Nunes:2016qyp}
and consequently, via (\ref{alphaLinder}), we obtain for the 
dimensionless parameter $\alpha=0.7302$. Additionally, these values are confirmed by 
different data sets too, and in general at 3$\sigma$ one obtains
$p\geq3$   \cite{Nesseris:2013jea,Nunes:2016plz,Qi:2017xzl}.

We close the analysis by mentioning that as can be observed from
Figs. \ref{Fig:phasespace1} and \ref{Fig:phasespace2}, $f_{1}$CDM and $f_{2}$CDM
models exhibit common features in the $H>0$ patch, if we make the interchange $p=-1/b$
for $p<0$, or $p=1/b$ for $0<p<1$. However, in the contraction patch $H<0$
they present different behaviours.

\subsection{$f_{3}$CDM model: $f(T)=T+\alpha
T_{0}\left(1-e^{-pT/T_{0}}\right)$}\label{Sec4.3}
In this model the $f(T)$ has an exponential form
\cite{BGLL2011}
\begin{equation}\label{Bamba_fT}
f(T)=T+\alpha T_{0}\left(1-e^{-pT/T_{0}}\right),
\end{equation}
where $\alpha$ and $p$ are the two model parameters.
Inserting (\ref{Bamba_fT})
into (\ref{Tor-density}) and then into the first Friedmann equation (\ref{MFR1})
at current time we acquire
\begin{equation}\label{alphaBamba}
\alpha=\frac{1-\Omega_{m,0}}{1-(1+2p)e^{-p}}.
\end{equation}
This model reduces to $\Lambda$CDM cosmology for $p \rightarrow +\infty$.
Inserting (\ref{Bamba_fT}) into the $f(T)$ phase portrait
(\ref{ps}) we obtain
\begin{equation}\label{Bamba_phase_space}
\dot{H}=-3(1+w )H_{0}^{2}
\left\{\frac{H^2-\alpha\left[H_{0}^{2}-(H_{0}^{2}+2pH^{2})e^{\frac{-pH^{2}}{H_{0}^{2}}}
\right]}
{2H_{0}^{2}+2\alpha p (H_{0}^2-p H^{2})e^{\frac{-pH^{2}}{H_{0}^{2}}}}\right\}.
\end{equation}

In Fig. \ref{Fig:phasespace3} we present the
phase space portrait of the exponential model (\ref{Bamba_fT}) for some
choices of the parameter $p$, for the case of dust matter $w =0$.
The fixed points are obtained for
$H^{*}=\pm 92.22,~0$ and $\pm 58.34$ km/s/Mpc, for $p=-2/3,~3$ and $1/2$, respectively.
As we can see, the
choice $p=-2/3$ cannot lead to a universe evolution in agreement with the observed one.
For $p=1/2$ and $p=3$ we can see that the deceleration-to-acceleration transition occurs
at $H=134.93$ and $134.14$ km/s/Mpc respectively. However, for $p=3$ the phase
portrait indicates another transition back to deceleration at $H\sim 84.1$ km/s/Mpc, with
a finite time singularity, while for $p=1/2$ the universe can evolve towards a future 
fixed point. In general, the larger the value of $p$ is, the more realistic is the 
cosmological behavior. 
This is in agreement with the fact that $\Lambda$CDM cosmology is obtained for $p 
\rightarrow +\infty$.  Moreover, this is confirmed by detailed confrontation with 
observations, using combined CC + H$_0$ + SNIa + BAO 
data sets, which yield a best-fit value  $b=0.03207$ for  $b=1/p$, and consequently, 
through   (\ref{alphaBamba}) to $\alpha=0.7352$ \cite{Nunes:2016qyp}. In general, at 
3$\sigma$ one obtains
$p\geq3$   \cite{Nesseris:2013jea,Nunes:2016qyp,Nunes:2016plz,Qi:2017xzl}.

\begin{figure}[ht]
\begin{center}
\includegraphics[scale=0.7]{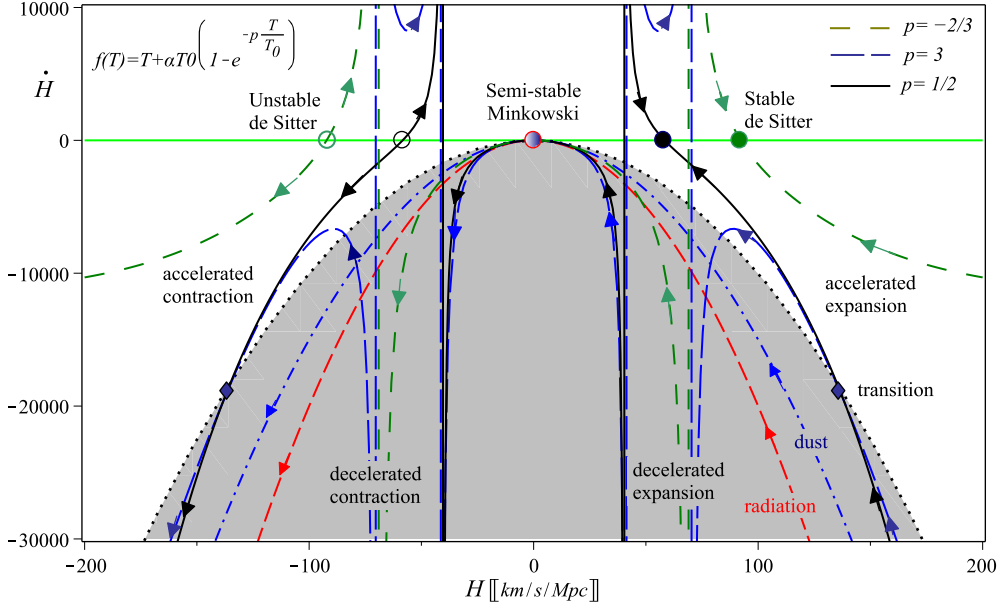}
\caption{
{\it{
Phase space portraits for the exponential $f_{3}$CDM model of
(\ref{Bamba_fT}), according to (\ref{Bamba_phase_space}), for three values of
the model parameter $p$. Similarly to the previous figures, the zero acceleration
boundary and the dust and radiation
curves, are drawn for convenience. The
scale of the graph is determined by the value of $T_0=-6H_0^2$, and thus by
the imposition of $H_{0}=76.11$ km/s/Mpc.
}}}
\label{Fig:phasespace3}
\end{center}
\end{figure}

Finally, examining the asymptotic behaviour of the phase portrait 
(\ref{Bamba_phase_space}) at 
early times, namely at $H \gg H_{tr}$,
for $p< 0$ we find that $\dot{H}=\frac{3}{2}\frac{(1+w )}{p}H_{0}^2$, while for $p>0$ we 
obtain
$ \dot{H}=-\frac{3}{2}(1+w ) H^2$, where the equation of state has to be set to $w=0$ 
since we focus
on the dust case. Hence, in the parameter region $p< 0$ the phase portrait asymptotically 
evolves 
towards a constant value and thus the model does not accept the
correct matter era. On the other hand, for $p>0$, apart from late-time acceleration, the 
model 
exhibits the correct matter epoch at early times.

\section{A new viable $f(T)$ model}\label{Sec5}
In the previous section we used the basic advantage that in $f(T)$ cosmology in a flat
FRW geometry all quantities can be expressed as functions of the Hubble function $H$, and
we transformed the cosmological equation into a one-dimensional autonomous system whose
phase space portraits could reveal the basic cosmological features and behaviours.
Additionally, after we explored the general properties of $f(T)$ cosmology, we studied
three specific viable $f(T)$ models characterized by two parameters, which are in the
best agreement with observations. In the present section we use as a guide the basic
features arisen from the above investigation of the phase space portraits, in order to
construct a new model of $f(T)$ gravity. This model proves to be efficient in the
description of the cosmological history of the universe.
\subsection{The model: $f(T)=Te^{~\beta T_{0}/T}$}\label{Sec5.1}
The simple model that we propose in this work is
\begin{equation}\label{f(T)}
f(T)=Te^{~\beta T_{0}/T},
\end{equation}
where $\beta$ is the single dimensionless model parameter.
Inserting this relation into (\ref{Tor-density}), (\ref{Tor-press}), and using
(\ref{TorHubble}), namely that $T=-6H^2$, we find
\begin{eqnarray}
\label{rhoTourmodel}
&&\rho_{T}(H)=\frac{3}{\kappa^2} \left[ H^2-( H^2-2\beta H_{0}^{2})
e^{\beta \frac{H_{0}^{2}}{H^2}}\right],\label{rhoT}
\\
&&p_{T}(H)=
- \frac{3\beta H_0^2 H^2 }{\kappa^2}\left[ \frac{H^2+2\beta H_0^2}{H^4-\beta H_0^2
H^2+2\beta^2
H_0^4}
\right]
,\label{pTourmodel}
\end{eqnarray}
and thus the (torsion originated) dark energy equation-of-state parameter
becomes
\begin{equation}\label{Tor_EoS}
w_{T}(H)=
\frac{ - \beta H_0^2 H^2 \left(H^2+2\beta H_0^2\right)}{\left[H^4-\beta H_0^2
H^2+2\beta^2
H_0^4\right] \left[ H^2-( H^2-2\beta H_{0}^{2})
e^{\beta \frac{H_{0}^{2}}{H^2}}\right]}
.
\end{equation}
Inserting (\ref{rhoTourmodel}) into the first Friedmann equation
(\ref{MFR1}) at current time we express $\beta$ in terms of the current value of the
matter density parameter, namely
\begin{equation}\label{beta2}
\beta = \frac{1}{2} + W\left(-\frac{1}{2} e^{-1/2}\, \Omega_{m,0}\right),
\end{equation}
where $W(x)$ denotes the Lambert-$W$ function, which is the solution of the
transcendental equation $We^{W}=x$.
Using $\Omega_{m,0}=0.318$, we acquire $\beta\sim 0.393$, or $\beta\sim
-3.127$.

\subsection{Phase space portraits}
Let us now explore the phase
space portrait of the new $f(T)$ model (\ref{f(T)}). Inserting (\ref{rhoTourmodel}) and
(\ref{pTourmodel}) into the conservation equation (\ref{continuity1}) (or inserting it
straightaway into (\ref{ps})) we obtain
the one-dimensional phase-portrait equation
\begin{equation}\label{Hdot}
\dot{H}=-\frac{3}{2}(1+w)\frac{(H^2-2\beta H_{0}^{2})H^4}{
H^4 -\beta H_{0}^2 H^2+ 2\beta^2H_{0}^{4}}.
\end{equation}
\begin{figure}[ht]
\begin{center}
\includegraphics[scale=0.7]{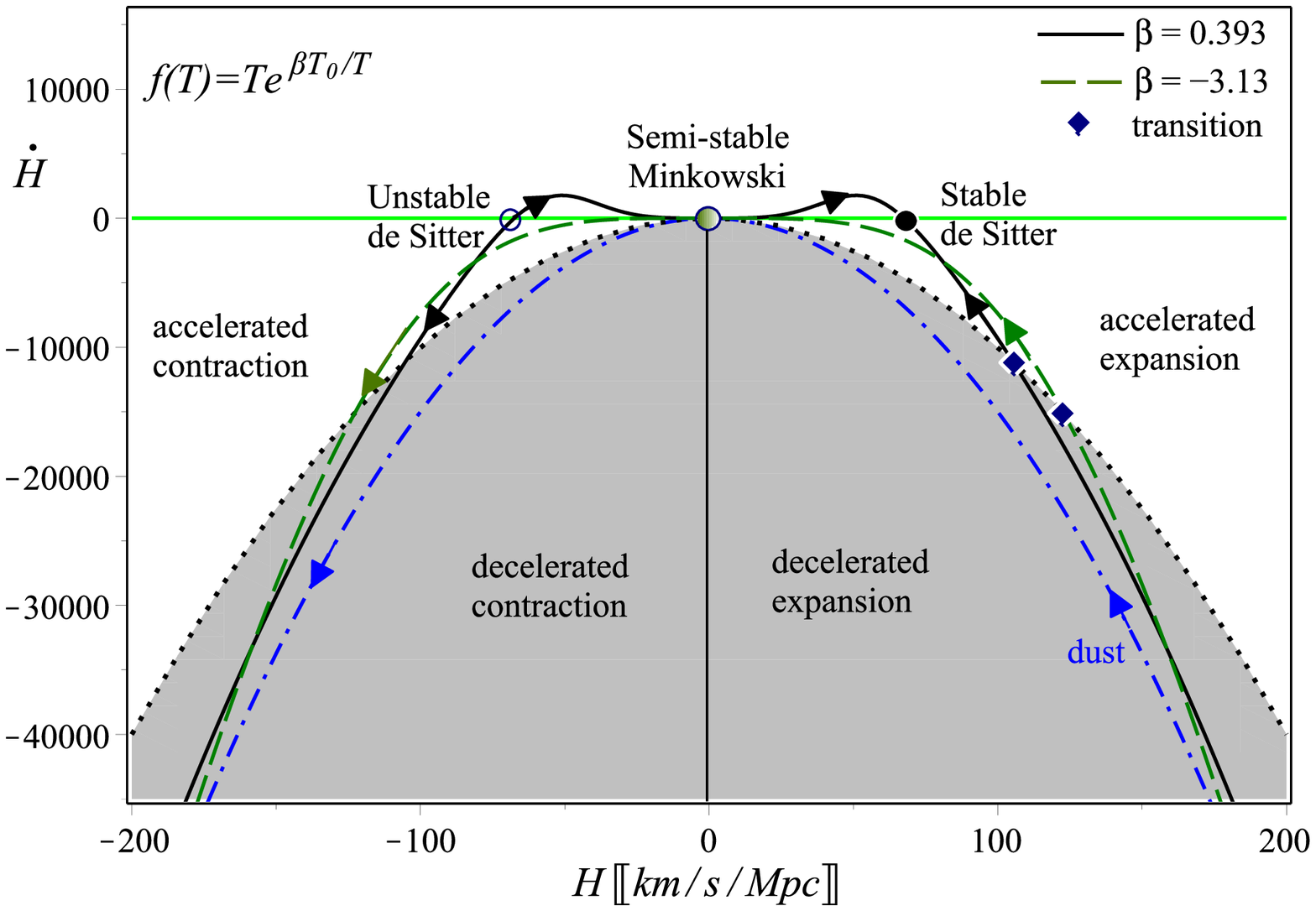}
\caption{
{\it{
Phase space portraits for the new model of
(\ref{f(T)}), according to (\ref{Hdot}), for two values of
the model parameter $\beta$. Similarly to the previous figures, the zero acceleration
boundary and the dust curve are drawn for convenience. The
scale of the graph is determined by the value of $T_0$, and thus by
the imposition of $H_{0}=76.11$ km/s/Mpc.
}}}
\label{Fig:phasespace4}
\end{center}
\end{figure}

In Fig. \ref{Fig:phasespace4} we present the corresponding phase portrait for the two
values
of $\beta$ calculated above, corresponding to $\Omega_{m,0}=0.318$ through (\ref{beta2}).
Note that for $\beta<0$ the system has exactly one fixed point at $H=0$, while for
$\beta>0$ there exist three fixed points at
$H=0,\pm\sqrt{2\beta}~ |H_{0}|$.
Nevertheless, both $\beta$ cases exhibit a transition from deceleration to
acceleration at late times. Additionally, at early times, namely at large $H$, both
$\beta$ cases can describe the correct matter era, since in this regime we
asymptotically have
$ \dot{H}=-\frac{3}{2}(1+w)H^2$.

Let us now discuss in more details the phase-space portrait features for the two $\beta$
cases separately.
\begin{itemize}
\item {\boldmath $\beta<0$}. As we observe from Fig. \ref{Fig:phasespace4}, the
phase portrait splits the phase space into two patches: (i) When $H<0$ the universe has
no initial singularity but it contracts towards a future Big Crunch, namely towards a
finite-time singularity. (ii) When $H>0$, the universe begins with a Big
Bang singularity, it matches   standard
cosmology during the intermediate-time region, and it exhibits a late-time acceleration
phase. However, unlike $\Lambda$CDM model, the universe evolves towards a Minkowskian
future fixed point.

\item {\boldmath $\beta>0$}. As we can see from Fig. \ref{Fig:phasespace4}, the phase
portrait
splits the phase
space into four distinguishable patches: (i) When $H<-\sqrt{2\beta}~ |H_{0}|$ the
universe does
not have a past singularity, it is contracting, and it evolves
towards a future Big Crunch singularity. (ii) When $-\sqrt{2\beta}~ |H_{0}|<H<0$, the
universe
is contracting and non-singular, interpolating smoothly between de Sitter and Minkowski
solutions in a phantom-like regime ($\dot{H}>0$). (iii) When
$0<H<\sqrt{2\beta}~ |H_{0}|$, we obtain an expanding accelerating universe, which is
non-singular and lies in the phantom regime. (iv) When $H>\sqrt{2\beta}~ |H_{0}|$,
the universe begins with a finite time singularity (Big Bang) with a decelerated
expansion phase, it matches   standard cosmology during the intermediate-time region,
then at late times it enters into a non-phantom (since $\dot{H}<0$) accelerated
expansion, and finally in the future the universe results to a de Sitter solution. In
this procedure the value of $\beta$ controls the various transitions points, and in
particular the transition from deceleration to acceleration. In particular,
a larger $\beta$ value corresponds to a later transition to acceleration
phase. In summary, this last region is the one that is in agreement with the observed
universe evolution.
\end{itemize}
Having explored the basic features of the phase space portrait of the model at hand, let
us extract analytical relations for the deceleration-to-acceleration transition. From
Eq. (\ref{Hdot}) we can see that the zero acceleration curve, i.e
$\dot{H}=-H^{2}$, is crossed at
\begin{equation}\label{phase-transitions}
H_{tr}=\pm H_{0}\sqrt{\frac{\beta\left(2+3w \pm\sqrt{8+24w +9w ^2}\right)}{(1+3w )}}.
\end{equation}
Focusing on the expanding universe $H>0$ and assuming the matter to be dust ($w =0$) we
result to the simple relation
\begin{equation}\label{beta}
\beta= \left( \frac{- 1 \mp \sqrt{2}}{2}\right) \left(\frac{H_{tr}}{H_{0}}\right)^2.
\end{equation}
Hence, if the Hubble value at the transition $H_{tr}$ is given accurately from
observations then the
value of the parameter $\beta$ can be calculated. In Table
\ref{Table2} we give the estimated values of $\beta$ according to different choices
of $H_{tr}$ (they correspond to $z_{tr}\sim 0.5 - 0.7$ according to observations
\cite{Ade:2015lrj}).
The results show that $0.363 \lesssim \beta \lesssim 0.464$ or $-2.707 \lesssim \beta
\lesssim -2.
117$, which as expected are
consistent with the $\beta$ values that were used in Fig. \ref{Fig:phasespace4} and were
arisen from the requirement $\Omega_{m,0}=0.318$ (definitely the measurements of
$\Omega_{m,0}$ and $H_0$ are more accurate than the measured of $H_{tr}$ ).
\begin{table}[ht]
\begin{tabular*}
{\columnwidth}{@{\extracolsep{\fill}}ccc@{}}
\hline
$z_{tr}$ & $H_{tr}$ [km/s/Mpc] & $\beta$\\[5pt]
\hline
$0.5$& $\sim 100.8$ & $\sim -2.117$ or $0.363$\\
$0.6$& $\sim 107.18$ & $\sim -2.394$ or $0.411$\\
$0.7$& $\sim 113.97$ & $\sim -2.707$ or $0.464$\\
\hline
\end{tabular*}
\caption{
Values of the   $\beta$ parameter of the new model
(\ref{f(T)}), and of the corresponding Hubble parameter and redshift
at the transition, according to (\ref{beta}).}
\label{Table2}
\end{table}

\subsection{Cosmological evolution}\label{Sec5.2} 
In this subsection we explore some features of the cosmological evolution in the scenario
of the new $f(T)$ model (\ref{f(T)}). First of all,
inserting (\ref{f(T)}) into (\ref{FR1H})
and (\ref{FR2H}), using also (\ref{TorHubble}), we obtain
\begin{eqnarray}
\rho(H)&=&\frac{3}{\kappa^2}\left(H^2-2\beta H_0^2\right)e^{\beta
\frac{H_0^2}{H^2}},\label{density}\\
p(H)&=&-\frac{2\dot{H}}{\kappa^2}\left[1-\beta
\left(\frac{H_0}{H}\right)^{2}+2\beta^2 \left(\frac{H_0}
{H}\right)^{4} \right]e^{\beta \frac{H_0^2}{H^2}}-\rho(H).\label{pressure}
\end{eqnarray}
Now since integration of the continuity equation
(\ref{continuity1}) in the case of dust matter   gives
$\rho(H)=\rho_{0}/a(H)^{3}
$, using (\ref{density}) we obtain the scale factor as a function
of the Hubble parameter
as
\begin{equation}\label{scf}
a(H)=\left(\frac{\Omega_{m,0}H_0^2}{\Omega_{m}H^2}\right)^{1/3}=\frac{\Omega_{m,0}^{1/3}
H_0^{2/3}e^{\frac{-\beta H_0^2}{3 H^{2}}}}{(H^{2}-2 \beta H_0^2)^{1/3}},
\end{equation}
where
$\Omega_{m}=\frac{\kappa^2 \rho}{3 H^2}$ and $\Omega_{m,0}$ its value at present.

In the following we focus on the $\beta>0$ case, since as we discussed in the phase-space
portrait analysis of the previous subsection, it is the one that can lead to a universe
evolution in agreement with the observed one.
Using (\ref{scf}) we obtain the redshift as a function of the Hubble parameter $H$ as
\begin{equation}\label{redshift}
z(H)=\left(\frac{H^{2}-2 \beta H_0^2}{H_0^{2}\Omega_{m,0}}\right)^{1/3}e^{\frac{\beta
H_0^2}{3
H^{2}}}-1.
\end{equation}
The inverse relation of (\ref{redshift}) gives $H\equiv H(z)$, which can be shown
graphically in Fig. \ref{Fig:cosm-parameters}\subref{fig:Hubble-parameter}.
\begin{figure}[ht]
\centering
\subfigure[~Hubble
parameter]{\label{fig:Hubble-parameter}\includegraphics[scale=.25]{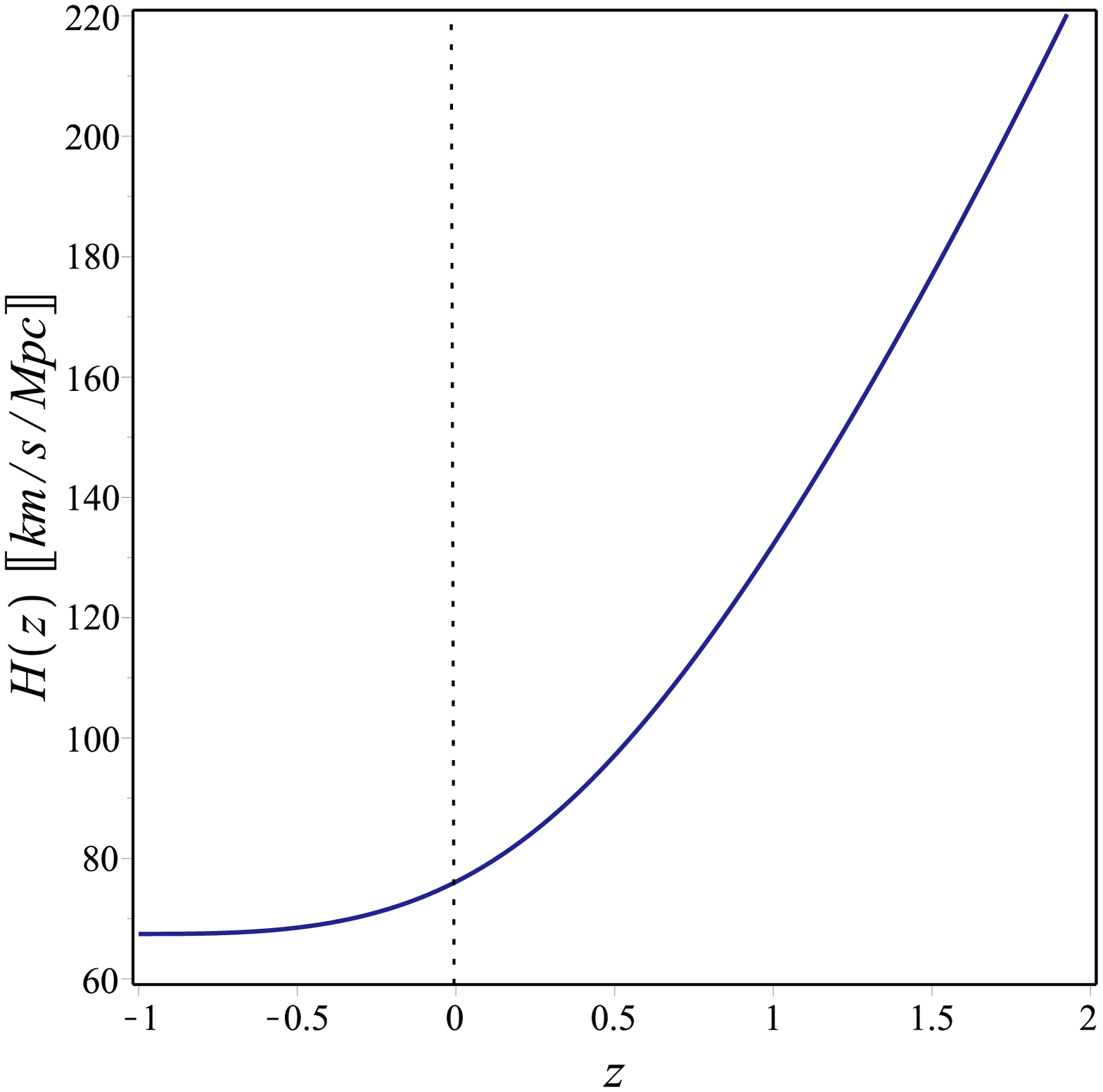}}
\subfigure[~Density
parameters]{\label{fig:dens-parameters}\includegraphics[scale=.25]{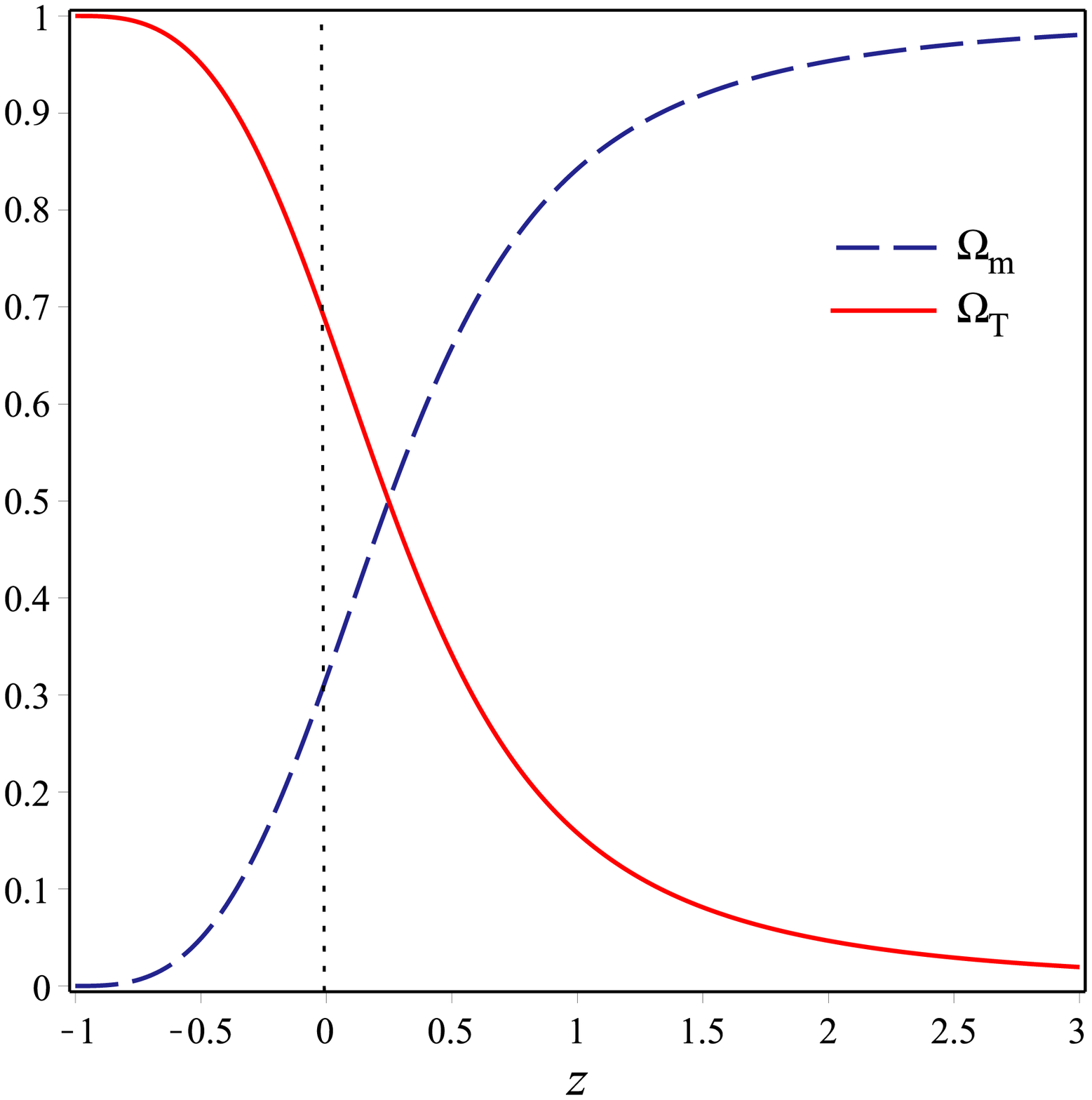}
}
\subfigure[~Deceleration
parameter]{\label{fig:decel-parameter}\includegraphics[scale=.25]{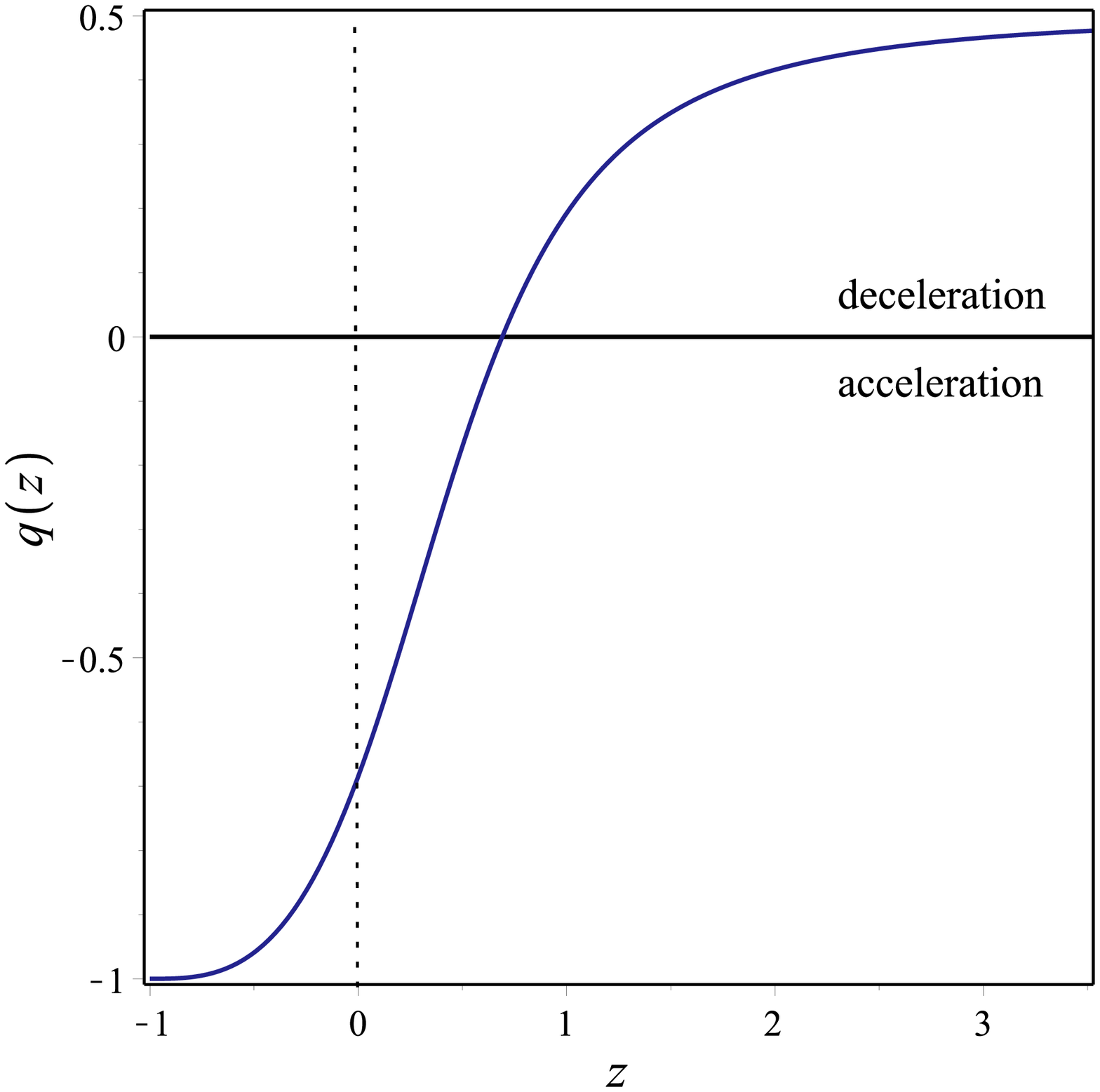}}
\caption[figtopcap]{
{\it{ Cosmological evolution of various quantities, in the new model of
(\ref{f(T)}), as a function of the redshift. Left graph: The evolution of the Hubble
parameter $H(z)$ from the inverse relation of
(\ref{redshift}). Middle graph: The evolution of the matter and (torsional) dark
energy density parameters, $\Omega_m(z)$ and $\Omega_T(z)$, from (\ref{Omega_m}) and
(\ref{Omega_T}) respectively. Right graph:
The evolution of the deceleration parameter $q(z)$ from (\ref{q}).
We have taken $\beta=0.393$, and we have set the present values as $H_{0}
=76.11$ km/s/Mpc and $\Omega_{m,0}=0.318$.}}}
\label{Fig:cosm-parameters}
\end{figure}

Similarly, from (\ref{density}) we extract the matter density parameter as
\begin{equation}\label{Omega_m}
\Omega_{m}(H)=\frac{(H^2-2 \beta H_0^2)}{H^2}e^{~\beta\frac{ H_0^2}{H^{2}}},
\end{equation}
while from (\ref{rhoTourmodel}) the (torsional) dark energy density parameter reads as
\begin{equation}\label{Omega_T}
\Omega_{T}(H)= 1-\frac{(H^2-2 \beta H_0^2)}{H^2}e^{~\beta\frac{H_0^2}{H^{2}}}.
\end{equation}
Thus, using the inverse of (\ref{redshift}), as well as (\ref{Omega_m}) and
(\ref{Omega_T}), in Fig. \ref{Fig:cosm-parameters}\subref{fig:dens-parameters} we depict
$ \Omega_{m}$ and $ \Omega_{T}$ as functions of $z$. The current
value of $ \Omega_{m}$ is $\Omega_{m}(z=0)=\Omega_{m,0}\approx 0.318$, while it
approaches   unity at larger redshift values.
On the other hand, $ \Omega_{T}$ is almost
zero at large $z$, while its current value is $\Omega_{T}(z=0)\approx 0.682$. These
behaviours are in a very good agreement with
observations. Additionally, extending the graphs to the far future, namely at
$z\rightarrow-1$, we can see that $\Omega_{m}$ drops to zero while $\Omega_{T}\to 1$, i.e
the universe results in a de Sitter phase.

Concerning the deceleration parameter (\ref{deccelmodI}), in the case of dust matter
using (\ref{scf}) it becomes
\begin{equation}\label{q}
q(H)= \frac{1}{2}\,\frac{H^4-4 \beta H_0^2 H^2-4 \beta^2 H_0^4}{H^4-\beta H_0^2 H^2+2
\beta^2
H_0^4}.
\end{equation}
Therefore, using the inverse relation of (\ref{redshift}) in Fig.
\ref{Fig:cosm-parameters}\subref{fig:decel-parameter} we present
$q(z)$. As we observe, the universe was decelerating at early times, ($q\to 0.5$ at
larger $z$ as expected for a dust-matter
dominated phase), it experienced the deceleration-to-acceleration transition at redshift
$z_{tr}\approx 0.63$ (in agreement with the expected range $0.6
\lesssim
z_{tr} \lesssim 0.8$), and at present it has a value $q(z=0)\approx -0.649$.

Let us now focus on the (torsional) dark energy equation-of-state parameter $w_{T}$ of
(\ref{Tor_EoS}), which using the inverse relation of (\ref{redshift}) can be expressed as
a function of the redshift. In Fig. \ref{Fig:EoS}\subref{fig:DE-EoS}
we depict $w_{T}(z)$ for $\beta=0.393$, which is consistent with $\Omega_{m,0}=0.318$ and
$H_{0}=76.11$ km/s/Mpc. At large $z$ the graph shows that $w_T(z)\to -1$, i.e
the (torsional) dark energy behaves as a cosmological constant, nevertheless since at
the same time $\Omega_{T}(z)$ is almost zero (see Fig.
\ref{Fig:cosm-parameters}\subref{fig:dens-parameters}) we do not expect any
deviation from standard cosmology and the galaxy formation.
However, at smaller redshifts $w_{T}(z)$ evolves in the phantom regime, still inside the
observational bounds \cite{Ade:2015lrj,Cepa:2004bc,DiValentino:2016hlg}.
Moreover, at the limit $z\to -1$,
$w_{T}(z)$
evolves towards $-1$, consistently with a dark-energy dominated de Sitter universe.

\begin{figure}[ht]
\centering
\subfigure[Dark-energy equation-of-state
parameter]{\label{fig:DE-EoS}\includegraphics[scale=.37]{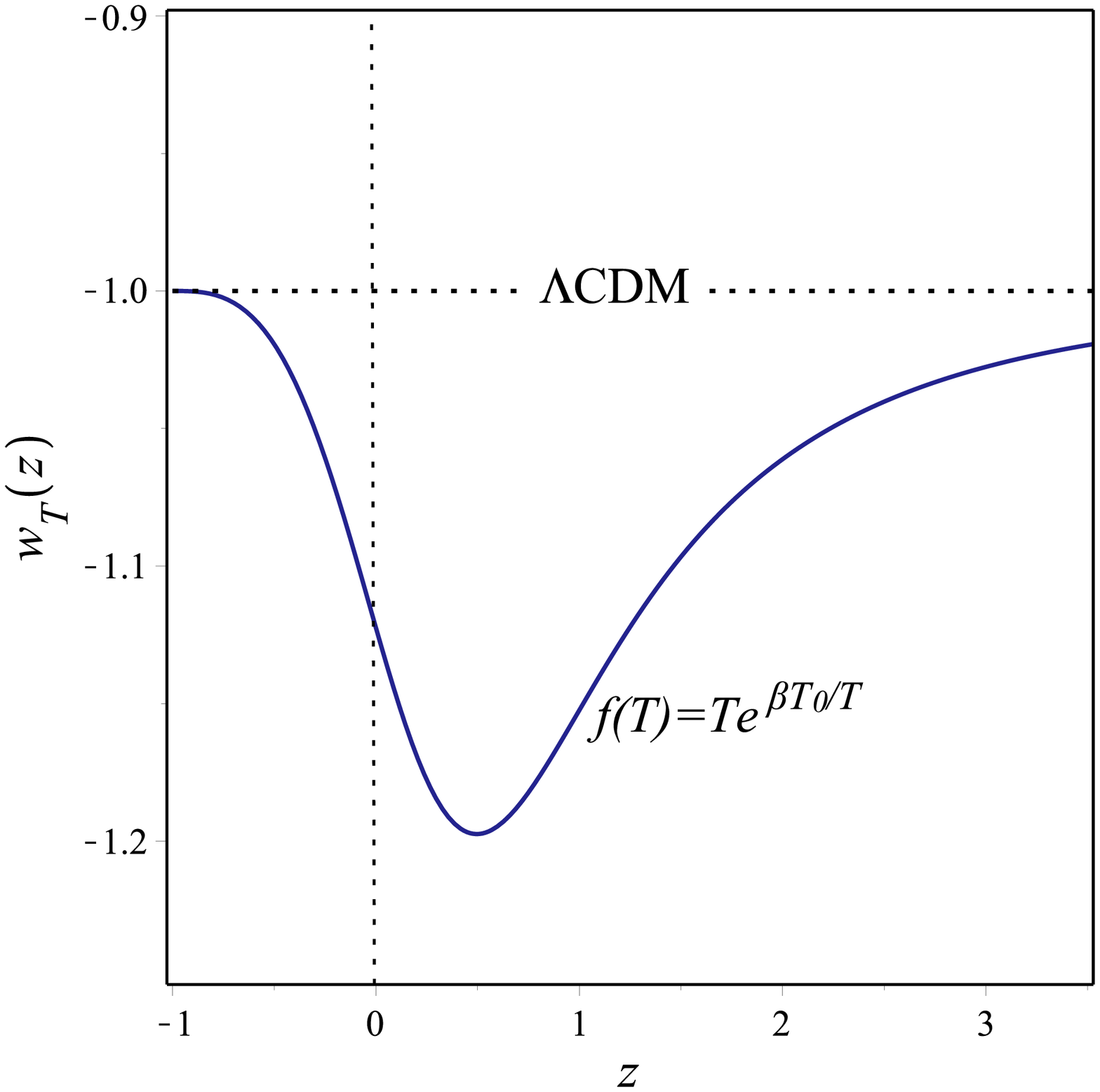}}\hspace{0.4cm}
\subfigure[Effective
equation-of-state parameter
]{\label{fig:Eff-EoS}\includegraphics[scale=.37]{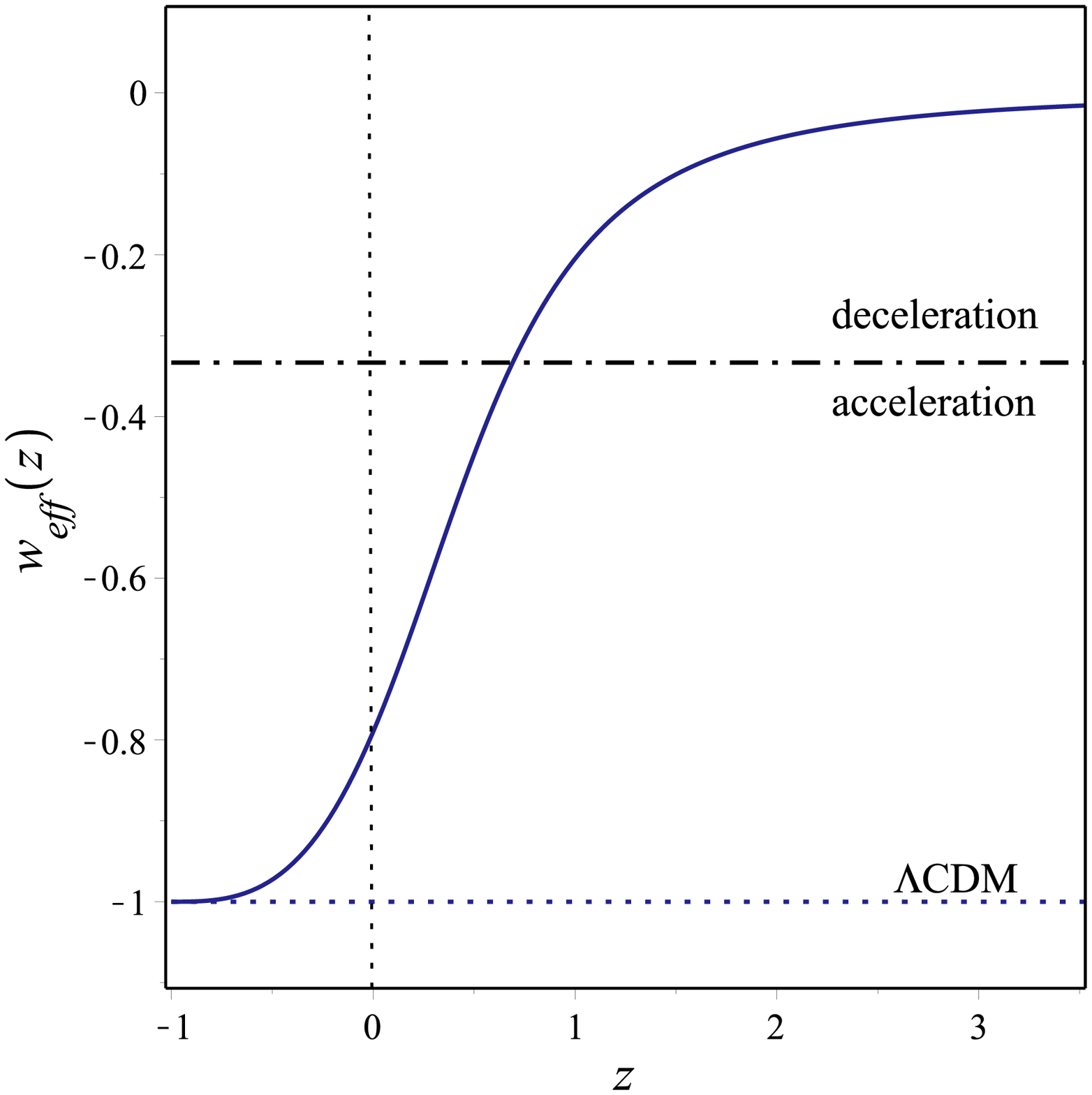}}
\caption[figtopcap]
{
{\it{ Left graph:
Evolution of the dark energy equation-of-state parameter $w_T(z)$ as a function of the
redshift, from (\ref{Tor_EoS}) using the inverse of
(\ref{redshift}), in the new model of
(\ref{f(T)}). Right graph: the corresponding evolution of the effective (total)
equation-of-state parameter $w_{eff}(z)$, from (\ref{eff_EoS2}) using the inverse of
(\ref{redshift}).
We have taken $\beta=0.393$, and we have set the present values as $H_{0}
=76.11$ km/s/Mpc and $\Omega_{m,0}=0.318$.}}}
\label{Fig:EoS}
\end{figure}

Similarly, let us examine the behaviour of the total equation-of-state parameter of the
universe $w_{eff}$. Inserting (\ref{Hdot}) into (\ref{eff_EoS0}) we can express it as a
function of $H$ as
\begin{equation}\label{eff_EoS2}
w _{eff}(H)=-1+\frac{(1+w ) (H^2-2\beta H_0^2)H^2}{(H^4-\beta H_0^2 H^2+2 \beta^2
H_0^4)},
\end{equation}
which for the case of dust matter ($w=0$) and using the inverse relation of
(\ref{redshift}), can result in a relation $w_{eff}(z)$. In Fig.
\ref{Fig:EoS}\subref{fig:Eff-EoS} we depict $w_{eff}(z)$ for the same $\beta$ values
with Fig. \ref{Fig:EoS}\subref{fig:DE-EoS}. As we observe, $w_{eff}$ behave as dust
at large $z$ (as expected during matter domination), it crosses $w_{eff}=-1/3$ (which
corresponds to the deceleration-to-acceleration transition) at redshift $z_{tr}\approx
0.63$, and it drops to $w_{eff}(z)\approx -0.766$ at present ($z=0$). These behaviours are
in agreement with observations \cite{Ade:2015lrj,Mukherjee:2016eqj}. Finally, in the
far future, namely at the limit $z\to -1$, $w_{eff}$
evolves towards $-1$, consistently with a de Sitter universe.

Lastly, from the above analysis we have all the materials to estimate the age of the
universe
\begin{equation}t_{age}=-\int_{H_{0}}^{\infty}\dot{H}^{-1} dH.
\end{equation}
Imposing $H_{0}=76.11$ km/s/Mpc, $\beta=0.426$ and
using the phase portrait of the model at hand, namely (\ref{Hdot}), we find $t_{age}\sim
13.7$
billion years. Although the value $\beta=0.426$ is consistent with the results of Table
\ref{Table2}, the model in this case gives a lower matter density parameter 
$\Omega_{m,0}\sim 0.227$
as
predicted by (\ref{beta2}). However, by setting $\beta=0.393$ (which gives
$\Omega_{m,0}=0.318$)
and $H_{0}=70$ km/s/Mpc, the model predicts an age of $\sim 13.6$ billion years. Hence, we
conclude
that even if the current Hubble
constant $H_{0}$ and the matter density parameter $\Omega_{m,0}$ have large values, the
model can
predict an age consistent with the WMAP and Planck results \cite{Ade:2015xua}.

\subsection{Two diagnostic tests}\label{Sec:5.4} 
We close this section by performing two diagnostics tests on the new $f(T)$ model
proposed in (\ref{f(T)}).
The first test is
the $Om(z)$ diagnostic test \cite{Sahni:2008xx}, and it is useful in order to resolve the
known degeneracy that exists between the dark-energy equation-of-state parameter, which
in the present work is of torsional origin, namely $w_{T}$, with the dark-matter density
parameter $\Omega_{m}$.
The second study concerns the sound speed of the scalar perturbations of the theory.

The $Om(z)$ diagnostic
test is a useful tool to distinguish a specific dark energy
model amongst others, as well as from the $\Lambda$CDM cosmology. The $Om(z)$ diagnostic
is defined by \cite{Sahni:2008xx}
\begin{equation}\label{Om}
Om(z)=\frac{\left(H(z)/H_{0}\right)^2-1}{(1+z)^3-1},
\end{equation}
which has less dependence on the matter density parameter and can be determined by the
value of
$H_{0}$. For $\Lambda$CDM paradigm it simply gives a straight line in ($z,Om(z)$) plane,
since
$H^2\propto (1+z)^3$, while the dynamical dark energy models   lead to curves. In
phantom dark energy models ($w_{DE}<-1$), $Om(z)$
has a negative slope, while in models where dark energy lies in the quintessence regime
($w_{DE}>-1$) it has a positive slope.

In the model at hand we can use the inverse relation of (\ref{redshift}) in order to
extract the corresponding $Om(z)$ through (\ref{Om}), and in Fig.
\ref{Fig:diagnostics}\subref{fig:Om-diagnostic} we depict it.
This graph confirms that in the present model the torsional dark energy lies in the
phantom regime, since the slope is negative. This is in agreement
with the combination of Supernovae Type I (SNIa), Baryon Acoustic Oscillations (BAO) and
Cosmic Microwave Background (CMB) data at 1$\sigma$ confidence level
\cite{Sahni:2008xx,Lonappan:2017hok}.
At large $z$ the graph matches $\Lambda$CDM cosmology, as we also found in the previous
subsections.
\begin{figure}[ht]
\centering
\subfigure[~$Om(z)$ diagnostic
test]{\label{fig:Om-diagnostic}\includegraphics[scale=0.38]{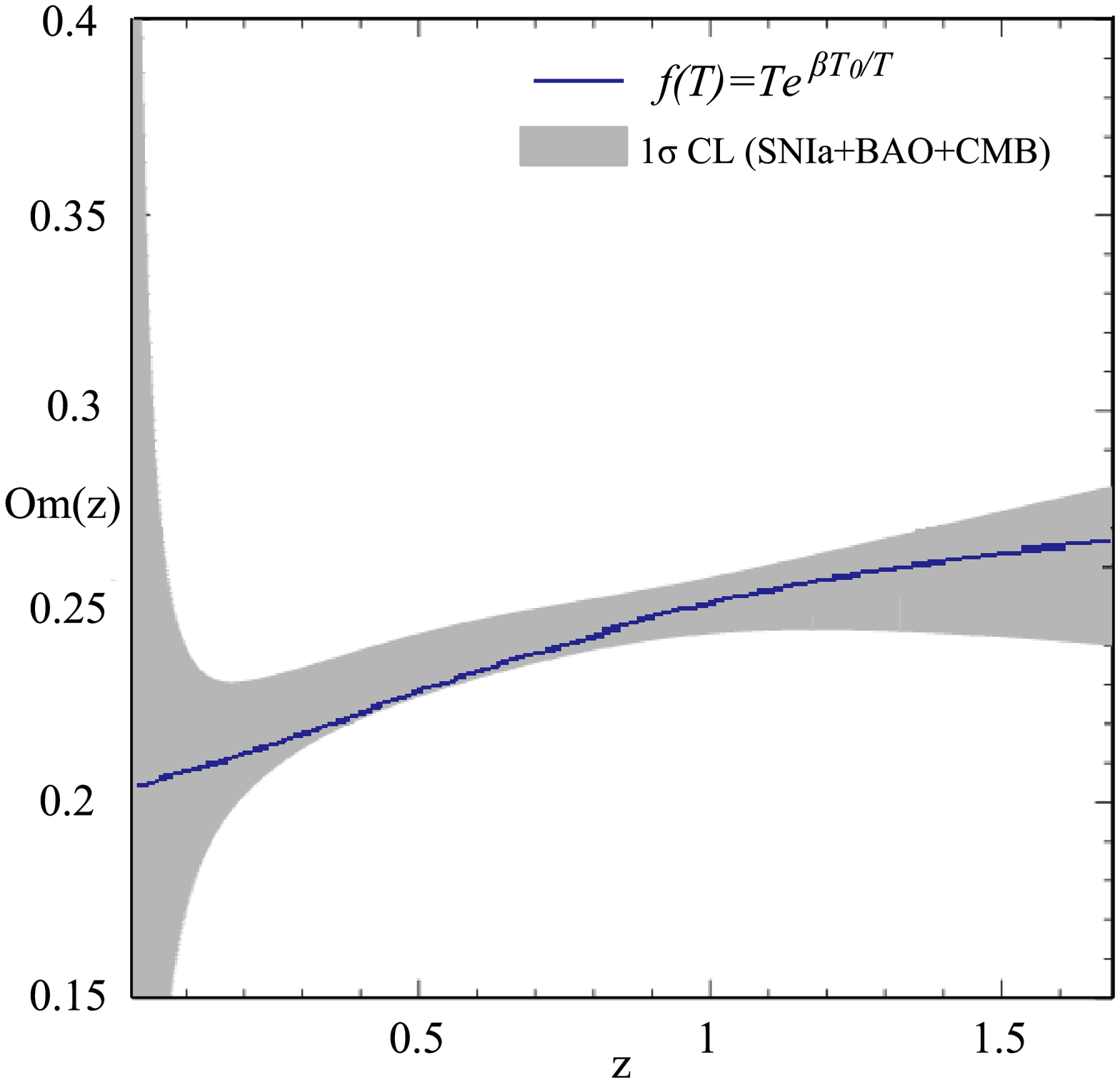}
}
\subfigure[~Squared sound speed
of scalar perturbations]{\label{fig:sound-speed}\includegraphics[scale=0.37]{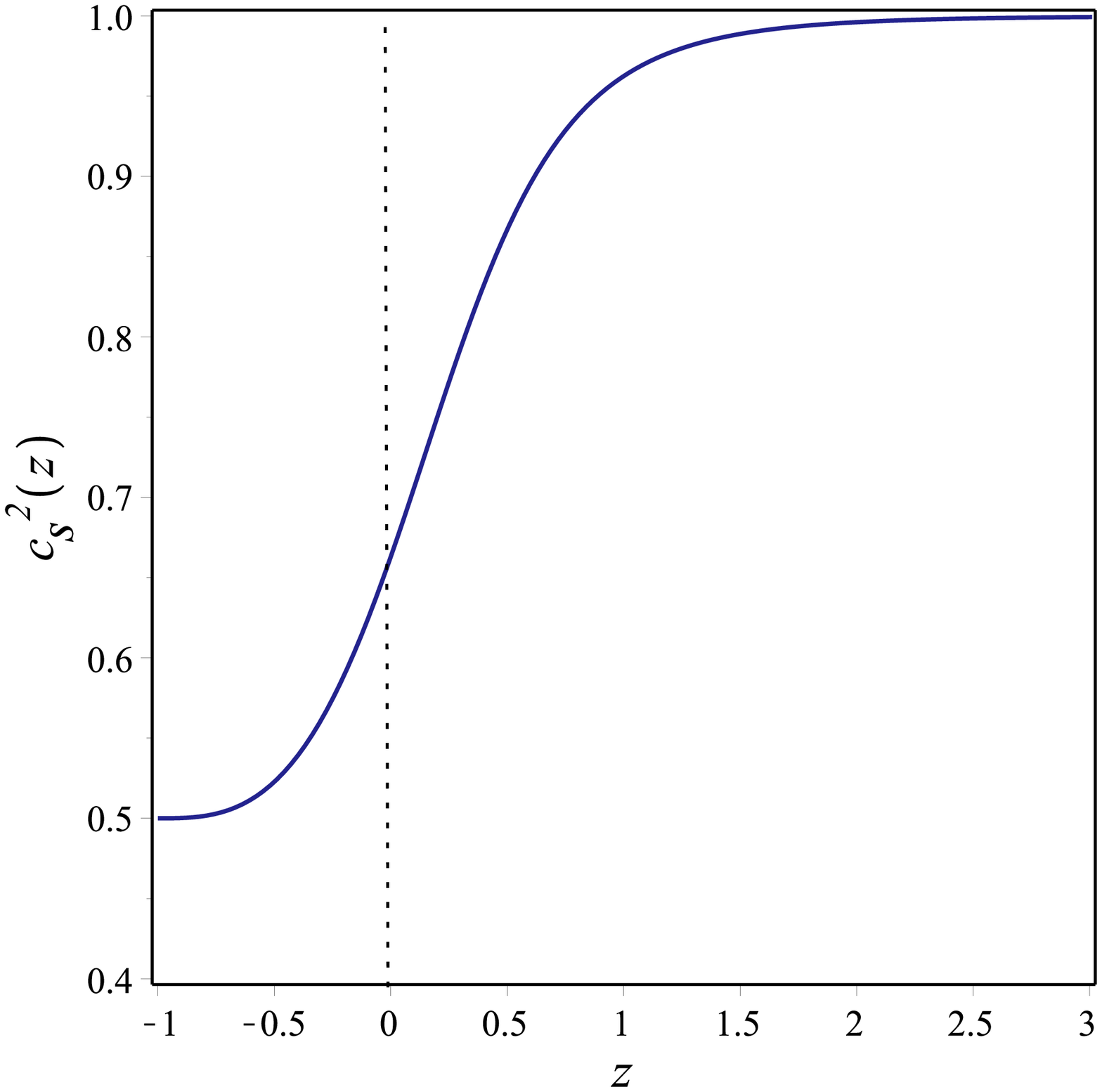}}
\caption[figtopcap]{
{\it{ Left graph:
Evolution of the $Om(z)$ parameter (\ref{Om}) for the $Om(z)$ diagnostic
test, using also the inverse of
(\ref{redshift}), for the new model of
(\ref{f(T)}). The gray-shaded region defines the $1\sigma$ confidence level fittings
using the combined SNIa, BAO and CMB data
\cite{Sahni:2008xx}. Right graph: Evolution of the squared sound speed
of the scalar perturbations using (\ref{sound-speed}), using also the inverse of
(\ref{redshift}).
We have taken $\beta=0.393$, and we have set the present values as $H_{0}
=76.11$ km/s/Mpc and $\Omega_{m,0}=0.318$.}}}
\label{Fig:diagnostics}
\end{figure}

Let us now come to the behaviour of scalar perturbations in the theory, since for every
gravitational theory it provides an important test for its stability and validity. In
viable models the square of the sound speed of these perturbations should be $0\leq
c_{s}^{2}\leq 1$ in order to maintain the causality and stability conditions. For
$f(T)$ gravity in an FRW background the squared sound speed  of
  scalar perturbation is given by \cite{Chen:2010va,Cai:2011tc}
\begin{equation}\label{sound-speed0}
c_s^2 = \frac{f_{T}}{f_T+2T f_{TT}},
\end{equation}
which for the $f(T)$ model (\ref{f(T)}), and using that $T=-6H^2$, it becomes
\begin{equation}\label{sound-speed}
c_s^2(H) = \frac{(H^2-\beta H_0^2) H^2}{H^4-\beta H_0^2 H^2+2 \beta^2 H_0^4}.
\end{equation}
Hence, using the inverse relation of (\ref{redshift}) we extract
$c_{s}^{2}(z)$ as a function of the redshift, and we depict it in
Fig. \ref{Fig:diagnostics}\subref{fig:sound-speed}. From this graph it is clear that the
sound-speed square is close to $1$ (still with $c_{s}^{2}\leq 1$) at large $z$,
then it drops to
$\approx
0.66$ at present ($z=0$), while $c_{s}^{2} \to 0.5$ in the far future $z\to
-1$.
Therefore, the model at hand does not suffer from ghost instabilities or acausality at
any time of the evolution.

In summary, the $f(T)$ ansatz (\ref{f(T)}) fulfills the basic requirements on
both background and perturbation levels, and thus it can be considered as a viable $f(T)$
model.

\section{Conclusions}\label{Sec6}
In the present work we used dynamical system methods in order to explore the general
behaviour of $f(T)$ cosmology. These methods
allow to bypass the complications and non-linearities of the
equations in a cosmological scenario, and obtain information about their global behaviour
and dynamics without the need of extracting complete analytical
solutions.
In contrast to the standard applications of dynamical
system methods in given cosmological scenarios, in which one
results to a multi-dimensional system whose investigation is in general complicated
and it is restricted only to specific forms of the involved functions, in our analysis we
presented a way to transform the $f(T)$ cosmological equations into a one-dimensional
autonomous system. In particular, we took advantage of the crucial
property that the torsion scalar in flat FRW geometry is just a function of the Hubble
function $H$, and thus in a general $f(T)$ cosmological
scenario every quantity is
expressed
as a function of $H$. Hence, we were able to embed all the information of $f(T)$
cosmology in one phase portrait equation of the form $\dot{H}=\mathcal{F}(H)$.

As a first step, we investigated in detail the phase space portraits that arise from the
above equation in the case of general $f(T)$ cosmology, exploring the basic features
and possible behaviours. As we showed, $f(T)$ cosmology can describe the universe
evolution in agreement with observations, namely starting from a Big Bang singularity,
evolving into the subsequent thermal history and the matter domination, entering into a
late-time accelerated expansion, and resulting to the de Sitter phase in the far future.
Nevertheless, $f(T)$ cosmology exhibits a rich class of more exotic behaviours, such as
decelerated and accelerated contraction, cosmological (singular or non-singular) bounce
and turnaround, the realization of the phantom-divide crossing, and the appearance of
the Big Brake and the Big Crunch. Moreover, it can exhibit various singularities,
including the non-harmful ones of type II and type IV.

As a next step, we investigated the phase space portraits of three specific viable $f(T)$
forms, that pass the basic cosmological requirements. Using the advanced method of the
one-dimensional analysis, we were able to reproduce the results that were previously
extracted in the literature, providing additionally extra information on the
different solution branches and possible
behaviours, and offering a more complete picture. Using our method,
we were able to extract the required bounds on the model parameters in order to obtain
the correct early and intermediate behaviour too, apart from the correct late-time
asymptotic one. Interestingly enough, we obtained the same parameter bounds with those
acquired through detailed observational confrontation.

Taking into account the information obtained from the investigation of the phase space
portraits, we presented a new model of $f(T)$ gravity, namely $f(T)=Te^{\beta T_0/T}$,
that can lead to a universe in agreement with observations. In particular, this model can
describe the required thermal history of the universe, exhibit the
deceleration-to-acceleration
transition at the expected redshift, and result today in the correct percentage of dark
matter and
dark energy. Furthermore, the
effective dark energy behaves as cosmological constant at large redshifts, while it
behaves as phantom in the epoch of late-time acceleration. In
the future the universe is attracted by the de Sitter solution, and it is
dominated by the torsional dark energy which behaves as cosmological constant. Moreover,
concerning the basic cosmological quantities, these are in agreement with their observed
values. Finally, we performed the $Om(z)$ diagnostic test, which can
differentiate various dark energy models since it has less dependence on the matter
density parameter. Additionally, we examined the behaviour of the scalar perturbations in
the scenario at hand, and we showed that their sound speed square is always in the
interval $0.5 \leq c_{s}^{2} \leq 1$, which fulfills the stability and
causality conditions at all times.

In summary, the method of one-dimensional phase space portraits, that is applicable in
$f(T)$ cosmology, can reveal the rich structure and the capabilities of the theory.
$f(T)$ gravity proves to be efficient as a candidate for the description of nature.

\acknowledgments
This work is partially supported by the Egyptian Ministry of Scientific Research under
project No. 24-2-12. This article is
based upon work from COST Action ``Cosmology and Astrophysics Network
for Theoretical Advances and Training Actions'', supported by COST (European Cooperation
in Science and Technology). ENS wishes to thank Institut de Physique Th\'{e}orique,
Universit\'e Paris-Saclay and Laboratoire de Physique Th\'eorique, Univ. Paris-Sud,
Universit\'e Paris-Saclay, for the hospitality during the last stages of this project.


\begin{thebibliography}{100}

\bibitem{Smoot:1992td}
{\scshape COBE} collaboration, G.~F. Smoot et~al., \emph{{Structure in the COBE
  differential microwave radiometer first year maps}},
  \href{https://doi.org/10.1086/186504}{\emph{Astrophys. J.} {\bfseries 396}
  (1992) L1--L5}.

\bibitem{Spergel:2003ApJS}
D.~N. {Spergel}, L.~{Verde}, H.~V. {Peiris}, E.~{Komatsu}, M.~R. {Nolta}, C.~L.
  {Bennett} et~al., \emph{{First-Year Wilkinson Microwave Anisotropy Probe
  (WMAP) Observations: Determination of Cosmological Parameters}},
  \href{https://doi.org/10.1086/377226}{\emph{Astrophys. J. S.} {\bfseries 148}
  (Sept., 2003) 175--194},
  [\href{https://arxiv.org/abs/astro-ph/0302209}{{\ttfamily
  astro-ph/0302209}}].

\bibitem{Bennett:2013ApJS}
C.~L. {Bennett}, D.~{Larson}, J.~L. {Weiland}, N.~{Jarosik}, G.~{Hinshaw},
  N.~{Odegard} et~al., \emph{{Nine-year Wilkinson Microwave Anisotropy Probe
  (WMAP) Observations: Final Maps and Results}},
  \href{https://doi.org/10.1088/0067-0049/208/2/20}{\emph{Astrophys. J.}
  {\bfseries 208} (Oct., 2013) 20},
  [\href{https://arxiv.org/abs/1212.5225}{{\ttfamily 1212.5225}}].

\bibitem{Ade:2015lrj}
{\scshape Planck} collaboration, P.~A.~R. Ade et~al., \emph{{Planck 2015
  results. XX. Constraints on inflation}},
  \href{https://doi.org/10.1051/0004-6361/201525898}{\emph{Astron. Astrophys.}
  {\bfseries 594} (2016) }, [\href{https://arxiv.org/abs/1502.02114}{{\ttfamily
  1502.02114}}].

\bibitem{Adam:2015rua}
{\scshape Planck} collaboration, R.~Adam et~al., \emph{{Planck 2015 results. I.
  Overview of products and scientific results}},
  \href{https://doi.org/10.1051/0004-6361/201527101}{\emph{Astron. Astrophys.}
  {\bfseries 594} (2016) }, [\href{https://arxiv.org/abs/1502.01582}{{\ttfamily
  1502.01582}}].

\bibitem{Rubin:1978ApJL}
V.~C. {Rubin}, N.~{Thonnard} and W.~K. {Ford}, Jr., \emph{{Extended rotation
  curves of high-luminosity spiral galaxies. IV - Systematic dynamical
  properties, SA through SC}},
  \href{https://doi.org/10.1086/182804}{\emph{Astrophys. J. L.} {\bfseries 225}
  (Nov., 1978) L107--L111}.

\bibitem{Faber:1979ARAA}
S.~M. {Faber} and J.~S. {Gallagher}, \emph{{Masses and mass-to-light ratios of
  galaxies}},
  \href{https://doi.org/10.1146/annurev.aa.17.090179.001031}{\emph{Annual
  review of astronomy and astrophysics} {\bfseries 17} (1979) 135--187}.

\bibitem{Fabricant:1980ApJ}
D.~{Fabricant}, M.~{Lecar} and P.~{Gorenstein}, \emph{{X-ray measurements of
  the mass of M87}}, \href{https://doi.org/10.1086/158369}{\emph{Astrophys. J.}
  {\bfseries 241} (Oct., 1980) 552--560}.

\bibitem{Riess:1998AJ}
A.~G. {Riess}, A.~V. {Filippenko}, P.~{Challis}, A.~{Clocchiatti},
  A.~{Diercks}, P.~M. {Garnavich} et~al., \emph{{Observational Evidence from
  Supernovae for an Accelerating Universe and a Cosmological Constant}},
  \href{https://doi.org/10.1086/300499}{\emph{The Astronomical Journal}
  {\bfseries 116} (Sept., 1998) 1009--1038},
  [\href{https://arxiv.org/abs/astro-ph/9805201}{{\ttfamily
  astro-ph/9805201}}].

\bibitem{Perlmutter:1999ApJ}
S.~{Perlmutter}, G.~{Aldering}, G.~{Goldhaber}, R.~A. {Knop}, P.~{Nugent},
  P.~G. {Castro} et~al., \emph{{Measurements of {$\Omega$} and {$\Lambda$} from
  42 High-Redshift Supernovae}},
  \href{https://doi.org/10.1086/307221}{\emph{Astrophys. J.} {\bfseries 517}
  (June, 1999) 565--586},
  [\href{https://arxiv.org/abs/astro-ph/9812133}{{\ttfamily
  astro-ph/9812133}}].

\bibitem{Farooq:2016zwm}
O.~Farooq, F.~R. Madiyar, S.~Crandall and B.~Ratra, \emph{{Hubble Parameter
  Measurement Constraints on the Redshift of the Deceleration-acceleration
  Transition, Dynamical Dark Energy, and Space Curvature}},
  \href{https://doi.org/10.3847/1538-4357/835/1/26}{\emph{Astrophys. J.}
  {\bfseries 835} (2017) 26},
  [\href{https://arxiv.org/abs/1607.03537}{{\ttfamily 1607.03537}}].

\bibitem{DiValentino:2017gzb}
E.~Di~Valentino, \emph{{Crack in the cosmological paradigm}}, {\emph{Nat.
  Astron.} {\bfseries 1} (2017) 569},
  [\href{https://arxiv.org/abs/1709.04046}{{\ttfamily 1709.04046}}].

\bibitem{Freedman:2017yms}
W.~L. Freedman, \emph{{Cosmology at at Crossroads: Tension with the Hubble
  Constant}}, {\emph{Nat. Astron.} {\bfseries 1} (2017) 0169},
  [\href{https://arxiv.org/abs/1706.02739}{{\ttfamily 1706.02739}}].

\bibitem{DiValentino:2016hlg}
E.~Di~Valentino, A.~Melchiorri and J.~Silk, \emph{{Reconciling Planck with the
  local value of $H_0$ in extended parameter space}},
  \href{https://doi.org/10.1016/j.physletb.2016.08.043}{\emph{Phys. Lett. B}
  {\bfseries 761} (2016) 242--246},
  [\href{https://arxiv.org/abs/1606.00634}{{\ttfamily 1606.00634}}].

\bibitem{Zhao:2017cud}
G.-B. Zhao et~al., \emph{{Dynamical dark energy in light of the latest
  observations}}, \href{https://doi.org/10.1038/s41550-017-0216-z}{\emph{Nat.
  Astron.} {\bfseries 1} (2017) 627--632},
  [\href{https://arxiv.org/abs/1701.08165}{{\ttfamily 1701.08165}}].

\bibitem{DiValentino:2017zyq}
E.~Di~Valentino, A.~Melchiorri, E.~V. Linder and J.~Silk, \emph{{Constraining
  Dark Energy Dynamics in Extended Parameter Space}},
  \href{https://doi.org/10.1103/PhysRevD.96.023523}{\emph{Phys. Rev. D}
  {\bfseries 96} (2017) 023523},
  [\href{https://arxiv.org/abs/1704.00762}{{\ttfamily 1704.00762}}].

\bibitem{Copeland:2006wr}
E.~J. Copeland, M.~Sami and S.~Tsujikawa, \emph{{Dynamics of dark energy}},
  \href{https://doi.org/10.1142/S021827180600942X}{\emph{Int. J. Mod. Phys.}
  (2006) 1753--1936}, [\href{https://arxiv.org/abs/hep-th/0603057}{{\ttfamily
  hep-th/0603057}}].

\bibitem{Cai:2009zp}
Y.-F. Cai, E.~N. Saridakis, M.~R. Setare and J.-Q. Xia, \emph{{Quintom
  Cosmology: Theoretical implications and observations}},
  \href{https://doi.org/10.1016/j.physrep.2010.04.001}{\emph{Phys. Rept.}
  {\bfseries 493} (2010) 1--60},
  [\href{https://arxiv.org/abs/0909.2776}{{\ttfamily 0909.2776}}].

\bibitem{Nojiri:2006ri}
S.~Nojiri and S.~D. Odintsov, \emph{{Introduction to modified gravity and
  gravitational alternative for dark energy}},
  \href{https://doi.org/10.1142/S0219887807001928}{\emph{eConf} (2006) 06},
  [\href{https://arxiv.org/abs/hep-th/0601213}{{\ttfamily hep-th/0601213}}].

\bibitem{Capozziello:2011et}
S.~Capozziello and M.~De~Laurentis, \emph{{Extended Theories of Gravity}},
  \href{https://doi.org/10.1016/j.physrep.2011.09.003}{\emph{Phys. Rept.}
  {\bfseries 509} (2011) 167--321},
  [\href{https://arxiv.org/abs/1108.6266}{{\ttfamily 1108.6266}}].

\bibitem{Cai:2015emx}
Y.-F. Cai, S.~Capozziello, M.~De~Laurentis and E.~N. Saridakis, \emph{{f(T)
  teleparallel gravity and cosmology}},
  \href{https://doi.org/10.1088/0034-4885/79/10/106901}{\emph{Rept. Prog.
  Phys.} {\bfseries 79} (2016) 106901},
  [\href{https://arxiv.org/abs/1511.07586}{{\ttfamily 1511.07586}}].

\bibitem{Pereira.book}
R.~{Aldrovandi} and J.~G. {Pereira}, \emph{{Teleparallel Gravity}}, vol.~173.
\newblock Springer Science+Business Media Dordrecht, 2013,
  \href{https://doi.org/10.1007/978-94-007-5143-9}{10.1007/978-94-007-5143-9}.

\bibitem{Bengochea:2008gz}
G.~R. Bengochea and R.~Ferraro, \emph{{Dark torsion as the cosmic speed-up}},
  \href{https://doi.org/10.1103/PhysRevD.79.124019}{\emph{Phys. Rev. D}
  {\bfseries 79} (2009) 124019},
  [\href{https://arxiv.org/abs/0812.1205}{{\ttfamily 0812.1205}}].

\bibitem{Linder:2010py}
E.~V. Linder, \emph{{Einstein's Other Gravity and the Acceleration of the
  Universe}}, \href{https://doi.org/10.1103/PhysRevD.81.127301,
  10.1103/PhysRevD.82.109902}{\emph{Phys. Rev. D} {\bfseries 81} (2010)
  127301}, [\href{https://arxiv.org/abs/1005.3039}{{\ttfamily 1005.3039}}].

\bibitem{Ferraro:2006jd}
R.~Ferraro and F.~Fiorini, \emph{{Modified teleparallel gravity: Inflation
  without inflaton}},
  \href{https://doi.org/10.1103/PhysRevD.75.084031}{\emph{Phys. Rev. D}
  {\bfseries 75} (2007) 084031},
  [\href{https://arxiv.org/abs/gr-qc/0610067}{{\ttfamily gr-qc/0610067}}].

\bibitem{Ferraro:2008ey}
R.~Ferraro and F.~Fiorini, \emph{{On Born-Infeld Gravity in Weitzenbock
  spacetime}}, \href{https://doi.org/10.1103/PhysRevD.78.124019}{\emph{Phys.
  Rev. D} {\bfseries 78} (2008) 124019},
  [\href{https://arxiv.org/abs/0812.1981}{{\ttfamily 0812.1981}}].

\bibitem{Bamba:2014zra}
K.~Bamba, S.~Nojiri and S.~D. Odintsov, \emph{{Trace-anomaly driven inflation
  in $f(T)$ gravity and in minimal massive bigravity}},
  \href{https://doi.org/10.1016/j.physletb.2014.02.041}{\emph{Phys. Lett. B}
  {\bfseries 731} (2014) 257--264},
  [\href{https://arxiv.org/abs/1401.7378}{{\ttfamily 1401.7378}}].

\bibitem{Bamba:2016wjm}
K.~Bamba, S.~D. Odintsov and E.~N. Saridakis, \emph{{Inflationary cosmology in
  unimodular $F(T)$ gravity}},
  \href{https://doi.org/10.1142/S0217732317501140}{\emph{Mod. Phys. Lett. A}
  {\bfseries 32} (2017) 1750114},
  [\href{https://arxiv.org/abs/1605.02461}{{\ttfamily 1605.02461}}].

\bibitem{Chen:2010va}
S.-H. Chen, J.~B. Dent, S.~Dutta and E.~N. Saridakis, \emph{{Cosmological
  perturbations in f(T) gravity}},
  \href{https://doi.org/10.1103/PhysRevD.83.023508}{\emph{Phys. Rev. D}
  {\bfseries 83} (2011) 023508},
  [\href{https://arxiv.org/abs/1008.1250}{{\ttfamily 1008.1250}}].

\bibitem{Tsyba:2010ji}
P.~{\relax Yu}. Tsyba, I.~I. Kulnazarov, K.~K. Yerzhanov and R.~Myrzakulov,
  \emph{{Pure kinetic k-essence as the cosmic speed-up}},
  \href{https://doi.org/10.1007/s10773-011-0703-4}{\emph{Int. J. Theor. Phys.}
  {\bfseries 50} (2011) 1876--1886},
  [\href{https://arxiv.org/abs/1008.0779}{{\ttfamily 1008.0779}}].

\bibitem{Yang:2010hw}
R.-J. Yang, \emph{{New types of $f(T)$ gravity}},
  \href{https://doi.org/10.1140/epjc/s10052-011-1797-9}{\emph{Eur. Phys. J. C}
  {\bfseries 71} (2011) 1797},
  [\href{https://arxiv.org/abs/1007.3571}{{\ttfamily 1007.3571}}].

\bibitem{Bamba:2010iw}
K.~Bamba, C.-Q. Geng and C.-C. Lee, \emph{{Comment on `Einstein's Other Gravity
  and the Acceleration of the Universe''}},
  \href{https://arxiv.org/abs/1008.4036}{{\ttfamily 1008.4036}}.

\bibitem{Myrzakulov:2010tc}
R.~{Myrzakulov}, \emph{{F(T) gravity and k-essence}},
  \href{https://doi.org/10.1007/s10714-012-1439-z}{\emph{Gen. Rel. Grav.}
  {\bfseries 44} (Dec., 2012) 3059--3080},
  [\href{https://arxiv.org/abs/1008.4486}{{\ttfamily 1008.4486}}].

\bibitem{Bamba:2010wb}
K.~Bamba, C.-Q. Geng, C.-C. Lee and L.-W. Luo, \emph{{Equation of state for
  dark energy in $f(T)$ gravity}},
  \href{https://doi.org/10.1088/1475-7516/2011/01/021}{\emph{JCAP} {\bfseries
  1101} (2011) 021}, [\href{https://arxiv.org/abs/1011.0508}{{\ttfamily
  1011.0508}}].

\bibitem{Dent:2011zz}
J.~B. Dent, S.~Dutta and E.~N. Saridakis, \emph{{f(T) gravity mimicking
  dynamical dark energy. Background and perturbation analysis}},
  \href{https://doi.org/10.1088/1475-7516/2011/01/009}{\emph{JCAP} {\bfseries
  1101} (2011) 009}, [\href{https://arxiv.org/abs/1010.2215}{{\ttfamily
  1010.2215}}].

\bibitem{Cai:2011tc}
Y.-F. Cai, S.-H. Chen, J.~B. Dent, S.~Dutta and E.~N. Saridakis, \emph{{Matter
  Bounce Cosmology with the $f(T)$ Gravity}},
  \href{https://doi.org/10.1088/0264-9381/28/21/215011}{\emph{Class. Quant.
  Grav.} {\bfseries 28} (2011) 215011},
  [\href{https://arxiv.org/abs/1104.4349}{{\ttfamily 1104.4349}}].

\bibitem{Sharif:2011bi}
M.~Sharif and S.~Rani, \emph{{F(T) Models within Bianchi Type I Universe}},
  \href{https://doi.org/10.1142/S0217732311036127}{\emph{Mod. Phys. Lett. A}
  {\bfseries 26} (2011) 1657--1671},
  [\href{https://arxiv.org/abs/1105.6228}{{\ttfamily 1105.6228}}].

\bibitem{Wei:2011mq}
H.~Wei, H.-Y. Qi and X.-P. Ma, \emph{{Constraining $f(T)$ Theories with the
  Varying Gravitational Constant}},
  \href{https://doi.org/10.1140/epjc/s10052-012-2117-8}{\emph{Eur. Phys. J. C}
  {\bfseries 72} (2012) 2117},
  [\href{https://arxiv.org/abs/1108.0859}{{\ttfamily 1108.0859}}].

\bibitem{Wu:2011xa}
P.~Wu and H.~Yu, \emph{{The Stability of the Einstein static state in $f(T)$
  gravity}}, \href{https://doi.org/10.1016/j.physletb.2011.07.087}{\emph{Phys.
  Lett. B} {\bfseries 703} (2011) 223--227},
  [\href{https://arxiv.org/abs/1108.5908}{{\ttfamily 1108.5908}}].

\bibitem{Karami:2011np}
K.~Karami, A.~Abdolmaleki, S.~Asadzadeh and Z.~Safari, \emph{{Holographic
  $f(T)$-gravity model with power-law entropy correction}},
  \href{https://doi.org/10.1103/PhysRevD.88.084034}{\emph{Phys. Rev. D}
  {\bfseries 88} (2013) 084034},
  [\href{https://arxiv.org/abs/1111.7269}{{\ttfamily 1111.7269}}].

\bibitem{Jamil:2011mc}
M.~{Jamil}, D.~{Momeni}, N.~S. {Serikbayev} and R.~{Myrzakulov}, \emph{{FRW and
  Bianchi type I cosmology of f-essence}},
  \href{https://doi.org/10.1007/s10509-011-0964-7}{\emph{Astrophys. Space Sci.}
  {\bfseries 339} (May, 2012) 37--43},
  [\href{https://arxiv.org/abs/1112.4472}{{\ttfamily 1112.4472}}].

\bibitem{Karami:2012if}
K.~Karami, A.~Abdolmaleki, S.~Asadzadeh and Z.~Safari, \emph{{QCD ghost
  $f(T)$-gravity model}},
  \href{https://doi.org/10.1140/epjc/s10052-013-2565-9}{\emph{Eur. Phys. J. C}
  {\bfseries 73} (2013) 2565},
  [\href{https://arxiv.org/abs/1202.2278}{{\ttfamily 1202.2278}}].

\bibitem{Setare:2012vs}
M.~R. Setare and M.~J.~S. Houndjo, \emph{{Finite-time future singularities
  models in $f(T)$ gravity and the effects of viscosity}},
  \href{https://doi.org/10.1139/cjp-2012-0533}{\emph{Can. J. Phys.} {\bfseries
  91} (2013) 260--267}, [\href{https://arxiv.org/abs/1203.1315}{{\ttfamily
  1203.1315}}].

\bibitem{Dong:2012en}
H.~Dong, Y.-b. Wang and X.-h. Meng, \emph{{Extended Birkhoff's Theorem in the
  f(T) Gravity}},
  \href{https://doi.org/10.1140/epjc/s10052-012-2002-5}{\emph{Eur. Phys. J. C}
  {\bfseries 72} (2012) 2002},
  [\href{https://arxiv.org/abs/1203.5890}{{\ttfamily 1203.5890}}].

\bibitem{Tamanini:2012hg}
N.~Tamanini and C.~G. Boehmer, \emph{{Good and bad tetrads in f(T) gravity}},
  \href{https://doi.org/10.1103/PhysRevD.86.044009}{\emph{Phys. Rev. D}
  {\bfseries 86} (2012) 044009},
  [\href{https://arxiv.org/abs/1204.4593}{{\ttfamily 1204.4593}}].

\bibitem{Daouda:2012wt}
M.~E. Rodrigues, M.~H. Daouda, M.~J.~S. Houndjo, R.~Myrzakulov and M.~Sharif,
  \emph{{Inhomogeneous Universe in f(T) Theory}},
  \href{https://doi.org/10.1134/S0202289314020108}{\emph{Grav. Cosmol.}
  {\bfseries 20} (2014) 80--89},
  [\href{https://arxiv.org/abs/1205.0565}{{\ttfamily 1205.0565}}].

\bibitem{Banijamali:2012nx}
A.~Banijamali and B.~Fazlpour, \emph{{Tachyonic Teleparallel Dark Energy}},
  \href{https://doi.org/10.1007/s10509-012-1140-4}{\emph{Astrophys. Space Sci.}
  {\bfseries 342} (2012) 229--235},
  [\href{https://arxiv.org/abs/1206.3580}{{\ttfamily 1206.3580}}].

\bibitem{Darabi:2012zh}
F.~Darabi, \emph{{Reconstruction of $f(R)$, $f(T)$ and $f(G)$ models inspired
  by variable deceleration parameter}},
  \href{https://doi.org/10.1007/s10509-012-1250-z}{\emph{Astrophys. Space Sci.}
  {\bfseries 343} (2013) 499--504},
  [\href{https://arxiv.org/abs/1207.0212}{{\ttfamily 1207.0212}}].

\bibitem{Liu:2012fk}
D.~Liu and M.~J. Reboucas, \emph{{Energy conditions bounds on f(T) gravity}},
  \href{https://doi.org/10.1103/PhysRevD.86.083515}{\emph{Phys. Rev. D}
  {\bfseries 86} (2012) 083515},
  [\href{https://arxiv.org/abs/1207.1503}{{\ttfamily 1207.1503}}].

\bibitem{Setare:2012ry}
M.~R. Setare and N.~Mohammadipour, \emph{{Cosmological viability conditions for
  $f(T)$ dark energy models}},
  \href{https://doi.org/10.1088/1475-7516/2012/11/030}{\emph{JCAP} {\bfseries
  1211} (2012) 030}, [\href{https://arxiv.org/abs/1211.1375}{{\ttfamily
  1211.1375}}].

\bibitem{Cardone:2012xq}
V.~F. Cardone, N.~Radicella and S.~Camera, \emph{{Accelerating f(T) gravity
  models constrained by recent cosmological data}},
  \href{https://doi.org/10.1103/PhysRevD.85.124007}{\emph{Phys. Rev.} (2012)
  124007}, [\href{https://arxiv.org/abs/1204.5294}{{\ttfamily 1204.5294}}].

\bibitem{Chattopadhyay:2012eu}
S.~{Chattopadhyay} and A.~{Pasqua}, \emph{{Reconstruction of f( T) gravity from
  the Holographic Dark Energy}},
  \href{https://doi.org/10.1007/s10509-012-1315-z}{\emph{Astrophys. Space Sci.}
  {\bfseries 344} (Mar., 2013) 269--274},
  [\href{https://arxiv.org/abs/1211.2707}{{\ttfamily 1211.2707}}].

\bibitem{Bamba:2012ka}
K.~Bamba, J.~de~Haro and S.~D. Odintsov, \emph{{Future Singularities and
  Teleparallelism in Loop Quantum Cosmology}},
  \href{https://doi.org/10.1088/1475-7516/2013/02/008}{\emph{JCAP} {\bfseries
  1302} (2013) 008}, [\href{https://arxiv.org/abs/1211.2968}{{\ttfamily
  1211.2968}}].

\bibitem{Setare:2013xh}
M.~R. Setare and N.~Mohammadipour, \emph{{Can $f(T)$ gravity theories mimic
  $\Lambda$CDM cosmic history}},
  \href{https://doi.org/10.1088/1475-7516/2013/01/015}{\emph{JCAP} {\bfseries
  1301} (2013) 015}, [\href{https://arxiv.org/abs/1301.4891}{{\ttfamily
  1301.4891}}].

\bibitem{Li:2013xea}
J.-T. Li, C.-C. Lee and C.-Q. Geng, \emph{{Einstein Static Universe in
  Exponential $f(T)$ Gravity}},
  \href{https://doi.org/10.1140/epjc/s10052-013-2315-z}{\emph{Eur. Phys. J. C}
  {\bfseries 73} (2013) 2315},
  [\href{https://arxiv.org/abs/1302.2688}{{\ttfamily 1302.2688}}].

\bibitem{Camera:2013bwa}
S.~Camera, V.~F. Cardone and N.~Radicella, \emph{{Detectability of Torsion
  Gravity via Galaxy Clustering and Cosmic Shear Measurements}},
  \href{https://doi.org/10.1103/PhysRevD.89.083520}{\emph{Phys. Rev.} (2014)
  083520}, [\href{https://arxiv.org/abs/1311.1004}{{\ttfamily 1311.1004}}].

\bibitem{Bamba:2013fta}
K.~Bamba, S.~Nojiri and S.~D. Odintsov, \emph{{Effective $F(T)$ gravity from
  the higher-dimensional Kaluza-Klein and Randall-Sundrum theories}},
  \href{https://doi.org/10.1016/j.physletb.2013.07.052}{\emph{Phys. Lett. B}
  {\bfseries 725} (2013) 368--371},
  [\href{https://arxiv.org/abs/1304.6191}{{\ttfamily 1304.6191}}].

\bibitem{Basilakos:2013rua}
S.~Basilakos, S.~Capozziello, M.~De~Laurentis, A.~Paliathanasis and
  M.~Tsamparlis, \emph{{Noether symmetries and analytical solutions in
  f(T)-cosmology: A complete study}},
  \href{https://doi.org/10.1103/PhysRevD.88.103526}{\emph{Phys. Rev. D}
  {\bfseries 88} (2013) 103526},
  [\href{https://arxiv.org/abs/1311.2173}{{\ttfamily 1311.2173}}].

\bibitem{Qi:2014yxa}
J.-Z. Qi, R.-J. Yang, M.-J. Zhang and W.-B. Liu, \emph{{Transient acceleration
  in $f(T)$ gravity}},
  \href{https://doi.org/10.1088/1674-4527/16/2/022}{\emph{Res. Astron.
  Astrophys.} {\bfseries 16} (2016) 022},
  [\href{https://arxiv.org/abs/1403.7287}{{\ttfamily 1403.7287}}].

\bibitem{Nashed:2014baa}
G.~G.~L. Nashed, \emph{Exact homogenous anisotropic solution in f(t) gravity
  theory}, \href{https://doi.org/10.1140/epjp/i2014-14188-9}{\emph{Eur. Phys.
  J. P} {\bfseries 129} (Sep, 2014) 188}.

\bibitem{Darabi:2014dla}
F.~Darabi, M.~Mousavi and K.~Atazadeh, \emph{{Geodesic deviation equation in
  f(T) gravity}}, \href{https://doi.org/10.1103/PhysRevD.91.084023}{\emph{Phys.
  Rev. D} {\bfseries 91} (2015) 084023},
  [\href{https://arxiv.org/abs/1501.00103}{{\ttfamily 1501.00103}}].

\bibitem{Nashed:2015pda}
G.~L. Nashed, \emph{{FRW in quadratic form of f(T) gravitational theories}},
  \href{https://doi.org/10.1007/s10714-015-1917-1}{\emph{Gen. Rel. Grav.}
  {\bfseries 47} (2015) 75},
  [\href{https://arxiv.org/abs/1506.08695}{{\ttfamily 1506.08695}}].

\bibitem{Fayaz:2015yka}
V.~Fayaz, H.~Hossienkhani, A.~Pasqua, M.~Amirabadi and M.~Ganji, \emph{f(t)
  theories from holographic dark energy models within bianchi type i universe},
  \href{https://doi.org/10.1140/epjp/i2015-15028-2}{\emph{The European Physical
  Journal Plus} {\bfseries 130} (Feb, 2015) 28}.

\bibitem{Abedi:2015cya}
H.~Abedi and M.~Salti, \emph{Multiple field modified gravity and localized
  energy in teleparallel framework},
  \href{https://doi.org/10.1007/s10714-015-1935-z}{\emph{Gen. Rel. Grav.}
  {\bfseries 47} (Jul, 2015) 93}.

\bibitem{Krssak:2015oua}
M.~Kr\v{s}\v{s}\'{a}k and E.~N. Saridakis, \emph{{The covariant formulation of
  f(T) gravity}},
  \href{https://doi.org/10.1088/0264-9381/33/11/115009}{\emph{Class. Quant.
  Grav.} {\bfseries 33} (2016) 115009},
  [\href{https://arxiv.org/abs/1510.08432}{{\ttfamily 1510.08432}}].

\bibitem{Pan:2016jli}
S.~Pan, J.~de~Haro, A.~Paliathanasis, R.~J. Slagter and S.~R.~J.,
  \emph{{Evolution and Dynamics of a Matter creation model}},
  \href{https://doi.org/10.1093/mnras/stw1034}{\emph{Mon. Not. Roy. Astron.
  Soc.} {\bfseries 460} (2016) 1445--1456},
  [\href{https://arxiv.org/abs/1601.03955}{{\ttfamily 1601.03955}}].

\bibitem{Paliathanasis:2016vsw}
A.~Paliathanasis, J.~D. Barrow and P.~G.~L. Leach, \emph{{Cosmological
  Solutions of $f(T)$ Gravity}},
  \href{https://doi.org/10.1103/PhysRevD.94.023525}{\emph{Phys. Rev. D}
  {\bfseries 94} (2016) 023525},
  [\href{https://arxiv.org/abs/1606.00659}{{\ttfamily 1606.00659}}].

\bibitem{Geng:2011aj}
C.-Q. Geng, C.-C. Lee, E.~N. Saridakis and Y.-P. Wu, \emph{{Teleparallel dark
  energy}}, \href{https://doi.org/10.1016/j.physletb.2011.09.082}{\emph{Phys.
  Lett. B} {\bfseries 704} (2011) 384--387},
  [\href{https://arxiv.org/abs/1109.1092}{{\ttfamily 1109.1092}}].

\bibitem{Xu:2012jf}
C.~Xu, E.~N. Saridakis and G.~Leon, \emph{{Phase-Space analysis of Teleparallel
  Dark Energy}},
  \href{https://doi.org/10.1088/1475-7516/2012/07/005}{\emph{JCAP} {\bfseries
  1207} (2012) 005}, [\href{https://arxiv.org/abs/1202.3781}{{\ttfamily
  1202.3781}}].

\bibitem{Kofinas:2014owa}
G.~Kofinas and E.~N. Saridakis, \emph{{Teleparallel equivalent of Gauss-Bonnet
  gravity and its modifications}},
  \href{https://doi.org/10.1103/PhysRevD.90.084044}{\emph{Phys. Rev. D}
  {\bfseries 90} (2014) 084044},
  [\href{https://arxiv.org/abs/1404.2249}{{\ttfamily 1404.2249}}].

\bibitem{Haro:2014wha}
J.~Haro and J.~Amoros, \emph{{Viability of the matter bounce scenario in $F(T)$
  gravity and Loop Quantum Cosmology for general potentials}},
  \href{https://doi.org/10.1088/1475-7516/2014/12/031}{\emph{JCAP} {\bfseries
  1412} (2014) 031}, [\href{https://arxiv.org/abs/1406.0369}{{\ttfamily
  1406.0369}}].

\bibitem{Hanafy:2014ica}
W.~El~Hanafy and G.~G.~L. Nashed, \emph{{The hidden flat like universe}},
  \href{https://doi.org/10.1140/epjc/s10052-015-3501-y}{\emph{Eur. Phys. J. C}
  {\bfseries 75} (2015) 279},
  [\href{https://arxiv.org/abs/1409.7199}{{\ttfamily 1409.7199}}].

\bibitem{Sharif:2014CoTPh}
M.~{Sharif} and A.~{Sehrish}, \emph{{Dark Energy Models and Cosmic Acceleration
  with Anisotropic Universe in f(T) Gravity}},
  \href{https://doi.org/10.1088/0253-6102/61/4/13}{\emph{Communications in
  Theoretical Physics} {\bfseries 61} (Apr., 2014) 482--490}.

\bibitem{Hanafy:2015lda}
W.~El~Hanafy and G.~G.~L. Nashed, \emph{{The hidden flat like universe II:
  Quasi inverse power law inflation by $f(T)$ gravity}},
  \href{https://doi.org/10.1007/s10509-016-2853-6}{\emph{Astrophys. Space Sci.}
  {\bfseries 361} (2016) 266},
  [\href{https://arxiv.org/abs/1510.02337}{{\ttfamily 1510.02337}}].

\bibitem{Capozziello:2015rda}
S.~Capozziello, O.~Luongo and E.~N. Saridakis, \emph{{Transition redshift in
  $f(T)$ cosmology and observational constraints}},
  \href{https://doi.org/10.1103/PhysRevD.91.124037}{\emph{Phys. Rev. D}
  {\bfseries 91} (2015) 124037},
  [\href{https://arxiv.org/abs/1503.02832}{{\ttfamily 1503.02832}}].

  \bibitem{Nunes:2016qyp}
R.~C. Nunes, S.~Pan and E.~N. Saridakis, \emph{{New observational constraints
  on f(T) gravity from cosmic chronometers}},
  \href{https://doi.org/10.1088/1475-7516/2016/08/011}{\emph{JCAP} {\bfseries
  1608} (2016) 011}, [\href{https://arxiv.org/abs/1606.04359}{{\ttfamily
  1606.04359}}].
  
\bibitem{Nunes:2016plz}
R.~C. Nunes, A.~Bonilla, S.~Pan and E.~N. Saridakis, \emph{{Observational
  Constraints on $f(T)$ gravity from varying fundamental constants}},
  \href{https://doi.org/10.1140/epjc/s10052-017-4798-5}{\emph{Eur. Phys. J. C}
  {\bfseries 77} (2017) 230},
  [\href{https://arxiv.org/abs/1608.01960}{{\ttfamily 1608.01960}}].


\bibitem{Qi:2017xzl}
J.-Z. Qi, S.~Cao, M.~Biesiada, X.~Zheng and H.~Zhu, \emph{{New observational
  constraints on $f(T)$ cosmology from radio quasars}},
  \href{https://doi.org/10.1140/epjc/s10052-017-5069-1}{\emph{Eur. Phys. J. C}
  {\bfseries 77} (2017) 502},
  [\href{https://arxiv.org/abs/1708.08603}{{\ttfamily 1708.08603}}].

\bibitem{Awad:2017ign}
A.~Awad, W.~El~Hanafy, G.~G.~L. Nashed, S.~D. Odintsov and V.~K. Oikonomou,
  \emph{{Constant-roll Inflation in $f(T)$ Teleparallel Gravity}},
  \href{https://arxiv.org/abs/1710.00682}{{\ttfamily 1710.00682}}.

\bibitem{Oikonomou:2017isf}
V.~K. Oikonomou, \emph{{Viability of the intermediate inflation scenario with
  F(T) gravity}}, \href{https://doi.org/10.1103/PhysRevD.95.084023}{\emph{Phys.
  Rev. D} {\bfseries 95} (2017) 084023},
  [\href{https://arxiv.org/abs/1703.10515}{{\ttfamily 1703.10515}}].

\bibitem{Jawad:2017ApSS}
A.~{Jawad}, S.~{Rani} and M.~{Saleem}, \emph{{Cosmological study of
  reconstructed f(T) models}},
  \href{https://doi.org/10.1007/s10509-017-3040-0}{\emph{Astrophys. Space Sci.}
  {\bfseries 362} (Apr., 2017) 63}.


\bibitem{Capozziello:2017bxm}
S.~Capozziello, G.~Lambiase and E.~N. Saridakis, \emph{{Constraining f(T)
  teleparallel gravity by Big Bang Nucleosynthesis}},
  \href{https://doi.org/10.1140/epjc/s10052-017-5143-8}{\emph{Eur. Phys. J. C}
  {\bfseries 77} (2017) 576},
  [\href{https://arxiv.org/abs/1702.07952}{{\ttfamily 1702.07952}}].

\bibitem{Karpathopoulos:2017arc}
L.~Karpathopoulos, S.~Basilakos, G.~Leon, A.~Paliathanasis and M.~Tsamparlis,
  \emph{{Cartan symmetries and global dynamical systems analysis in a
  higher-order modified teleparallel theory}},
  \href{https://arxiv.org/abs/1709.02197}{{\ttfamily 1709.02197}}.

\bibitem{Miao:2011ki}
R.-X. Miao, M.~Li and Y.-G. Miao, \emph{{Violation of the first law of black
  hole thermodynamics in $f(T)$ gravity}},
  \href{https://doi.org/10.1088/1475-7516/2011/11/033}{\emph{JCAP} {\bfseries
  1111} (2011) 033}, [\href{https://arxiv.org/abs/1107.0515}{{\ttfamily
  1107.0515}}].

\bibitem{Capozziello:2012zj}
S.~Capozziello, P.~A. Gonzalez, E.~N. Saridakis and Y.~Vasquez, \emph{{Exact
  charged black-hole solutions in D-dimensional f(T) gravity: torsion vs
  curvature analysis}},
  \href{https://doi.org/10.1007/JHEP02(2013)039}{\emph{JHEP} {\bfseries 02}
  (2013) 039}, [\href{https://arxiv.org/abs/1210.1098}{{\ttfamily 1210.1098}}].

\bibitem{Bhadra:2013IJTP}
J.~{Bhadra} and U.~{Debnath}, \emph{{Primordial Black Holes Evolution in f(T)
  Gravity}},
  \href{https://doi.org/10.1007/s10773-013-1852-4}{\emph{International Journal
  of Theoretical Physics} (Oct., 2013) }.

\bibitem{Rodrigues:2013IJMPD}
M.~E. {Rodrigues}, M.~J.~S. {Houndjo}, D.~{Momeni} and R.~{Myrzakulov},
  \emph{{Planar Symmetry in f(T) Gravity}},
  \href{https://doi.org/10.1142/S0218271813500430}{\emph{International Journal
  of Modern Physics D} {\bfseries 22} (July, 2013) 1350043},
  [\href{https://arxiv.org/abs/1302.4372}{{\ttfamily 1302.4372}}].

\bibitem{Aftergood:2014wla}
J.~Aftergood and A.~DeBenedictis, \emph{{Matter conditions for regular black
  holes in $f(T)$ gravity}},
  \href{https://doi.org/10.1103/PhysRevD.90.124006}{\emph{Phys. Rev. D}
  {\bfseries 90} (2014) 124006},
  [\href{https://arxiv.org/abs/1409.4084}{{\ttfamily 1409.4084}}].

\bibitem{Paliathanasis:2014iva}
A.~Paliathanasis, S.~Basilakos, E.~N. Saridakis, S.~Capozziello, K.~Atazadeh,
  F.~Darabi et~al., \emph{{New Schwarzschild-like solutions in f(T) gravity
  through Noether symmetries}},
  \href{https://doi.org/10.1103/PhysRevD.89.104042}{\emph{Phys. Rev. D}
  {\bfseries 89} (2014) 104042},
  [\href{https://arxiv.org/abs/1402.5935}{{\ttfamily 1402.5935}}].

\bibitem{Nashed:2015JPSJ}
G.~G.~L. {Nashed}, \emph{{Kerr-Newman-NUT Black Hole in f(T) Gravity Theory and
  Its Thermodynamical Quantities}},
  \href{https://doi.org/10.7566/JPSJ.84.044006}{\emph{Journal of the Physical
  Society of Japan} {\bfseries 84} (Apr., 2015) 044006}.

\bibitem{Junior:2015dga}
E.~L.~B. Junior, M.~E. Rodrigues and M.~J.~S. Houndjo, \emph{{Born-Infeld and
  Charged Black Holes with non-linear source in $f(T)$ Gravity}},
  \href{https://doi.org/10.1088/1475-7516/2015/06/037}{\emph{JCAP} {\bfseries
  1506} (2015) 037}, [\href{https://arxiv.org/abs/1503.07427}{{\ttfamily
  1503.07427}}].

\bibitem{Nashed:2015EPJP}
G.~G.~L. {Nashed}, \emph{{Kerr-NUT black hole thermodynamics in f(T) gravity
  theories}}, \href{https://doi.org/10.1140/epjp/i2015-15124-3}{\emph{European
  Physical Journal Plus} {\bfseries 130} (July, 2015) 124}.

\bibitem{Junior:2015fya}
E.~L.~B. Junior, M.~E. Rodrigues and M.~J.~S. Houndjo, \emph{{Regular black
  holes in $f(T)$ Gravity through a nonlinear electrodynamics source}},
  \href{https://doi.org/10.1088/1475-7516/2015/10/060}{\emph{JCAP} {\bfseries
  1510} (2015) 060}, [\href{https://arxiv.org/abs/1503.07857}{{\ttfamily
  1503.07857}}].

\bibitem{Nashed:2016tbj}
G.~G.~L. Nashed and W.~El~Hanafy, \emph{{Analytic rotating black hole solutions
  in $N$-dimensional $f(T)$ gravity}},
  \href{https://doi.org/10.1140/epjc/s10052-017-4663-6}{\emph{Eur. Phys. J. C}
  {\bfseries 77} (2017) 90},
  [\href{https://arxiv.org/abs/1612.05106}{{\ttfamily 1612.05106}}].

\bibitem{Ahmed:2016cuy}
A.~K. Ahmed, M.~Azreg-Aïnou, S.~Bahamonde, S.~Capozziello and M.~Jamil,
  \emph{{Astrophysical flows near $f\,\,(T)$ gravity black holes}},
  \href{https://doi.org/10.1140/epjc/s10052-016-4118-5}{\emph{Eur. Phys. J. C}
  {\bfseries 76} (2016) 269},
  [\href{https://arxiv.org/abs/1602.03523}{{\ttfamily 1602.03523}}].

\bibitem{Rodrigues:2016uor}
M.~E. Rodrigues and E.~L.~B. Junior, \emph{{Spherical Accretion of Matter by
  Charged Black Holes on f(T) Gravity}},
  \href{https://arxiv.org/abs/1606.04918}{{\ttfamily 1606.04918}}.

\bibitem{Nashed:2017GrCo}
G.~G.~L. {Nashed}, \emph{{(1+4)-dimensional spherically symmetric black holes
  in f(T)}}, \href{https://doi.org/10.1134/S0202289317010121}{\emph{Gravitation
  and Cosmology} {\bfseries 23} (Jan., 2017) 63--69}.

\bibitem{Mai:2017riq}
Z.-F. Mai and H.~Lu, \emph{{Black Holes, Dark Wormholes and Solitons in f(T)
  Gravities}}, \href{https://doi.org/10.1103/PhysRevD.95.124024}{\emph{Phys.
  Rev. D} {\bfseries 95} (2017) 124024},
  [\href{https://arxiv.org/abs/1704.05919}{{\ttfamily 1704.05919}}].

\bibitem{Awad:2017tyz}
A.~M. Awad, S.~Capozziello and G.~G.~L. Nashed, \emph{{$D$-dimensional charged
  Anti-de-Sitter black holes in $f(T)$ gravity}},
  \href{https://doi.org/10.1007/JHEP07(2017)136}{\emph{JHEP} {\bfseries 07}
  (2017) 136}, [\href{https://arxiv.org/abs/1706.01773}{{\ttfamily
  1706.01773}}].

\bibitem{Capozziello:2017uam}
S.~Capozziello, R.~D'Agostino and O.~Luongo, \emph{{Model-independent
  reconstruction of $f(T)$ teleparallel cosmology}},
  \href{https://doi.org/10.1007/s10714-017-2304-x}{\emph{General Relativity and
  Gravitation} {\bfseries 49} (2017) 141},
  [\href{https://arxiv.org/abs/1706.02962}{{\ttfamily 1706.02962}}].

\bibitem{Copeland:1997et}
E.~J. Copeland, A.~R. Liddle and D.~Wands, \emph{{Exponential potentials and
  cosmological scaling solutions}},
  \href{https://doi.org/10.1103/PhysRevD.57.4686}{\emph{Phys. Rev. D}
  {\bfseries 57} (1998) 4686--4690},
  [\href{https://arxiv.org/abs/gr-qc/9711068}{{\ttfamily gr-qc/9711068}}].

\bibitem{Ferreira:1997au}
P.~G. Ferreira and M.~Joyce, \emph{{Structure formation with a selftuning
  scalar field}},
  \href{https://doi.org/10.1103/PhysRevLett.79.4740}{\emph{Phys. Rev. Lett.}
  {\bfseries 79} (1997) 4740--4743},
  [\href{https://arxiv.org/abs/astro-ph/9707286}{{\ttfamily
  astro-ph/9707286}}].

\bibitem{Coley:2003mj}
A.~A.~C. (auth.), \emph{Dynamical Systems and Cosmology}, vol.~291 of
  \emph{Astrophysics and Space Science Library}.
\newblock Springer Netherlands, 1~ed., 2003,
  \href{https://doi.org/10.1007/978-94-017-0327-7}{10.1007/978-94-017-0327-7}.

\bibitem{Leon2011}
G.~Leon and C.~R. Fadragas, \emph{{Cosmological dynamical systems}}.
\newblock LAP Lambert Academic Publishing, 2012.

\bibitem{Coley:2003ASSL}
A.~A. {Coley}, ed., \emph{{Dynamical Systems and Cosmology}}, vol.~291 of
  \emph{Astrophysics and Space Science Library}, Oct., 2003.
\newblock 10.1007/978-94-017-0327-7.

\bibitem{Ellis:2005CUP}
J.~{Wainwright} and G.~F.~R. {Ellis}, \emph{Dynamical Systems in Cosmology}.
\newblock Cambridge University Press, June, 2005,
  \href{https://doi.org/10.1017/CBO9780511524660}{10.1017/CBO9780511524660}.

\bibitem{Chen:2008ft}
X.-m. Chen, Y.-g. Gong and E.~N. Saridakis, \emph{{Phase-space analysis of
  interacting phantom cosmology}},
  \href{https://doi.org/10.1088/1475-7516/2009/04/001}{\emph{JCAP} {\bfseries
  0904} (2009) 001}, [\href{https://arxiv.org/abs/0812.1117}{{\ttfamily
  0812.1117}}].

\bibitem{awad2013}
A.~{Awad}, \emph{{Fixed points and FLRW cosmologies: Flat case}},
  \href{https://doi.org/10.1103/PhysRevD.87.103001}{\emph{Phys. Rev. D}
  {\bfseries 87} (May, 2013) 103001},
  [\href{https://arxiv.org/abs/1303.2014}{{\ttfamily 1303.2014}}].

\bibitem{Boehmer:2014vea}
C.~G. Boehmer and N.~Chan, \emph{{Dynamical systems in cosmology}},
  \href{https://arxiv.org/abs/1409.5585}{{\ttfamily 1409.5585}}.

\bibitem{Kofinas:2014aka}
G.~Kofinas, G.~Leon and E.~N. Saridakis, \emph{{Dynamical behavior in
  $f(T,T_G)$ cosmology}},
  \href{https://doi.org/10.1088/0264-9381/31/17/175011}{\emph{Class. Quant.
  Grav.} {\bfseries 31} (2014) 175011},
  [\href{https://arxiv.org/abs/1404.7100}{{\ttfamily 1404.7100}}].

\bibitem{Odintsov:2017icc}
S.~D. Odintsov, V.~K. Oikonomou and P.~V. Tretyakov, \emph{{Phase space
  analysis of the accelerating multifluid Universe}},
  \href{https://doi.org/10.1103/PhysRevD.96.044022}{\emph{Phys. Rev. D}
  {\bfseries 96} (2017) 044022},
  [\href{https://arxiv.org/abs/1707.08661}{{\ttfamily 1707.08661}}].

\bibitem{Wu:2010xk}
P.~Wu and H.~W. Yu, \emph{{The dynamical behavior of $f(T)$ theory}},
  \href{https://doi.org/10.1016/j.physletb.2010.07.038}{\emph{Phys. Lett. B}
  {\bfseries 692} (2010) 176--179},
  [\href{https://arxiv.org/abs/1007.2348}{{\ttfamily 1007.2348}}].

\bibitem{Skugoreva:2014ena}
M.~A. Skugoreva, E.~N. Saridakis and A.~V. Toporensky, \emph{{Dynamical
  features of scalar-torsion theories}},
  \href{https://doi.org/10.1103/PhysRevD.91.044023}{\emph{Phys. Rev. D}
  {\bfseries 91} (2015) 044023},
  [\href{https://arxiv.org/abs/1412.1502}{{\ttfamily 1412.1502}}].

\bibitem{Carloni:2015lsa}
S.~Carloni, F.~S.~N. Lobo, G.~Otalora and E.~N. Saridakis, \emph{{Dynamical
  system analysis for a nonminimal torsion-matter coupled gravity}},
  \href{https://doi.org/10.1103/PhysRevD.93.024034}{\emph{Phys. Rev. D}
  {\bfseries 93} (2016) 024034},
  [\href{https://arxiv.org/abs/1512.06996}{{\ttfamily 1512.06996}}].

\bibitem{Bamba:2016gbu}
K.~Bamba, G.~G.~L. Nashed, W.~El~Hanafy and S.~K. Ibraheem, \emph{{Bounce
  inflation in $f(T)$ Cosmology: A unified inflaton-quintessence field}},
  \href{https://doi.org/10.1103/PhysRevD.94.083513}{\emph{Phys. Rev. D}
  {\bfseries 94} (2016) 083513},
  [\href{https://arxiv.org/abs/1604.07604}{{\ttfamily 1604.07604}}].

\bibitem{Awad:2017sau}
A.~Awad and G.~Nashed, \emph{{Generalized teleparallel cosmology and initial
  singularity crossing}},
  \href{https://doi.org/10.1088/1475-7516/2017/02/046}{\emph{JCAP} {\bfseries
  1702} (2017) 046}, [\href{https://arxiv.org/abs/1701.06899}{{\ttfamily
  1701.06899}}].

\bibitem{ElHanafy:2017sih}
W.~El~Hanafy and G.~G.~L. Nashed, \emph{{Lorenz Gauge Fixing of $f(T)$
  Teleparallel Cosmology}},
  \href{https://doi.org/10.1142/S0218271817501541}{\emph{Int. J. Mod. Phys. D}
  {\bfseries 26} (2017) 1750154},
  [\href{https://arxiv.org/abs/1707.01802}{{\ttfamily 1707.01802}}].

\bibitem{ElHanafy:2017xsm}
W.~El~Hanafy and G.~G.~L. Nashed, \emph{{Generic Phase Portrait Analysis of the
  Finite-time Singularities and Generalized Teleparallel Gravity}},
  \href{https://doi.org/10.1088/1674-1137/41/12/125103}{\emph{Chin. Phys.}
  (2017) 125103}, [\href{https://arxiv.org/abs/1702.05786}{{\ttfamily
  1702.05786}}].

\bibitem{Hohmann:2017jao}
M.~Hohmann, L.~Jarv and U.~Ualikhanova, \emph{{Dynamical systems approach and
  generic properties of $f(T)$ cosmology}},
  \href{https://doi.org/10.1103/PhysRevD.96.043508}{\emph{Phys. Rev. D}
  {\bfseries 96} (2017) 043508},
  [\href{https://arxiv.org/abs/1706.02376}{{\ttfamily 1706.02376}}].

\bibitem{Mirza:2017vrk}
B.~Mirza and F.~Oboudiat, \emph{{Constraining f(T) gravity by dynamical system
  analysis}},  \href{https://arxiv.org/abs/1704.02593}{{\ttfamily 1704.02593}}.

\bibitem{BF09}
G.~R. {Bengochea} and R.~{Ferraro}, \emph{{Dark torsion as the cosmic
  speed-up}}, \href{https://doi.org/10.1103/PhysRevD.79.124019}{\emph{Phys.
  Rev. D} {\bfseries 79} (June, 2009) 124019},
  [\href{https://arxiv.org/abs/0812.1205}{{\ttfamily 0812.1205}}].

\bibitem{L10}
E.~V. {Linder}, \emph{{Einstein's other gravity and the acceleration of the
  Universe}}, \href{https://doi.org/10.1103/PhysRevD.81.127301}{\emph{Phys.
  Rev. D} {\bfseries 81} (June, 2010) 127301},
  [\href{https://arxiv.org/abs/1005.3039}{{\ttfamily 1005.3039}}].

\bibitem{Li:2011rn}
M.~Li, R.-X. Miao and Y.-G. Miao, \emph{{Degrees of freedom of $f(T)$
  gravity}}, \href{https://doi.org/10.1007/JHEP07(2011)108}{\emph{JHEP}
  {\bfseries 07} (2011) 108},
  [\href{https://arxiv.org/abs/1105.5934}{{\ttfamily 1105.5934}}].

\bibitem{book:Steven}
S.~H. Strogatz, \emph{Nonlinear Dynamics And Chaos: With Applications To
  Physics, Biology, Chemistry And Engineering}, vol.~1 of \emph{Studies in
  Nonlinearity}.
\newblock 1994.

\bibitem{Nojiri:2005sx}
S.~Nojiri, S.~D. Odintsov and S.~Tsujikawa, \emph{{Properties of singularities
  in (phantom) dark energy universe}},
  \href{https://doi.org/10.1103/PhysRevD.71.063004}{\emph{Phys. Rev. D}
  {\bfseries 71} (2005) 063004},
  [\href{https://arxiv.org/abs/hep-th/0501025}{{\ttfamily hep-th/0501025}}].

\bibitem{Barrow:1986}
J.~D. {Barrow}, G.~J. {Galloway} and F.~J. {Tipler}, \emph{{The closed-universe
  recollapse conjecture}},
  \href{https://doi.org/10.1093/mnras/223.4.835}{\emph{Mon. Not. Roy. Astr.
  Soc.} {\bfseries 223} (Dec., 1986) 835--844}.

\bibitem{Barrow:2004xh}
J.~D. Barrow, \emph{{Sudden future singularities}},
  \href{https://doi.org/10.1088/0264-9381/21/11/L03}{\emph{Class. Quant. Grav.}
  {\bfseries 21} (2004) L79--L82},
  [\href{https://arxiv.org/abs/gr-qc/0403084}{{\ttfamily gr-qc/0403084}}].

\bibitem{Keresztes:2012zn}
Z.~Keresztes, L.~A. Gergely and A.~{\relax Yu}. Kamenshchik, \emph{{The paradox
  of soft singularity crossing and its resolution by distributional
  cosmological quantitities}},
  \href{https://doi.org/10.1103/PhysRevD.86.063522}{\emph{Phys. Rev. D}
  {\bfseries 86} (2012) 063522},
  [\href{https://arxiv.org/abs/1204.1199}{{\ttfamily 1204.1199}}].

\bibitem{Awad:2015syb}
A.~Awad, \emph{{Weyl Anomaly and Initial Singularity Crossing}},
  \href{https://doi.org/10.1103/PhysRevD.93.084006}{\emph{Phys. Rev. D}
  {\bfseries 93} (2016) 084006},
  [\href{https://arxiv.org/abs/1512.06405}{{\ttfamily 1512.06405}}].

\bibitem{Cai:2011bs}
Y.-F. Cai and E.~N. Saridakis, \emph{{Non-singular Cyclic Cosmology without
  Phantom Menace}}, {\emph{J. Cosmol.} {\bfseries 17} (2011) 7238--7254},
  [\href{https://arxiv.org/abs/1108.6052}{{\ttfamily 1108.6052}}].

\bibitem{Nojiri:2013ru}
S.~Nojiri and E.~N. Saridakis, \emph{{Phantom without ghost}},
  \href{https://doi.org/10.1007/s10509-013-1509-z}{\emph{Astrophys. Space Sci.}
  {\bfseries 347} (2013) 221--226},
  [\href{https://arxiv.org/abs/1301.2686}{{\ttfamily 1301.2686}}].

\bibitem{Novello:2008ra}
M.~Novello and S.~E.~P. Bergliaffa, \emph{{Bouncing Cosmologies}},
  \href{https://doi.org/10.1016/j.physrep.2008.04.006}{\emph{Phys. Rept.}
  {\bfseries 463} (2008) 127--213},
  [\href{https://arxiv.org/abs/0802.1634}{{\ttfamily 0802.1634}}].

\bibitem{Cai:2009in}
Y.-F. Cai and E.~N. Saridakis, \emph{{Non-singular cosmology in a model of
  non-relativistic gravity}},
  \href{https://doi.org/10.1088/1475-7516/2009/10/020}{\emph{JCAP} {\bfseries
  0910} (2009) 020}, [\href{https://arxiv.org/abs/0906.1789}{{\ttfamily
  0906.1789}}].

\bibitem{Fern:2006AIPC}
L.~{Fern{\'a}ndez-Jambrina} and R.~{Lazkoz}, \emph{{Geodesic Completeness
  around Sudden Singularities}},  in \emph{A Century of Relativity Physics: ERE
  2005} (L.~{Mornas} and J.~{Diaz Alonso}, eds.), vol.~841 of \emph{American
  Institute of Physics Conference Series}, pp.~420--423, June, 2006,
  \href{https://arxiv.org/abs/0903.5529}{{\ttfamily 0903.5529}},
  \href{https://doi.org/10.1063/1.2218204}{DOI}.

\bibitem{Fern:2006PRD}
L.~{Fern{\'a}ndez-Jambrina} and R.~{Lazkoz}, \emph{{Classification of
  cosmological milestones}},
  \href{https://doi.org/10.1103/PhysRevD.74.064030}{\emph{Phys. Rev. D}
  {\bfseries 74} (Sept., 2006) 064030},
  [\href{https://arxiv.org/abs/gr-qc/0607073}{{\ttfamily gr-qc/0607073}}].

\bibitem{Barrow:2013PRD}
J.~D. {Barrow} and S.~{Cotsakis}, \emph{{Geodesics at sudden singularities}},
  \href{https://doi.org/10.1103/PhysRevD.88.067301}{\emph{Phys. Rev. D}
  {\bfseries 88} (Sept., 2013) 067301},
  [\href{https://arxiv.org/abs/1307.5005}{{\ttfamily 1307.5005}}].

\bibitem{Kamenshchik:2013ink}
A.~Kamenshchik, Z.~Keresztes and L.~A. Gergely, \emph{{The paradox of soft
  singularity crossing avoided by distributional cosmological quantities}},  in
  \emph{{Proceedings, 13th Marcel Grossmann Meeting on Recent Developments in
  Theoretical and Experimental General Relativity, Astrophysics, and
  Relativistic Field Theories (MG13): Stockholm, Sweden, July 1-7, 2012}},
  2015.

\bibitem{Gergely:2013via}
L.~A. Gergely, Z.~Keresztes and A.~{\relax Yu}. Kamenshchik,
  \emph{{Distributional cosmological quantities solve the paradox of soft
  singularity crossing}}, \href{https://doi.org/10.1063/1.4791740}{\emph{AIP
  Conf. Proc.} {\bfseries 1514} (2013) 132--135},
  [\href{https://arxiv.org/abs/1304.1415}{{\ttfamily 1304.1415}}].

\bibitem{Nesseris:2013jea}
S.~Nesseris, S.~Basilakos, E.~N. Saridakis and L.~Perivolaropoulos,
  \emph{{Viable $f(T)$ models are practically indistinguishable from
  $\Lambda$CDM}}, \href{https://doi.org/10.1103/PhysRevD.88.103010}{\emph{Phys.
  Rev. D} {\bfseries 88} (2013) 103010},
  [\href{https://arxiv.org/abs/1308.6142}{{\ttfamily 1308.6142}}].

\bibitem{Iorio:2012cm}
L.~Iorio and E.~N. Saridakis, \emph{{Solar system constraints on f(T)
  gravity}}, \href{https://doi.org/10.1111/j.1365-2966.2012.21995.x}{\emph{Mon.
  Not. Roy. Astron. Soc.} {\bfseries 427} (2012) 1555},
  [\href{https://arxiv.org/abs/1203.5781}{{\ttfamily 1203.5781}}].

\bibitem{Ade:2015xua}
{\scshape Planck} collaboration, P.~A.~R. Ade et~al., \emph{{Planck 2015
  results. XIII. Cosmological parameters}},
  \href{https://doi.org/10.1051/0004-6361/201525830}{\emph{Astron. Astrophys.}
  {\bfseries 594} (2016) }, [\href{https://arxiv.org/abs/1502.01589}{{\ttfamily
  1502.01589}}].

\bibitem{BGLL2011}
K.~{Bamba}, C.-Q. {Geng}, C.-C. {Lee} and L.-W. {Luo}, \emph{{Equation of state
  for dark energy in f(T) gravity}},
  \href{https://doi.org/10.1088/1475-7516/2011/01/021}{\emph{J. Cosmo.
  Astropart. Phys.} {\bfseries 1} (Jan., 2011) 21},
  [\href{https://arxiv.org/abs/1011.0508}{{\ttfamily 1011.0508}}].

\bibitem{Cepa:2004bc}
J.~Cepa, \emph{{Constraints on the cosmic equation of state: Age conflict
  versus phantom energy. Age - redshift relations in an accelerated universe}},
  \href{https://doi.org/10.1051/0004-6361:20035734}{\emph{Astron. Astrophys.}
  {\bfseries 422} (2004) 831--839},
  [\href{https://arxiv.org/abs/astro-ph/0403616}{{\ttfamily
  astro-ph/0403616}}].

\bibitem{Mukherjee:2016eqj}
A.~Mukherjee, \emph{{Acceleration of the universe: a reconstruction of the
  effective equation of state}},
  \href{https://doi.org/10.1093/mnras/stw964}{\emph{Mon. Not. Roy. Astron.
  Soc.} {\bfseries 460} (2016) 273--282},
  [\href{https://arxiv.org/abs/1605.08184}{{\ttfamily 1605.08184}}].

\bibitem{Sahni:2008xx}
V.~Sahni, A.~Shafieloo and A.~A. Starobinsky, \emph{{Two new diagnostics of
  dark energy}}, \href{https://doi.org/10.1103/PhysRevD.78.103502}{\emph{Phys.
  Rev. D} {\bfseries 78} (2008) 103502},
  [\href{https://arxiv.org/abs/0807.3548}{{\ttfamily 0807.3548}}].

\bibitem{Lonappan:2017hok}
A.~I. Lonappan, Ruchika and A.~A. Sen, \emph{{Is it time to go beyond
  $\Lambda$CDM universe?}},  \href{https://arxiv.org/abs/1705.07336}{{\ttfamily
  1705.07336}}.

\end{thebibliography}

\providecommand{\href}[2]{#2}\begingroup\raggedright\endgroup
\end{document}